

\magnification=\magstep1
\hsize = 33pc
\vsize = 46pc
\baselineskip=12pt
\tolerance 8000
\parskip=5pt

\font\grand=cmbx10 at 14.4truept
\font\grand=cmbx10 at 14.4truept\bigskip
\font\ninebf=cmbx9
\font\ninerm=cmr9

\def\q#1{[#1]}              
\def\bibitem#1{\parindent=8mm\item{\hbox to 6 mm{\q{#1}\hfill}}}
\font\extra=cmss10 scaled \magstep0
\setbox1 = \hbox{{{\extra R}}}
\setbox2 = \hbox{{{\extra I}}}
\setbox3 = \hbox{{{\extra C}}}
\def\RRR{{{\extra R}}\hskip-\wd1\hskip2.0 true pt{{\extra I}}\hskip-\wd2
\hskip-2.0 true pt\hskip\wd1}
\def\Real{\hbox{{\extra\RRR}}}    
\def\C{{{\extra C}}\hskip-\wd3\hskip2.5 true pt{{\extra I}}\hskip-\wd2
\hskip-2.5 true pt\hskip\wd3}
\def\Complex{\hbox{{\extra\C}}}   
\setbox4=\hbox{{{\extra Z}}}
\def\Z{{{\extra Z}}\hskip-\wd4\hskip 2.5 true pt{{\extra Z}}}
\def\Zed{\hbox{{\extra\Z}}}       

\def\pa{\partial}
\def\tr{{\rm tr\,}}
\def\Tr{{\rm Tr\,}}

\def\M{{\cal M}}
\def\G{{\cal G}}
\def\M{{\cal M}}
\def\W{{\cal W}}
\def\G{{\cal G}}
\def\L{{\cal L}}
\def\N{{\cal N}}
\def\mc{{\cal M}_{\rm c}}
\def\mr{{\cal M}_{\rm red}}
\def\mds{{\cal M}_{\rm DS}}\def\pa{\partial}
\def\paw{(\partial+w)^{-1}}
\def\tG{{\widetilde{\cal G}}}

\def\hatj{{\hat \jmath}}
\def\half{{1/2}}
\def\vt{{\vartheta}}
\def\cinf{{C^\infty}}
\def\wt{\widetilde}
\def\mat{{\rm mat}}
\def\hatj{{\hat \jmath}}
\def\half{{1/2}}

\def\bs{\baselineskip=14.5pt}

\pageno=0
\def\folio{
\ifnum\pageno<1 \footline{\hfil} \else\number\pageno \fi}

\baselineskip=12pt
\phantom{not-so-FUNNY}
\rightline{ SWAT-95-61\break}
\rightline{ March 1995\break}
\rightline{ hep-th/9503217\break}
\vskip 1.0truecm

\bs


\vskip 0.8truecm

\centerline
{\grand  Extensions of the matrix Gelfand-Dickey hierarchy}

\centerline{\grand from generalized Drinfeld-Sokolov reduction}

\vskip 0.8truecm

\centerline{
L\'aszl\'o Feh\'er${}^{a,}$\footnote*{\ninerm
On leave from Theoretical Physics Department  of Szeged University,
 H-6720 Szeged, Hungary.}
\  and \  Ian Marshall${}^{b}$}

\bigskip

\centerline{\it ${}^{a}$Department of Physics, University of Wales  Swansea,}
\centerline{\it Singleton Park, Swansea SA2 8PP, U.K.}
\centerline{\it e-mail: L.Feher@swansea.ac.uk}

\medskip

\centerline{\it ${}^{b}$Department of Mathematics, Leeds University,}
 \centerline{\it Leeds LS2 9JT, U.K.}
\centerline{\it e-mail: amt6im@amsta.leeds.ac.uk}

\vskip 0.8cm

{\parindent=25pt
\narrower\smallskip\noindent
\ninerm
{\ninebf Abstract.}\quad
The $p\times p$ matrix version of the  $r$-KdV hierarchy has been
recently treated as the reduced system arising in  a
Drinfeld-Sokolov type Hamiltonian  symmetry reduction applied to
a Poisson submanifold in the dual of the Lie algebra
$\widehat{gl}_{pr}\otimes {\Complex}[\lambda, \lambda^{-1}]$.
Here a series of extensions of this
matrix Gelfand-Dickey  system is derived
by means of a  generalized Drinfeld-Sokolov
reduction defined for the Lie algebra
$\widehat{gl}_{pr+s}\otimes {\Complex}[\lambda,\lambda^{-1}]$
 using the natural embedding
$gl_{pr}\subset gl_{pr+s}$ for $s$ any positive integer.
The hierarchies  obtained admit a
description in terms of a $p\times p$ matrix pseudo-differential operator
comprising an $r$-KdV type positive part and a non-trivial negative part.
This system has been investigated previously  in the $p=1$ case
as a constrained  KP system.
In this paper  the previous results are considerably extended
and a systematic study is presented on the basis of
the Drinfeld-Sokolov approach that
has the advantage that it leads to local Poisson brackets
and makes clear the conformal ($\cal W$-algebra) structures
related to the KdV type  hierarchies.
Discrete reductions  and modified versions of the extended
$r$-KdV hierarchies  are also discussed.
}

\vfill\eject

\centerline{\bf Contents}
\vskip 7mm

\item{0.}
Introduction
\dotfill 2

\vskip 2.5 mm
\item{1.}
A Hamiltonian reduction approach to KdV type hierarchies
\dotfill 6

\vskip 2.5 mm
\item{2.}
A generalized Drinfeld-Sokolov reduction
\dotfill 10

\vskip 2.5mm
\item{3.}
Residual symmetries
\dotfill 15

\vskip 2.5mm
\item{4.}
The Poisson brackets on the reduced phase space
\dotfill 16

\vskip 2.5mm
\item{5.}
Local monodromy invariants and residues of fractional powers
\dotfill 26
\item{}
5.1. Local monodromy invariants and solutions of exponential type
\dotfill 26
\item{}
5.2. Solutions of exponential type and residues of fractional powers
\dotfill 31

\vskip 2.5mm
\item{6.}
Examples for $r=2$ and  factorizations of the AKNS factor
\dotfill 36
\item{}
6.1. The reduced second PB, Miura maps and factorizations
\dotfill 37
\item{}
6.2. Some explicit examples
\dotfill 42

\vskip 2.5mm
\item{7.} Discussion: discrete reductions and generalized KdV hierarchies
\dotfill 44

\vskip 2.5mm
\noindent
\hskip 2mm
Appendices
\item{}
A: The Poisson submanifold $M_K\subset {\cal A}$
\dotfill 49
\item{}
B: The formula of the reduced PB in the DS gauge
\dotfill 52

\vskip 2.5mm
\noindent
\hskip 2mm
References
\dotfill 57

\vfill\eject


\def\FHM{{\bf FHM}}
\def\GD{{\bf GD}}
\def\Man{{\bf Man}}
\def\DS{{\bf DS}}
\def\GHM{{\bf GHM}}
\def\BGHM{{\bf BGHM}}
\def\RSTS{{\bf RSTS}}
\def\FORT{{\bf FORT}}
\def\Cheng{{\bf Cheng}}
\def\OS{{\bf OS}}
\def\Di{{\bf Di}}
\def\BEHHH{{\bf BEHHH}}
\def\Kac{{\bf Kac}}
\def\KP{{\bf KP}}
\def\FGMG{{\bf FGMG}}
\def\DF{{\bf DF}}
\def\FORTW{{\bf FORTW}}
\def\BTvD{{\bf BTvD}}
\def\A{{\bf A}}
\def\Dic{{\bf Dic}}
\def\De{{\bf De}}
\def\FK{{\bf FK}}
\def\FM+{{\bf FM+}}
\def\FW{{\bf FW}}
\def\Wi{{\bf Wi}}
\def\Che{{\bf Che}}
\def\Fla{{\bf Fla}}
\def\BF{{\bf BF}}
\def\Sp{{\bf Sp}}
\def\Wil{{\bf Wil}}
\def\McI{{\bf McI}}
\def\Car{{\bf Car}}

\def\FHM{1}
\def\GD{2}
\def\Man{3}
\def\DS{4}
\def\Wil{5}
\def\McI{6}
\def\GHM{7}
\def\BGHM{8}
\def\FGMG{9}
\def\RSTS{10}
\def\FORT{11}
\def\Cheng{12}
\def\OS{13}
\def\Di{14}
\def\BEHHH{15}
\def\Kac{16}
\def\KP{17}
\def\Sp{18}
\def\DF{19}
\def\FORTW{20}
\def\BTvD{21}
\def\A{22}
\def\Dic{23}
\def\De{24}
\def\FW{25}
\def\FK{26}
\def\Wi{27}
\def\Wils{28}
\def\Che{29}
\def\Fla{30}
\def\BF{31}
\def\Car{32}

\bs

\centerline{\bf 0.~Introduction}
\medskip

This paper is a continuation of \q{\FHM}, where it was shown
how the matrix  Gelfand-Dickey hierarchy \q{\GD,\Man} fits
into the Drinfeld-Sokolov approach \q{\DS}
(see also \q{\Wil,\McI,\GHM,\BGHM,\FGMG}) to generalized
KdV hierarchies.
The phase space of this  hierarchy is the space
of $p\times p$ matrix Lax operators
$$
L_{p,r}= P \pa^r + u_1 \pa^{r-1}+ \cdots +u_{r-1} \pa+  u_r,
\quad
u_i\in \cinf(S^1,gl_p),
\eqno(0.1)$$
where $P$ is a diagonal constant matrix with distinct,
non-zero entries.
This phase space has two compatible Poisson brackets:
the linear and quadratic matrix Gelfand-Dickey Poisson brackets.
The Hamiltonians generating a commuting hierarchy
of bihamiltonian flows are obtained from
the residues of the componentwise fractional powers of the
 $p\times p$ diagonal matrix  pseudo-differential operator $\hat L_{p,r}$
determined by diagonalizing $L_{p,r}$ in the algebra of matrix
pseudo-differential operators.
This system arises from a Drinfeld-Sokolov type Hamiltonian
symmetry reduction applied to a Poisson submanifold in
the dual of the
Lie algebra $\widehat{gl}_{pr}\otimes {\Complex}[\lambda, \lambda^{-1}]$ ---
where $\widehat{gl}_{pr}$ is the central extension of the loop algebra
$\widetilde{gl}_{pr}=\cinf(S^1,{gl}_{pr})$ ---
endowed with the family of compatible Poisson brackets and commuting
Hamiltonians provided by the r-matrix  (AKS)
construction (see e.g.~\q{\RSTS}).
The corresponding reduced phase space is identified with the set of first
order matrix
differential operators ${\cal L}_{p,r}$ of the form
$$
{\cal L}_{p,r}={\bf 1}_{pr} \pa + j_{p,r} + \Lambda_{p,r},
\eqno(0.2)$$
where $j_{p,r}\in\cinf(S^1,gl_{pr})$ and
$\Lambda_{p,r} \in gl_{pr}\otimes {\Complex}[\lambda,\lambda^{-1}]$
are written  as $r\times r$ matrices
 with $p\times p$ matrix entries as follows:
$$
j_{p,r}=
\left(\matrix{
0&\cdots&\cdots&0\cr
\vdots&\phantom{\cdots}&\phantom{\cdots}&\vdots\cr
0&\cdots&\cdots &0\cr
v_r&\cdots&\cdots&v_1\cr}\right),
\qquad \quad
\Lambda_{p,r}
=\left(\matrix{
0&\Gamma&0&\cdots&0\cr
\vdots&0&\Gamma&\ddots&\vdots\cr
\vdots&{}&\ddots&\ddots&0\cr
0&{}&{}&\ddots&\Gamma\cr
\lambda \Gamma&0&\cdots&\cdots&0\cr}\right),
\eqno(0.3)$$
where $\Gamma$ is a $p\times p$ diagonal constant matrix for
which $P =(-\Gamma^{-1})^r$; ${\bf1}_{pr}$ is the $pr\times pr$ identity
matrix.
The correspondence between $L_{p,r}$ in (0.1) and ${\cal L}_{p,r}$ in (0.2)
is established through the relation
$$
{\cal L}_{p,r} (\psi_1^t, \psi_2^t, \ldots, \psi_r^t)^t=0\quad
\Leftrightarrow \quad L_{p,r}\psi_1 =\lambda \psi_1,
\eqno(0.4)$$
where the $\psi_i$ are $p$-component column vectors,
yielding $u_i= \Delta v_i \Delta^{r-i}$ with $\Delta:=-\Gamma^{-1}$.

The main purpose of this paper is to derive a series of extensions
of the above  system  using the natural embedding of the Lie algebra
$gl_{pr}$ into $gl_{pr+s}$ for any positive integer $s$.
This embedding is given by writing the general element $m\in gl_{pr+s}$
in the block form
$$
m=\left(\matrix{A&B\cr C&D\cr}\right),
\eqno(0.5)$$
where $A\in gl_{pr}$ is written as
an $r\times r$ matrix with $p\times p$ matrix entries,
$D$ is an $s\times s$ matrix and $B$ (respectively  $C$) is an
$r$-component column (row) vector with $p\times s$ ($s\times p$)
matrix entries.
In particular, the image of $\Lambda_{p,r}$ under this embedding is
$\Lambda_{p,r,s}\in gl_{pr+s}\otimes {\Complex}[\lambda, \lambda^{-1}]$,
$$
\Lambda_{p,r,s}:=\left(\matrix{\Lambda_{p,r}&0\cr 0&{\bf 0}_s\cr}\right).
\eqno(0.6)$$
A generalized  Drinfeld-Sokolov reduction based
on the Lie algebra
$\widehat{gl}_{pr+s}\otimes {\Complex}[\lambda, \lambda^{-1}]$
will be defined in such a way
that the  matrix Gelfand-Dickey system is recovered
when setting all fields outside the
$gl_{pr}\subset gl_{pr+s}$ block to zero.
The corresponding reduced phase space
will turn out to be the set of first order matrix differential operators
${\cal L}_{p,r,s}$ of the form
$$
{\cal L}_{p,r,s}={\bf 1}_{pr+s} \pa + j_{p,r,s} + \Lambda_{p,r,s},
\eqno(0.7)$$
where $j_{p,r,s}$ reads
$$
j_{p,r,s}=
\left(\matrix{
0&\cdots&\cdots &0&0\cr
\vdots&\phantom{\cdots}&\phantom{\cdots}&\vdots&\vdots\cr
0&\cdots&\cdots &0&0\cr
v_r&\cdots&\cdots&v_1&\zeta_+\cr
\zeta_-&0&\cdots&0&w}\right).
\eqno(0.8)$$
The dynamical variables encoded in $j_{p,r,s}$ reduce to those in
$j_{p,r}$ (0.3) upon setting the   fields $\zeta_\pm$,  $w$ to zero.
The phase space whose points are the operators ${\cal L}_{p,r,s}$
will be shown to carry two compatible local Poisson brackets,
which are  naturally
induced by the reduction procedure,
and an infinite family of commuting
local Hamiltonians  defined by  the local monodromy
invariants of ${\cal L}_{p,r,s}$.
One of the Poisson brackets will be identified as a classical
${\cal W}$-algebra (see e.g.~\q{\FORT}).
This phase space can be mapped into the space of $p\times p$
matrix pseudo-differential operators with the aid of the
usual   elimination procedure:
$$
{\cal L}_{p,r,s} (\psi_1^t, \psi_2^t, \ldots, \psi_r^t, \phi^t)^t=0
\quad
\Leftrightarrow \quad
L_{p,r,s}\psi_1 =\lambda \psi_1,
\eqno(0.9)$$
which yields
$$
L_{p,r,s}= P \pa^r + u_1 \pa^{r-1}+ \cdots +u_{r-1} \pa+  u_r
+  z_+ ({\bf 1}_s \pa + w)^{-1} z_-,
\eqno(0.10)$$
where $u_i$ is related to $v_i$ as in (0.4) and
$z_+=\Gamma^{-1}\zeta_+$, $z_-=\zeta_-$.
In contrast to the standard case,
the operator $L_{p,r,s}$ attached to ${\cal L}_{p,r,s}$ is now
not a differential operator but contains a non-trivial negative part,
and the mapping ${\cal L}_{p,r,s}\mapsto L_{p,r,s}$ is not a
one-to-one mapping.   In fact, this mapping corresponds to factoring
out a residual $\widehat{gl}_{s}$ symmetry of the hierarchy
resulting from the generalized  Drinfeld-Sokokov reduction,
which is generated by the current $w\in\cinf(S^1,gl_s)$ through
the ${\cal W}$-algebra Poisson bracket.

\vfill\eject

\noindent
Our main results are the following:

First, we shall prove that the set of operators $L_{p,r,s}$  is
a Poisson submanifold
in the space of $p\times p$ matrix pseudo-differential operators
with respect to the compatible Gelfand-Dickey Poisson brackets,
and that the mapping ${\cal L}_{p,r,s}\mapsto L_{p,r,s}$ is
a Poisson mapping from the bihamiltonian manifold
obtained as the reduced phase space in
the Drinfeld-Sokolov reduction onto this Poisson submanifold.
We shall also present the explicit form of the Poisson brackets on
the reduced phase space.

Second, we shall show that the
mapping ${\cal L}_{p,r,s}\mapsto L_{p,r,s}$  converts the commuting
 Hamiltonians determined by
the local monodromy invariants of ${\cal L}_{p,r,s}$ into the Hamiltonians
generated by the residues of the componentwise fractional powers
of the diagonalized form ${\hat L}_{p,r,s}$ of $L_{p,r,s}$.

Third, we shall derive  a ``modified'' version of the
generalized KdV  hierarchy carried by the manifold of operators
${\cal L}_{p,r,s}$, which via the mapping
${\cal L}_{p,r,s}\mapsto L_{p,r,s}$ corresponds to an
interesting factorization of $L_{p,r,s}$ (given in eq.~(4.21)).
One of the factors in this factorization  (the factor $\Delta K$ in (4.21))
arises also independently in the $r=1$ case of our construction,
when there is no Drinfeld-Sokolov  reduction and we are dealing
with a generalized AKNS hierarchy.
By further factorizing this ``AKNS factor'' we shall
obtain a second modification.

In the $p=1$, $s=1$ special case, the system based on the
Lax operator $L_{p,r,s}$ (0.10) has been considered in several
recent papers (see \q{\Cheng,\OS,\Di} and references therein) from
the point of view of
constrained KP hierarchies.
Specifically, the system considered in the literature
may be obtained from (0.10) by setting $w=0$.
Setting $w=0$ is consistent with the flows of the hierarchy, but has
the inconvenient feature  that the resulting Dirac
brackets turn out to be {\it non-local} for the  second
Hamiltonian structure.
It is interesting that one can in this way recover the second Hamiltonian
structure {\it postulated} by Oevel and Strampp \q{\OS},
for $p=s=1$,  as a non-local reduction of the local second Hamiltonian
structure that automatically results from the Hamiltonian reduction in the
Drinfeld-Sokolov approach.
The Drinfeld-Sokolov approach to these systems that we shall present
leads to  a systematic understanding and for this reason we think
it is of interest in its own right. This approach also has clear advantages
in that it leads to  local Poisson brackets and it
makes clear the conformal, $\cal W$-algebra
structures (see e.g.~\q{\FORT})   related to these hierarchies.
It is worth noting that in the $p=1$ case  the quantum mechanical versions
of these $\W$-algebras have been recently found to have
interesting applications in conformal field theory \q{\BEHHH}.

The organization of the paper is apparent from the table of contents.
Section 1 is a brief review of the version of the Drinfeld-Sokolov approach
that will be used.  In Section 2 the generalized
Drinfeld-Sokolov reduction relevant in our case is defined and  the
resulting reduced system is analyzed in terms of convenient gauge slices.
In Section 3 the residual symmetries of the reduced system are pointed
out. Section 4 is devoted to describing the mapping
of the reduced system into the space of matrix pseudo-differential
operators and to deriving the form of the Poisson brackets in terms of
the reduced space variables.
This mapping  will be considered  both in
the ``Drinfeld-Sokolov  gauge''
(0.8) and in a ``block-diagonal gauge''  which gives rise to  a factorization
of the Lax operator
$L_{p,r,s}$ and to a modified version of the generalized KdV hierarchy.
Section 5 contains the pseudo-differential
operator description of the local monodromy invariants that in the
first order matrix differential operator setting generate the natural
family of commuting Hamiltonians.
A few simple  examples
are presented in Section 6 as an illustration.
As a byproduct from analyzing the examples,
we shall also derive here a second modification
of the hierarchy related to a further factorization
of the pseudo-differential Lax operator.
The connection of the results of this paper to the
group theoretic classification of generalized Drinfeld-Sokolov
reductions is further discussed in Section 7.
There are two appendices, Appendix A and Appendix B, containing the
technical details of certain proofs.

Those readers who  are
more interested in the concrete description of the reduced
system than its derivation by means of the Hamiltonian
reduction could directly turn to study Theorem 4.4 in Section 4 and
Theorem 5.2 together with Corollary 5.3 in Section 5, which
give the form of the reduced Poisson brackets and the commuting
Hamiltonians together with the compatible evolution equations, respectively.

\bigskip
\noindent
{\bf Notational conventions.}
Throughout the paper,   $\widetilde N=\cinf(S^1,N)$ will denote the space of
smooth loops in $N$ for $N$ a Lie group,
a Lie algebra, a vector space, or $\mat(m\times n)$:
the algebra of $m\times n$
matrices over the field of complex numbers $\Complex$.
All algebraic operations such as addition,
multiplication, Lie bracket, are extended to $\widetilde N$ in
the standard  pointwise fashion.
For  a finite dimensional vector space $V_i$ (or for $V$
a finite or infinite direct sum  of such vector spaces)
if we write $f_i: S^1 \rightarrow V_i$ it is  understood
that $f_i\in \widetilde {V}_i$ (or so for the components of
$f:S^1\rightarrow V$).
The symbol $e_{ik}$ will stand for
the element of $\mat(m\times n)$
containing a single non-zero entry $1$ at the $ik$ position;
${\bf 0}_n$ and ${\bf 1}_n\in gl_n=\mat(n\times n)$ will denote
the zero and identity matrices, respectively.
The (smooth) dual of the vector space  $\widetilde{\mat}(m\times n)$
will be identified with $\widetilde{\mat}(n\times m)$ by means of
the pairing $<\ ,\ >: \widetilde{\mat}(n\times m) \times
\widetilde{\mat}(m\times n) \rightarrow {\Complex}$  given by
$$
<\alpha,v>:=\int_{S^1} \tr(\alpha v)
\qquad\hbox{for}\qquad
\alpha\in \widetilde{\mat}(n\times m),\quad
v\in \widetilde{\mat}(m\times n).
\eqno(0.11)$$
For a (suitably   smooth) complex  function $F$
on $\widetilde{\mat}(m\times n)$, the functional derivative
${\delta F\over \delta u}\in \widetilde{\mat}(n\times m)$  at
$u\in \widetilde{\mat}(m\times n)$ is defined by
$$
{d\over dt}F(u+tv)\vert_{t=0} =<{\delta F\over \delta u},v> =
\int_0^{2\pi}dx\, \tr\left({\delta F\over \delta u(x)} v(x)\right)
\qquad\forall v\in\widetilde{\mat}(m\times n).
\eqno(0.12)$$
For $F$ depending on several arguments we shall use
partial functional derivatives
defined by extending the above formula in a natural manner.
For instance, if $F$ depends on
$u\in \widetilde{gl}_n$ for $n=n_1+n_2$ and
$u=\pmatrix{u_{11}&u_{12}\cr u_{21}&u_{22}\cr}$ with
$u_{ij}\in \widetilde{\mat}(n_i\times n_j)$ then
we have ${\delta F\over \delta u}=\pmatrix{\alpha_{11}&\alpha_{12}\cr
 \alpha_{21}&\alpha_{22}\cr}$
in terms of the partial functional derivatives
${\delta F\over \delta u_{ij}}=\alpha_{ji}$.

\noindent
Finally, for a  Lie algebra $\G$, we let
$\ell(\G):=\G\otimes {\Complex}[\lambda,\lambda^{-1}]$  denote the
space  of  Laurent polynomials in the spectral
parameter $\lambda$ with coefficients in $\G$.
The Lie bracket is extended to $\ell(\G)$ in the standard way.
The reason for which  $\ell(\G)$  has to be carefully distinguished from
$\widetilde \G=\cinf(S^1,\G)$ is that
$x\in [0,2\pi]$ parametrizing $S^1$ has the r\^ole of
 the  physical space variable
(we adopt periodic boundary condition for definiteness),
while  $\lambda$ will appear essentially as a bookkeeping device.

\medskip


\bigskip
\centerline{\bf
1.~A Hamiltonian reduction approach to KdV type hierarchies}
\medskip

This section is a short review of the Drinfeld-Sokolov approach
to  KdV type hierarchies.
{}From our viewpoint, the main idea of this approach is to combine
Hamiltonian symmetry reduction with the Adler-Kostant-Symes approach to
constructing commuting Hamiltonian flows.
We concentrate on a special case that will be used  in this paper.
More general expositions can be found  in~\q{\McI,\GHM,\BGHM,\FGMG}.

Let $\G$ be a finite dimensional Lie algebra with an invariant
scalar product ``$\rm tr$''.
(For most of this paper it may be assumed that
$\G={gl}_n$ for some $n$, in which case $\tr$ really is
the standard matrix trace.)
For $C_-\in\G$ arbitrarily fixed, consider the manifold
$$
{\cal M}:=
\bigl\{\, \L = \pa + J + \lambda C_- \,\vert\, J\in\tG\,\bigr\}
\eqno(1.1)$$
of first order differential operators for which $\lambda$ is a free
parameter, the so-called spectral parameter.
As is well known (see e.g.\q{\RSTS}),
${\cal M}$ can be identified with a subspace of
the dual of the Lie algebgra
$\widehat{\G}\otimes{\Complex}[\lambda,\lambda^{-1}]$,
where $\widehat{\G}$ is  the  central extension of the loop algebra
$\tG=\cinf(S^1,\G)$,
and there are two compatible Poisson brackets (PBs) on $\M$
 defined by the following formulae.
The current algebra PB, or affine Kac-Moody algebra  PB, is given by
$$
\{ f, h \}_2(J) = \int_{S^1}\tr\left(
J\left[{\delta f\over \delta J}, {\delta h \over \delta J}\right]
-{\delta f\over \delta J}\left( {\delta h\over \delta J}\right)'\right).
 \eqno(1.2)$$
This PB (or its reduction) is often referred to as the {\it second} PB.
The {\it first}   PB reads
$$
\{ f, h \}_1(J) = -\int_{S^1}\tr
C_-\left[{\delta f\over \delta J}, {\delta h \over \delta J}\right].
\eqno(1.3)$$
Notice that the first PB $\{\ ,\ \}_1$ is minus the Lie derivative
of the second PB $\{\ ,\ \}_2$ along translations of $J$ in
the direction of $C_-$.
The Hamiltonians of special  interest here
 are provided by evaluation of the invariants
(``eigenvalues'') of the monodromy matrix
$T(J,\lambda)$ of ${\cal L}$, which is given by the path ordered
exponential
$$
T(J,\lambda)={\cal P} \exp\left(-\int_0^{2\pi}dx \left(J(x)+\lambda
 C_-\right)\right),
\eqno(1.4)$$
where $x\in [0,2\pi ]$ parametrizes the space $S^1$.
The corresponding Hamiltonian flows commute as a special case
of the Adler-Kostant-Symes (r-matrix) construction (see e.g.~\q{\RSTS}).
We call the hierarchy of these bihamiltonian flows on ${\cal M}$
the ``AKS hierarchy''.
This hierarchy is {\it non-local} in general since the invariants
of the monodromy matrix (1.4) are  non-local functionals of $J$.
We wish  to  perform a Hamiltonian  reduction of the AKS hierarchy to
obtain a {\it local} hierarchy.  The locality refers
both to the commuting Hamiltonians and the reduced PBs.

The standard method to obtain commuting local Hamiltonians relies on
a perturbative procedure that uses some
 {\it graded}, {\it semisimple}  element $\Lambda$ of the Lie algebra
$\ell({\cal G})= {\cal G}\otimes {\Complex}[\lambda, \lambda^{-1}]$.
The grading in which $\Lambda$ is supposed to be homogeneous with a
non-zero grade,  say grade $k>0$,
is defined by the eigenspaces of a linear operator
$d_{N,H}: \ell({\cal G}) \rightarrow \ell({\cal G})$,
$$
d_{N,H} = N \lambda {d\over d\lambda} + {\rm ad} H\,,
\eqno(1.5)$$
where $N$ is a non-zero integer and $H\in {\cal G}$ is diagonalizable
 with (usually) integer eigenvalues in the adjoint representation.
Note that  ${\rm ad} H$ defines a  grading  of ${\cal G}$,
$$
{\cal G} = \oplus_i\, {\cal G}_i\,,
\qquad
[\G_i,\G_l]\subset \G_{i+l}.
\eqno(1.6)$$
In the cases of our interest the grade of $\Lambda$ is small
(usually $1$) and hence $\Lambda$ takes the form
$$
\Lambda =\left( C_+ + \lambda C_-\right)
\qquad\hbox{with some}\qquad
C_\pm \in \G.
\eqno(1.7)$$
The requirement that $\Lambda$ is {\it semisimple} means
that it defines a direct sum decomposition
$$
\ell(\G)={\rm Ker}({\rm ad\,} \Lambda) + {\rm Im}({\rm ad\,} \Lambda)\,,
\qquad
{\rm Ker}({\rm ad\,} \Lambda) \cap {\rm Im}({\rm ad\,} \Lambda) =\{ 0\},
\eqno(1.8)$$
where the {\it centralizer}  ${\rm Ker}({\rm ad\,} \Lambda)$ of $\Lambda$
is a subalgebra of $\ell(\G)$. Having chosen $\Lambda$ of the form given by
(1.7), thereby defining  $C_-$ in (1.1);
and having chosen also a compatible
grading operator $d_{N,H}$,
the construction then involves imposing constraints on ${\cal M}$
so that the constrained manifold ${\cal M}_c \subset {\cal M}$
consists of operators $\L$ of the form
$$
\L = \pa + (j + C_+) +\lambda C_- =\pa + j + \Lambda
\qquad\hbox{with}\quad
j: S^1 \rightarrow \sum_{i< k} \G_i.
\eqno(1.9)$$
That is to say, in addition to the semisimple leading term $\Lambda$,
$\L$ contains only terms
of strictly smaller grade than the grade of $\Lambda$.
In concrete applications there might be further constraints on
${\cal L}\in {\cal M}_c$, but the above grading and semisimplicity
assumptions  are already sufficient to obtain local Hamiltonians by the
subsequent procedure.
The crucial step is to transform ${\cal L}$  in (1.9) as follows:
$$
\left({\pa + j +\Lambda }\right)
\mapsto e^{{\rm ad\,} \Xi } \left({\pa + j +\Lambda }\right)
 := (\pa + h + \Lambda),
\eqno(1.10{\rm a})$$
where $\Xi=\sum_{i<0} \Xi_i$ and
$h=\sum_{i<k}h_k$ are (formal)  infinite series
consisting of terms that take their values
in homogeneous subspaces in the decomposition (1.8) according to
$$
\Xi: S^1 \rightarrow \left({\rm Im}({\rm ad\,} \Lambda)\right)_{<0}\,,
\qquad
h: S^1\rightarrow \left({\rm Ker}({\rm ad\,} \Lambda)\right)_{<k},
\eqno(1.10{\rm b})$$
with the subscripts  referring to the grading (1.5).
The above grading and semisimplicity  assumptions
ensure that (1.10) can be solved recursively, grade by grade, for
both $\Xi=\Xi(j)$ and $h=h(j)$ and the solution is given by
unique {\it differential polynomials} in the components of $j$.
The Hamiltonians of interest are associated to the
grade larger than $-k$ subspace in the {\it centre} of the subalgebra
${\rm Ker}({\rm ad\,} \Lambda)\subset \ell(\G)$ as follows.
Suppose that $\{ X_i \}$ is a basis of this linear space.
The corresponding Hamiltonians are defined by
$$
H_{X_i}(j) :=\int_{0}^{2\pi} dx\, \left( X_i , h(j(x)) \right)\,,
\eqno(1.11)$$
where
 we use the canonical scalar product
$$
\left( X, Y\right) :={1\over 2\pi i}\oint {d\lambda \over \lambda} {\rm tr}
\left( X(\lambda) Y(\lambda) \right)
\eqno(1.12)$$
for any $X, Y \in \ell(\G)$.
As we shall see in examples,
the local functionals $H_{X_i}(j)$ can be interpreted
as the coefficients in an asymptotic expansion  of eigenvalues of the
monodromy matrix of ${\cal L}$ (1.9) for $\lambda \approx \infty$.
They will inherit the property of the monodromy invariants
that they commute among themselves with respect to the
PBs (1.2-3) on $\M$
if the restriction to ${\cal M}_c\subset {\cal M}$ is implemented
by means of an appropriate Hamiltonian symmetry reduction.

Let $G$ be a finite dimensional Lie group corresponding to $\G$ and
let ${\rm Stab}(C_-)\subset G$ be the subgroup which stabilizes
the element $C_-\in\G$ appearing in the definition
(1.1) of ${\cal M}$.
Let $\widetilde{{\rm Stab}}(C_-)$ be the loop group based on
${\rm Stab}(C_-)$.
Consider the action of $\widetilde{{\rm Stab}}(C_-)$   on ${\cal M}$
given by
$$
(\pa + J + \lambda C_-) \mapsto g (\pa  + J + \lambda C_-)g^{-1}
= g (  \pa + J)g^{-1} + \lambda C_-\,,
\quad\, \forall\, g\in
\widetilde{{\rm Stab}}(C_-).
\eqno(1.13)$$
The action defined by (1.13) is a symmetry of the AKS hierarchy on
${\cal M}$. That is it leaves invariant the compatible PBs (1.2-3)
and the monodromy invariants. If $\{ T^i\}\subset \G$ is a basis
of the centralizer of $C_-$ in $\G$, $[T^i, C_-]=0$,
then the current
components $J^i=\tr(T^iJ)$ serve as the generators of
this symmetry (components of the appropriate momentum map) with
respect to the second PB (1.2). The same current
components are Casimir functions with respect to the
first PB (1.3).

The possibility  to apply Hamiltonian  reduction to the
AKS hierarchy on ${\cal M}$  rests on the action (1.13)
of the symmetry group $\widetilde{{\rm Stab}}(C_-)$.
The  aim is to perform a symmetry reduction using an
appropriate subgroup of this  group in such a way
to ensure the locality of the reduced AKS hierarchy.
In a Hamiltonian symmetry reduction the first step is to
introduce constraints (e.g.~by restricting to the inverse
image of a value of the momentum map).
To get local Hamiltonians,
the  constrained manifold should  consist of operators
satisfying the conditions in (1.9).
Typically, the  constraints also bring a gauge freedom into the system,
which is to be factored out to obtain the reduced
system.
Another requirement on the constraints is that the reduced PBs
inherited from
(1.2) and (1.3) should be given by local, differential polynomial formulae.
This is automatically ensured if i) the reduced PBs are given
by the original PBs of the gauge invariant differential
polynomials on $\M_c$ and ii) these invariants
 form a {\it freely generated differential ring}.
In practice, property ii) is satisfied
if the gauge orbits admit a global cross section
for which the components of the gauge fixed current
(regarded as function of $j$ in (1.9))
define a free generating set of the gauge invariant differential
polynomials on $\M_c$.
The existence of such gauges is a very strong
condition on the constraints and the gauge group.
Gauge slices of this type have been used by Drinfeld and
Sokolov \q{\DS}, and we refer to such gauges as {\it DS gauges}.
A detailed description of the notion of DS  gauges,
including various sufficient
conditions for their existence, can be found in \q{\FORT}.

We wish  to emphasize that
the presence of a non-trivial gauge freedom is
an important  characteristic of the construction of ``KdV type'' systems.
A system of ``modified KdV type'' that can be
 constructed without any
reference to a gauge freedom \q{\Wil,\McI,\GHM}  does not in general
admit a  KdV type system related to it by a ``generalized Miura map''.
The reason lies in the non-trivial conditions
required for the existence of  a DS gauge, which
is needed in order to obtain a  KdV type system and an associated Miura map.

We end this overview by recalling that in the original Drinfeld-Sokolov
case $\Lambda$ was chosen to be a grade $1$ {\it regular}
semisimple element from the principal Heisenberg subalgebra of
$\ell(\G)$  for $\G$ a complex simple Lie algebra \q{\DS,\Kac}.
Requiring $\Lambda$ to be a
{\it regular} semisimple element means by definition that the centralizer
${\rm Ker}({\rm ad\,} \Lambda)\subset\ell(\G)$ appearing in
the decomposition (1.8)
is an {\it abelian} subalgebra.
In the original case this centralizer is the principal Heisenberg
subalgebra (disregarding the central extension).
For $\G$ a complex simple Lie algebra or $gl_n$,
the graded regular semisimple elements taken from the other Heisenberg
subalgebras (graded, maximal abelian subalgebras) of $\ell(\G)$ \q{\KP}
can be  classified using the results of Springer \q{\Sp} on the regular
conjugacy classes of the Weyl group of $\G$
(see also
ref.~\q{\DF}\footnote{${}^{1}$}{\ninerm In ref.~\q{\DF} the graded
regular semisimple
elements of $\ell(\G)$ were classified for $\G$ a classical Lie algebra
or $G_2$  using elementary methods.
The results of Springer \q{\Sp}, which  imply this
classification as well as its extension to any untwisted or twisted
affine Kac-Moody algebra based on any simple Lie algebra, had been
 overlooked in \q{\DF}.}).
One advantage of choosing $\Lambda$ to be regular is that the
centre of the centralizer is then maximal, giving a large
set of commuting local Hamiltonians.
Another important point is that  one may  associate constraints to any
graded regular $\Lambda$ in such a way that  DS gauges are available
\q{\GHM,\BGHM,\FHM,\FGMG,\DF}.
If one implements the generalized Drinfeld-Sokolov
reduction procedure proposed in \q{\GHM} using a graded semisimple
but non-regular element $\Lambda$, then the existence of a DS gauge
must be separately imposed as a condition on $\Lambda$.
This was  noted also in the earlier work \q{\McI}, where
essentially the same reduction procedure was described.
In fact,  the set of $\Lambda$'s
permitted by DS gauge fixing appears  rather limited \q{\FHM,\FGMG}.
These developments raise the hope to
eventually be able to completely classify the KdV type
hierarchies that may be obtained in the Drinfeld-Sokolov
approach.
The analysis of the set of examples that follows is also
intended to contribute to this classification
(see  the discussion in Section 7).

\bigskip
\centerline{\bf 2.~A generalized Drinfeld-Sokolov reduction}
\medskip

We wish to apply the above formalism to ${\cal G}:=gl_n$.
The graded Heisenberg subalgebras of $\ell(gl_n)$ are
classified by the partitions of $n$ \q{\KP}.
We shall choose  $\Lambda$ to be a graded semisimple element of minimal
 positive grade from a Heisenberg subalgebra of $\ell(gl_n)$ associated
to a partition of $n$ into ``equal blocks plus singlets''
$$
n=pr+s=\overbrace{r+\cdots +r}^{p\;\rm times}+
\overbrace{1+\cdots +1}^{s\;\rm times}
\qquad\hbox{for some}\qquad
p\geq 1, \quad r>1,\quad s\geq 1.
\eqno(2.1)$$
It was explained in \q{\FHM} that
a graded {\it regular} semisimple  element only exists
for the partitions into equal blocks
$n=pr$ or equal blocks plus {\it one} singlet $n=pr+1$.
When continuing our previous study of the equal blocks case \q{\FHM}
with the equal blocks plus singlet case, it was realized
that DS gauge fixing is possible  also when
the partition contains an arbitrary number of singlets ($1$'s).
This is the  motivation for the study of the more general case $s\geq 1$ here;
and actually  the analysis will be  the same for any $s\geq 1$.
The construction that will be described in this section
may be also  viewed as a special case
of the generalized Drinfeld-Sokolov
reduction procedure proposed in \q{\McI,\GHM}.

Next we introduce the necessary notation.
Adapted to the partition in (2.1),  an element  $m\in gl_n$ will often be
 presented
in the following $(r+1)\times (r+1)$ block form:
$$
m=\sum_{i,j=1}^r e_{i,j}\otimes A_{i,j}
+\sum_{i=1}^r e_{i,r+1} \otimes B_i
+\sum_{i=1}^r e_{r+1,i} \otimes C_i
+e_{r+1,r+1} \otimes D,
\eqno(2.2)$$
where the $e_{i,j}\in gl_{r+1}$ are the usual elementary matrices,
$A_{i,j}\in gl_p$, $D\in gl_s$,  $B_i\in\mat(p\times s)$
and $C_i\in\mat(s\times p)$.
Alternatively, we may write
$$
m=\left(\matrix{A&B\cr C&D\cr}\right).
\eqno(2.3)$$
Introduce the $r\times r$  ``DS matrix'' $\Lambda_r\in \ell(gl_r)$
given by
$$
\Lambda_r :=\pmatrix{
0&1&0&\cdots&0\cr
\vdots&0&1&\ddots&\vdots\cr
\vdots&{}&\ddots&\ddots&0\cr
0&{}&{}&\ddots&1\cr
\lambda &0&\cdots&\cdots&0\cr}.
\eqno(2.4)$$
Recall that the Heisenberg subalgebra,
${\cal Z}_{p,r,s}\subset \ell(gl_n)$, associated
to the partition (2.1) is  the linear span of  the elements
$$
\left(\matrix{(\Lambda_{r})^l\otimes Y_p&0\cr 0&{\bf 0}_s\cr}\right),
\quad
\forall\, l\in {\Zed},\quad
\forall\,  Y_p={\rm diag}\left(y_1,\ldots, y_p\right),
\quad (y_i\in {\Complex},\ i=1,\ldots, p),
\eqno(2.5)$$
together with the elements
$$
\left(\matrix{{\bf 0}_p&0\cr 0&Y_s\lambda^l \cr}\right),
\qquad
\forall\, l\in {\Zed},\quad \forall\,  Y_s={\rm diag}\left(y_1,\ldots,
 y_s\right),
\quad (y_k\in {\Complex},\ k=1,\ldots,s).
\eqno(2.6)$$
${\cal Z}_{p,r,s}$
is a graded maximal abelian subalgebra of $\ell(gl_n)$
if we choose the grading defined by
$$
d:=d_{r,H} =r \lambda {d\over d\lambda} + {\rm ad\,} H
\eqno(2.7{\rm a})$$
(see (1.5) and (1.6)), where we take
$$
H:=\cases{
{\rm diag\,}\left(m {\bf 1}_p, (m-1){\bf 1}_p, \ldots, -m{\bf 1}_p,
{\bf 0}_s\right),
& if $r=(2m+1)$  odd;\cr
{}&{}\cr
{\rm  diag\,}\left(m {\bf 1}_p, (m-1) {\bf 1}_p, \ldots,  -(m-1){\bf 1}_p,
 {\bf 0}_s\right)
& if $r=2m$  even.\cr}
\eqno(2.7{\rm b})$$

Incidentally, note that $H$ equals  the Cartan generator $I_0$,
$$
I_{0}=
{\rm diag\,}\left({r-1\over 2} {\bf 1}_p, {r-3\over 2}{\bf 1}_p, \ldots,
 -{r-1\over 2}{\bf 1}_p,
{\bf 0}_s\right),
\eqno(2.8{\rm a})$$
 of the $sl_2$ subalgebra of $gl_n$ under which the defining
representation decomposes according to the partition (2.1)
if $r$ is odd, and $H$ is obtained from $I_0$ by adding
${1\over 2}$ to the half-integral eigenvalues of  $I_0$ if $r$ is even.
In the latter case an alternative grading, with respect to which
${\cal Z}_{p,r,s}\subset \ell(gl_n)$ is still graded,  is
defined by letting the grading operator be given by
$$
d':=d_{2r, 2I_0}= 2\left( r\lambda {d\over d\lambda} + {\rm ad\,} I_0\right),
\eqno(2.8{\rm b})$$
instead of by (2.7).
We have adopted the grading operator (2.7) since this
grading will permit us to treat
the case of even and odd $r$ without any difference and, in fact,
this choice of grading does not effect the final result.

We  choose  a grade $1$ element $\Lambda:=\Lambda_{p,r,s}$ (0.6) from
the Heisenberg subalgebra,
$$
\Lambda=\left(\matrix{\Lambda_{r}\otimes \Gamma&0\cr 0&{\bf 0}_s\cr}\right)
\quad\hbox{with}\quad
\Gamma:={\rm diag\,}\left(d_1,\ldots, d_p\right),
\quad
(d_i)^r \neq (d_j)^r,
\quad
d_i\neq 0.
\eqno(2.9)$$
For generic $\lambda$, all of the eigenvalues of $\Lambda$
except for  the eigenvalue $0$ are distinct.
The eigenvalue $0$ has multiplicity $s$.
It follows that  $\Lambda$ is a {\it regular} element
of $\ell(gl_n)$ if and only if $s=1$.
In addition to the linear span of the elements given in (2.5),
the {\it centralizer}   ${\rm Ker}({\rm ad\,} \Lambda)\subset \ell(gl_n)$
of $\Lambda$ contains the algebra $\ell(gl_s)\subset \ell(gl_n)$
spanned by the elements
$$
\left(\matrix{{\bf 0}_p&0\cr 0& D\lambda^l \cr}\right),
\qquad
\forall\, l\in {\Zed},\quad \forall\,
D\in gl_s.
\eqno(2.10)$$
Hence the {\it centre of the centralizer} of $\Lambda$
is spanned by the elements given in (2.5) together with the
set of elements of the form
$$
\left(\matrix{{\bf 0}_p&0\cr 0& {\bf 1}_s\lambda^l \cr}\right),
\qquad
\forall\, l\in {\Zed}.
\eqno(2.11)$$

Now we consider
the generalized Drinfeld-Sokolov reduction of the AKS hierarchy on
 the manifold  ${\cal M}$ defined in (1.1), with $C_-$ and $C_+$
given by writing $\Lambda$  as $\Lambda=\lambda C_- + C_+$,
i.e., from now on
$$
C_-= e_{r,1}\otimes \Gamma
\qquad\hbox{and}\qquad
C_+=\sum_{i=1}^{r-1} e_{i,i+1}\otimes \Gamma.
\eqno(2.12)$$
Using the grading of the Lie algebra ${\cal G}=gl_n$
defined by the eigenvalues of ${\rm ad\,}H$,
$$
{\cal G}={\cal G}_{-(r-1)}+\ldots + {\cal G}_{-1} + {\cal G}_0 +{\cal G}_1 +
\ldots +
{\cal G}_{r-1} = {\cal G}_{<0} + {\cal G}_0 + {\cal G}_{>0} ,
\eqno(2.13)$$
we  first write the current $J$ as
$$
J=J_{<0} + J_0 + J_{>0},
\eqno(2.14)$$
and impose the constraint
$$
J_{>0}=C_+,
\eqno(2.15)$$
which restricts the system to the submanifold
${\cal M}_c\subset {\cal M}$ given by
$$
\M_c:=\bigl\{\,
\L=\pa + j +\Lambda \,\vert\, j\in\cinf(S^1,{\cal G}_{\leq 0})\,\bigr\}.
\eqno(2.16)$$
Then we factorize the constrained manifold
${\cal M}_c$ by the group ${\cal N}$ of
{\it gauge transformations} $e^f$ acting according to
$$
e^f: {\cal L} \mapsto e^f{\cal L}e^{-f},
\qquad
f\in\cinf(S^1,{\cal G}_{<0}).
\eqno(2.17)$$
The Hamiltonian interpretation of this reduction procedure is the
same as in the case of the standard Drinfeld-Sokolov reduction for $gl_n$.
Briefly, from the point of view of the second  PB (1.2),
it is a Marsden-Weinstein type reduction with ${\cal N}$ being
the symmetry group and $C_+$ a character of ${\cal N}$
(which means that the constraints defining
${\cal M}_c\subset {\cal M}$ are {\it first class}). {}From the point of
view of the first PB (1.3), the reduction amounts to fixing
the values of Casimir functions and subsequently factoring by the
group of Poisson maps ${\cal N}$.
It follows that
the compatible PBs on ${\cal M}$
 induce compatible PBs on the space of the
gauge invariant functions on ${\cal M}_c$,
identified as
the space of functions on the reduced space
${\cal M}_{\rm red}:={\cal M}_c/{\cal N}$.
The reduced first PB is the Lie derivative of the
reduced second PB  with respect to the one parameter group
action on the reduced space induced by the following
one parameter group action on ${\cal M}_c$:
$$
{\cal L}\mapsto {\cal L} -\tau C_-\,,
\qquad
\forall \tau \in {\Real}.
\eqno(2.18)$$
This follows by combining  the fact that these translations
commute with the action of ${\cal N}$ on ${\cal M}_c$ with the fact
that the first PB (1.3) is the Lie derivative of the second
PB (1.2) with respect
to the respective one parameter group action on ${\cal M}$.
It is clear from (2.17) that the invariants (eigenvalues) of the
monodromy matrix
of ${\cal L}\in \M_c$ define gauge invariant functions on ${\cal M}_c$.
In conclusion,  we have
a local hierarchy of bihamiltonian flows on the reduced phase space
${\cal M}_{\rm red}$, which  is generated by the
 Hamiltonians provided  by the
local  monodromy invariants of ${\cal L}$ defined by the procedure
in (1.10) and (1.11).

In order to describe the reduced system in more detail,
one reverts to gauge slices.
We wish to mention two important  gauges.
The first is  the DS gauge whose gauge section
is  the manifold ${\cal M}_{\rm DS}\subset \M_c$ given by
$$
\eqalignno{
&{\cal M}_{\rm DS} := \Bigl\{\,
{\cal L} = \pa   + j_{\rm DS}+\Lambda \,\vert\, \cr
&\quad\,\,\, j_{\rm DS} = \sum_{i=1}^r e_{r,i}\otimes v_{r-i+1}
+ e_{r,r+1} \otimes \zeta_+ + e_{r+1,1} \otimes \zeta_-
+ e_{r+1,r+1}\otimes w\cr
&\quad\,\,{\hbox{ with }}\quad
v_i\in\wt{gl}_p,\quad w\in\wt{gl}_s,\quad
\zeta_+\in\wt{\mat}(p\times s), \quad \zeta_-\in\wt{\mat}(s\times p)\,
\Bigr\}.
&(2.19)\cr}$$
In explicit matrix notation $j_{\rm DS}:=j_{p,r,s}$ is given in (0.8).
The space ${\cal M}_{\rm DS}$ is
a {\it one-to-one} model of $\M_c/\N$
with  the property
that, when regarded as functions on ${\cal M}_c$,
the components of the gauge fixed current $j_{\rm DS}=j_{\rm DS}(j)$
 provide a basis
of the gauge invariant differential polynomials on $\M_c$.
This follows from a standard argument on DS gauge fixing
(see e.g.~\q{\FORTW}),
which relies on the grading and the  non-degeneracy condition
$$
{\rm Ker}\left({\rm ad\,} C_+\right) \cap {\cal G}_{<0} = \{ 0\},
\eqno(2.20)$$
which is satisfied in our case.
The  reduced AKS hierarchy on
the phase space ${\cal M}_{\rm red}\simeq {\cal M}_{\rm DS}$
is  a  generalization of the well-known
$r$-KdV hierarchy, as we shall see in Sections 4 and 5.

The other important gauge is what we call the ``$\Theta$-gauge''
(also called block-diagonal gauge),
which is defined by the following submanifold $\Theta\subset \M_c$,
$$
\Theta := \bigl\{ \,
{\cal L}=\pa + j_0  + \Lambda \,\vert\,
j_0\in\cinf(S^1,{\cal G}_0)\,\bigr\}.
\eqno(2.21)$$
Correspondingly to the grading operator $H$ in (2.7),
we parametrize $j_0$ as
$$
j_0={\rm diag}\left( \theta_m,  \ldots, \theta_1, a,
\theta_{-1}, \ldots, \theta_{-[{{r-1}\over 2}]}, d \,\right)
+ e_{m+1, r+1}\otimes b + e_{r+1, m+1}\otimes c \,,
\eqno(2.22{\rm a})$$
where $[{r-1\over 2}]$ is the integral part of ${{r-1}\over 2}$,
$m=[{r\over 2}]$.
The variables on $\Theta$ are
$\theta_i\in\widetilde{gl}_p$ for $i\neq 0$
and
$$
a\in\widetilde{gl}_p,
\quad
b\in\widetilde{{\rm mat}}(p\times s),
\quad
c\in\widetilde{{\rm mat}}(s\times p),
\quad
d\in\widetilde{gl}_s,
\eqno(2.22{\rm b})$$
which we collect into the matrix
$$
\theta_0:=\left( \matrix{a&b\cr c &d\cr} \right) \in\widetilde{gl}_{p+s}.
\eqno(2.23)$$
In the $\Theta$-gauge, in terms of the variables $\theta_i$,
the reduced second PB  becomes just the direct sum of
free current algebra PBs given by
$$
\{ {\bar f}, {\bar h}\}(\theta)=
\sum_{i=-[{{r-1}\over 2}]}^m \,\int_{S^1}\,
 \tr \left(\theta_i \left[{\delta {\bar f} \over \delta \theta_i},
{\delta {\bar h} \over \delta \theta_i}\right]
-{\delta {\bar f} \over \delta \theta_i}
\left({\delta {\bar h} \over \delta \theta_i}\right)' \right)
\eqno(2.24)$$
for ${\bar f}$, ${\bar h}$ smooth functions on $\Theta$.
In fact, the restriction
of the second PB of arbitrary  gauge invariant
functions $f$, $h$  on ${\cal M}_c$ to $\Theta$ has the form  (2.24)
with $\bar f=f\vert_{\Theta}$, $\bar h=h\vert_\Theta$.
This  can be proved in the same way as  Lemma 2.1 in \q{\FHM},
which is of course essentially the same as the proof found in \q{\DS}
in  the scalar $r$-KdV case.
Using DS gauge fixing we obtain a local, differential polynomial mapping
$\mu:\Theta\rightarrow {\cal M}_{\rm DS}$  yielding  a generalized
Miura transformation from the
modified KdV type hierarchy on $\Theta$ to the KdV type hierarchy
on ${\cal M}_{\rm DS}$.
As is expected from a Miura map, the inverse of $\mu$
is non-local  and is not single valued.
In other words, $\Theta$ cannot be reached by a local  gauge
fixing procedure and
the intersection of $\Theta\subset\M_c$ with a gauge orbit of $\N$
in $\M_c$ is not unique.

\smallskip
\noindent{\it Remark 2.1.}
The reduced current algebra PB on ${\cal M}_c/{\cal N}$
is known to provide  an example of a  ``classical ${\cal W}$-algebra''
\q{\FORT}.
A  basis of the
gauge invariant differential polynomials in $j$ (2.16)  consisting of
conformal tensors is furnished by
the so called lowest weight gauge.
In order to define this gauge one considers the $sl_2$ subalgebra
${\cal S}=\{ I_-, I_0, I_+\}\subset {\cal G}$ satisfying
$$
[I_0, I_\pm ]= \pm I_\pm,
\quad
[I_+, I_-]=2I_0
\eqno(2.25)$$
with $I_+:=C_+$, where $C_+$ is given in (2.12).
The gauge section ${\cal M}_{\rm l.w.}\subset {\cal M}_c$ of the lowest
weight gauge (a conformally distinguished DS type gauge)
is defined by
$$
{\cal M}_{\rm l.w.}:=\{\,
{\cal L}=\pa + j_{\rm l.w.} + \Lambda \,\vert\,
j_{\rm l.w.}: S^1 \rightarrow {\rm Ker}\left({\rm ad\,}I_-\right)\,\},
\eqno(2.26)$$
where ${\rm Ker}\left({\rm ad\,}I_-\right)$ is the space of lowest
weight vectors of ${\cal S}\subset {\cal G}$ in its representation
 on ${\cal G}$.
An analogue of this  lowest weight gauge may be associated
to any $sl_2$ subalgebra ${\cal S}\subset {\cal G}$,
and it has a PB algebra induced from the current algebra  which is
a $\W$-algebra \q{\BTvD,\FORTW}.

\smallskip
\noindent{\it Remark 2.2.}
In the   case $r=1$  it is natural to define
the element $\Lambda$ to be the diagonal matrix
$\Lambda:=\Lambda_{p,r=1,s}$ given by
$$
\Lambda_{p,r=1,s}:=\lambda\,\pmatrix{\Gamma &0\cr 0&{\bf 0}_s\cr}=
\lambda\, {\rm diag}\left(d_1,\ldots,d_p,0,\ldots,0\right),
\eqno(2.27)$$
which contains $s$ zeros and  distinct, non-zero
$d_i\in {\Complex}$ for $i=1,\ldots,p$.
This is a grade $1$ semisimple element of $\ell(gl_n)$, $n=p+s$,
with respect to the homogeneous grading.
In this case the Drinfeld-Sokolov reduction becomes trivial, i.e.,
 $\M=\M_c=\M_{\rm DS}=\Theta$.
For this reason  the assumption has been made so far
that $r>1$.
All  considerations in the rest of this paper apply to the $r=1$
case too and all of the result follow through.
There are  interesting consequences  for $r=1$ as well as for $r>1$.

\bigskip
\centerline{\bf 3.~Residual symmetries}
\medskip

We have performed a  Drinfeld-Sokolov type reduction on
the system ${\cal M}$ (1.1) using the  subgroup ${\cal N}$ (2.17)
of the symmetry group $\widetilde{{\rm Stab}}\left(C_-\right)$
defined by the stabilizer of the element $C_-$ in (2.12).
Here we wish to point out a residual symmetry of the reduced
system so obtained.
Consider the following transformations on ${\cal M}$:
$$
{\cal L} \mapsto
\exp\left(\pmatrix{{\bf 1}_{r}\otimes\alpha&0\cr 0&{\bf 0}_s\cr}\right)
{\cal L}
\exp\left(-\pmatrix{{\bf 1}_{r}\otimes \alpha&0\cr 0&{\bf 0}_s\cr}\right),
\quad
\alpha={\rm diag}(\alpha_1,\ldots,\alpha_p),
\eqno(3.1{\rm a})$$
with $\alpha_i\in\cinf(S^1,{\Complex})$ for $i=1,\ldots, p$, and
$$
{\cal L} \mapsto
\exp\left( \pmatrix{{\bf 0}_{pr}& 0\cr 0& D\cr}\right)
{\cal L}
\exp\left(-  \pmatrix{{\bf 0}_{pr}&0\cr 0& D\cr}\right),
\quad\hbox{with}
\quad
D\in\cinf(S^1,gl_s)\,.
\eqno(3.1{\rm b})$$
The group  generated by these transformations
is a
$\widetilde{GL}_1\times\cdots\times\widetilde{GL}_1\times \widetilde{GL}_s$
subgroup of
the symmetry group $\widetilde{{\rm Stab}}\left(C_-\right)$
acting on ${\cal M}$ according to (1.13).
We call it  the {\it group of residual symmetries} and denote it by $G_R$.
It is easily verified that these transformations map the
constrained manifold ${\cal M}_c\subset {\cal M}$ to itself.
For grading reasons,
$G_R\subset \widetilde{{\rm Stab}}\left(C_-\right)$
normalizes the gauge group
${\cal N}\subset \widetilde{{\rm Stab}}\left(C_-\right)$,
which implies that the transformations in  (3.1) induce
a corresponding action of $G_R$
on the space of gauge orbits ${\cal M}_c/{\cal N}$.
By construction,
this induced action  leaves invariant the commuting Hamiltonians as
well as the compatible PBs of the hierarchy on
${\cal M}_{\rm red}={\cal M}_c/ {\cal N}$.
Notice that
the gauge slices $\Theta$ in (2.21) and ${\cal M}_{\rm l.w.}$ in (2.26)
are mapped to themselves  by the transformations in (3.1).
Hence in terms of these gauge slices
the induced action of $G_R$ is simply  given by the restriction of (3.1).
The gauge slice ${\cal M}_{\rm DS}$ in (2.19) is mapped to itself by the
subgroup $\widetilde{GL}_s\subset G_R$ given by (3.1b).

Let us write the current $J\in\widetilde{gl}_{pr+s}$
defining  ${\cal L}=(\pa + J + \lambda C_-)\in {\cal M}$ in the
block form
$$
J=\pmatrix{J_{11}&J_{12}\cr J_{21}&J_{22}\cr},
\eqno(3.2)$$
where $J_{11}\in\widetilde{gl}_{pr}$, $J_{22}\in\widetilde{gl}_s$ etc.,
similarly to the matrix $m$ in (2.2), (2.3),
With respect to  the current algebra PB (1.2),
the infinitesimal generators (the momentum map)
of the transformations in (3.1a) and (3.1b)  on ${\cal M}$ are provided by
the current components $\Phi_i$ and $w$ given by
$$
\Phi_i(J):={\rm tr}\bigl(J_{11} ({\bf 1}_r\otimes e_{ii})\bigr),
\qquad  i=1,\ldots,p,
\eqno(3.3{\rm a})$$
where $e_{ii}\in gl_p$ is the usual elementary matrix,  and
$$
w(J):=J_{22},
\eqno(3.3{\rm b})$$
respectively.
The restrictions of these current components to ${\cal M}_c$
(by setting $J=(j+C_+)$ as in(2.16)) are gauge invariant.
These gauge invariant current components generate the
induced action of the  group of residual symmetries $G_R$
 on ${\cal M}_{\rm red}$ with respect to the reduced second PB.
Since the Hamiltonians and the compatible PBs
of the hierarchy on ${\cal M}_{\rm red}$ are invariant under this group,
it follows  that the current components $\Phi_i$ and $w$ are constants
along the flows of the hierarchy.
Of course  this also follows from the fact
that $\Phi_i$ and $w$ are Casimir functions with respect to the
first PB (1.3).

In conclusion,
the residual symmetry (3.1) may be used to perform
further reductions on the hierarchy obtained from the generalized
Drinfeld-Sokolov reduction.

\bigskip
\centerline{\bf 4.~The Poisson brackets on the reduced phase space}
\medskip

In this section we shall find  a Poisson mapping from
the  reduced phase space ${\cal M}_c/{\cal N}$, endowed with
the compatible PBs induced by the Drinfeld-Sokolov  reduction,
to the space of pseudo-differential operators (PDOs) with $gl_p$ valued
coefficients endowed with the usual Gelfand-Dickey Poisson brackets
\q{\GD,\Man,\A,\Dic}.
The image  includes the phase space of the matrix $r$-KdV hierarchy.
The mapping will be defined by means of the elimination procedure
similarly to the  $r$-KdV case \q{\DS,\FHM}.
In the present case the mapping will not be one-to-one,
but we shall be able to present the explicit form
of the reduced PBs on ${\cal M}_c/{\cal N}$ nonetheless.

Let ${\cal A}$ be the space of pseudo-differential operators with
$p\times p$ matrix coefficients:
$$
{\cal A} = \{\, L = \sum_{s=-\infty}^N L_s\partial^s\,
\vert\, \, L_s \in\widetilde{gl}_p ,
\,\,  N\in {\Zed}\,\}.
\eqno (4.1)
$$
Multiplication of matrix pseudo-differential operators is defined
in the usual way, i.e., the product rule is
given by matrix multiplication together with the formulae
$$
\pa\, \pa^{-1}=1 \quad{\hbox{ and }}\quad
\pa F=F\pa+F'\quad\hbox{for}\quad F\in \widetilde{gl_p},
\eqno(4.2{\rm a})$$
which engender the formula
$$
\pa^{-1}F=\sum_{i=0}^\infty(-1)^iF^{(i)}\pa^{-i-1}.
\eqno(4.2{\rm b})$$
The Adler trace \q{\A}, $\Tr: {\cal A} \to {\Complex}$, is given by
$$
\Tr L :=  \int_{S^1} \, \tr\,{\rm res}(L) =  \int_{S^1} \, \tr\, L_{-1}\,,
\eqno (4.3)$$
where $\tr$ is the ordinary matrix trace.
Let $P_\pm$ be the projectors on ${\cal A}$ onto the subalgebras
$$
{\cal A}_+ := \{\,L = \sum_{s=0}^N L_s\partial^s\,\},
\qquad
{\cal A}_- :=  \{\,L = \sum_{s=-\infty}^{-1} L_s\partial^s\,\},
\eqno (4.4)$$
respectively. Put $L_\pm:=P_\pm(L)$.
The space ${\cal A}$ is a bihamiltonian manifold.
For $f$, $h$ smooth functions on ${\cal A}$,
the quadratic (second) Gelfand-Dickey PB is given by
$$
\{f,h\}^{(2)}(L)=
\Tr\left({\delta h\over \delta L}L
\left({\delta f\over \delta L}L\right)_+ -
L{\delta h\over \delta L}
\left(L{\delta f\over \delta L} \right)_+\right)\,,
\eqno(4.5)
$$
where the gradient ${\delta f\over \delta L}\in {\cal A}$
of $f$ at $L\in {\cal A}$ is defined  by
$$
{d\over dt} f(L+t A){\vert_{t=0}} =\Tr\left(A
{\delta  f \over \delta L}\right)\,,
\qquad \forall\,A\in {\cal A}.
\eqno(4.6)$$
The Lie derivative of the quadratic bracket (4.5)
with respect to the one parameter group
of translations
$$
L\mapsto (L + \tau {\bf 1}_p),
\qquad
\forall\,\tau \in {\Real},
\eqno(4.7)$$
is the linear (first) Gelfand-Dickey PB:
$$
\{f,h\}^{(1)}(L)=
\Tr\left(L \left[
\left({\delta  f \over \delta L}\right)_+,
\left({\delta h\over \delta L}\right)_+\right] -
 L\left[
\left({\delta  f \over \delta L}\right)_-,
\left({\delta h\over \delta L}\right)_-\right]\right),
\eqno(4.8)$$
which is compatible (coordinated) with the quadratic PB.

In order to relate the Drinfeld-Sokolov reduction to the above formalism,
consider the  linear problem for
${\cal L}\in {\cal M}_c$ (2.16):
$$
{\cal L} \psi =0,
\qquad
\psi=( \psi_1^t, \psi_2^t, \ldots, \psi_r^t, \phi^t)^t\,,
\eqno(4.9)$$
where $\psi$ is a $(pr+s)$-component column vector
consisting of the $p$-component column vectors $\psi_i$ ($i=1,\ldots,r$) and
the $s$-component column vector $\phi$.
This system of equations is covariant under
the gauge transformation (2.17) accompanied by the transformation
$$
\psi \mapsto e^f \psi,
\qquad
f\in\cinf(S^1,{\cal G}_{<0}).
\eqno(4.10)$$
Observe that the  component $\psi_1$ is {\it invariant} under (4.10).
This implies that if we derive from (4.9)  an
equation on $\psi_1$, then the operator entering that equation
will be a gauge invariant function on ${\cal M}_c$.
The desired operator can be derived by the
usual elimination procedure.
This is particularly simple in the DS gauge
(2.19), where the explicit  form of (4.9) is
$$\eqalign{
\pa \psi_1 + \Gamma \psi_2&=0,\cr
&\vdots\cr
\pa \psi_{r-1} + \Gamma \psi_{r}&=0,\cr
\pa \psi_r +\sum_{i=1}^r v_i \psi_{r+1-i} + \zeta_+\phi &=
-\lambda \Gamma\psi_1,\cr
(\pa +w) \phi + \zeta_- \psi_1 &=0,\cr}
\eqno(4.11)$$
where $\Gamma$ is the diagonal matrix given in (2.9).
If we formally solve for $\psi_i$ ($i=2, \ldots, r$)
and for $\phi$ in terms of $\psi_1$  using
the first $(r-1)$ and the last equations in (4.11), respectively,
and re-insert the result into the remaining  equation,
we obtain
$$
L\psi_1 = \lambda \psi_1 ,
\eqno(4.12)$$
where
$$
L= \Delta^r \pa^r + u_1 \pa^{r-1} +u_2 \pa^{r-2}
+ \cdots
+u_{r-1} \pa+  u_r +  z_+ ({\bf 1}_s \pa +w )^{-1} z_- ,
\eqno(4.13)$$
with $\Delta=-\Gamma^{-1}$ and the variables being related to
those in (2.19) by
$$
u_i=\Delta v_i \Delta^{r-i},
\quad
z_+=-\Delta\zeta_+, \quad z_-=\zeta_-.
\eqno(4.14)$$

Let $M\subset {\cal A}$  denote the manifold of ``Lax operators''
$L$ of the form (4.13).
As discussed above, the elimination procedure gives rise
to a mapping
$$
\pi: {\cal M}_c \rightarrow M,
\qquad
\pi\left({\cal L}\right)=L,
\eqno(4.15{\rm a})$$
which is constant along the gauge orbits in ${\cal M}_c$.
Thus we have a corresponding induced mapping
$$
\bar \pi: {\cal M}_c/{\cal N}\rightarrow M.
\eqno(4.15{\rm b})$$
Observe that $L$ in (4.13) contains only
quadratic combinations of the fields $z_\pm$ that
parametrize the manifold ${\cal M}_{\rm DS}\simeq {\cal M}_c/{\cal N}$
(2.19),
but does not contain these fields in a linear manner.
This shows that (unlike in the usual $r$-KdV case)
the mapping $\bar \pi$ is {\it not} one-to-one.
The reason for this lies in the fact that the action
of the group  $\widetilde{GL}_s$ on ${\cal M}_c/{\cal N}$
defined by  (3.1b) is a symmetry of the mapping $\bar\pi$, i.e.,
every $\widetilde{GL}_s$ orbit in ${\cal M}_c/{\cal N}$ is mapped
to a single point.
To understand this, observe that the original linear problem
(4.9) is covariant not only with respect to the gauge group ${\cal N}$
but also with respect to the group of residual symmetries
$G_R$  acting according to
the formulae in (3.1a) and (3.1b) complemented with the formulae
$$
\psi\mapsto
\exp\left(\pmatrix{{\bf 1}_{r}\otimes \alpha&0\cr 0&{\bf 0}_s\cr}\right)
\psi
\eqno(4.16{\rm a})$$
and
$$
\psi \mapsto
\exp\left( \pmatrix{{\bf 0}_{pr}& 0\cr 0& D\cr}\right) \psi.
\eqno(4.16{\rm b})$$
Since the component $\psi_1$ is {\it invariant} under the
$\widetilde{GL}_s$ action (4.16b),  the operator $L=\pi({\cal L})$
entering (4.12)
must be also invariant with respect to the  $\widetilde{GL}_s$ action (3.1b).
One may directly verify that this is the case as follows.
If $v_i$, $\zeta_\pm$, $w$ are the coordinates of a point
${\cal L}\in {\cal M}_{\rm DS}\simeq {\cal M}_c/{\cal N}$, then the
 coordinates
$\tilde v_i$, $\tilde \zeta_\pm$, $\tilde w$ of the  $\widetilde{GL}_s$
transformed point $\tilde {\cal L}\in {\cal M}_{\rm DS}$ turn out to be
$$
\tilde v_i=v_i,
\quad
\tilde \zeta_+ =\zeta_+ e^{-D},
\quad
\tilde \zeta_-= e^{D}\zeta_-,
\quad
\tilde w=e^D w e^{-D} + e^D \left(e^{-D}\right)'.
\eqno(4.17{\rm a})$$
Hence
$$
\zeta_+ ({\bf 1}_s \pa + w)^{-1} \zeta_- =
\tilde \zeta_+({\bf 1}_s\pa +\tilde w)^{-1}
\tilde \zeta_-.
\eqno(4.17{\rm b})$$
Therefore the same Lax operator $L$
is attached to ${\cal L}$ and $\tilde {\cal L}$,
$$
\bar \pi (\tilde {\cal L})=\bar\pi \left({\cal L}\right),
\eqno(4.18)$$
showing that $\bar \pi$ (4.15b) maps
every $\widetilde{GL}_s$ orbit in ${\cal M}_c/{\cal N}$
to a single PDO as claimed above.
Correspondingly,
for the infinitesimal generator $w$ (3.3b) of the
$\widetilde{GL}_s$ symmetry and for  any function ${\cal F}$ of $L$, we  have
$$
\{ w , {\cal F} \}_2 =0.
\eqno(4.19)$$
In particular,  the expansion of $L$ in powers
of $\pa$ contains only such ${\cal N}$-invariant
 differential polynomials in the
components of ${\cal L}\in {\cal M}_c$ which commute with $w$
under the second PB.

The elimination procedure  may be performed on the linear
problem (4.9) in the
$\Theta$-gauge (2.21) analogously  as was done above in the DS gauge (2.19).
We obtain
$$
L_{\Theta} \psi_1 = \lambda \psi_1
\eqno(4.20)$$
with the factorized Lax operator
$$
L_{\Theta} = (\Delta (\pa + \theta_{-[{{r-1}\over 2}]}))\cdots
\left(\Delta (\pa + \theta_{-1})\right) \left(\Delta K\right)
\left(\Delta (\pa + \theta_1)\right) \cdots \left(\Delta
(\pa + \theta_m)\right),
\eqno(4.21)$$
where  $\Delta = -\Gamma^{-1}$ and  the operator $K$ is given by
$$
K=({\bf 1}_p \pa +a) - b({\bf 1}_s \pa +d)^{-1} c.
\eqno(4.22)$$
Here
$\theta_i$, $a, b, c, d$ are the fields parametrizing the $\Theta$-gauge
according to (2.22).
Since $\psi_1$ is gauge invariant, the operators $L$
and $L_{\Theta}$ attached to
such points of ${\cal M}_{\rm DS}$
and $\Theta$ that lie on the same gauge orbit are equal: That is
$L_{\Theta}$ (4.21) is a factorized form  of $L$ (4.13).
In particular,
$$
M = M_{\Theta},
\eqno(4.23)$$
 where  $M_{\Theta}:= \{\, L_{\Theta}\}$ is the set of
operators $L_{\Theta}$  (4.21).
Note that if $r=2m$ is  even,  there appear $(m-1)$ factors before
$(\Delta K)$ and $m$ factors after $(\Delta K)$ in (4.21).

\smallskip

\noindent
{\it Remark 4.1.}
Following Section 2, we  can define an
alternative version of the reduction procedure using the same
$\Lambda$ (2.9), but replacing $H$ in the grading operator (2.7) by
$$
H_k:=I_0 +k \,{\rm  diag\,}\left({\bf 1}_{pr}, {\bf 0}_s\right)
\qquad\hbox{with}\quad  \hbox{any}\qquad
k=0, \pm {1\over 2}, \pm 1,\dots, \pm {r-1\over 2},
\eqno(4.24)$$
where $I_0$ is the $sl_2$ generator (2.8a).
In fact, the reduced phase space would be the same for any $k$ since
$\M_{\rm DS}$ would still be a global gauge slice.
Different $\Theta$-type gauges would arise, giving rise
to factorizations of the Lax operator $L$ similar to (4.21),
but with a factor like  $K$ in (4.22) appearing in a different
position.
The associated modified KdV systems
would all be isomorphic by permutations of the variables,
except for the possibility of a further factorization of the factor $K$,
which will be discussed in Section 6.
Modifications of the $I_0$-grading like those given by the $H_k$
above  were introduced in  \q{\FORTW} in connection with
gradings compatible with a given $sl_2$ embedding.
This method was also used in \q{\De}, where
the factorization (4.21)  (but not the results below)
was derived  in the case $p=s=1$.
\smallskip

We  now consider the relationship between the set
$M_K$ of operators $K$ of the form (4.22)
and the space $\Theta_0:=\widetilde{gl}_{p+s}$.
Parametrizing the general element $\theta_0\in \Theta_0$
as  $\theta_0=\pmatrix{a&b\cr c&d\cr}$ like  in (2.23),
we have the mapping
$$
\eta: \Theta_0 \rightarrow M_K,
\qquad
\eta(\theta_0):={\bf 1}_p \pa + a - b( {\bf 1}_s\pa +d)^{-1} c.
\eqno(4.25)$$
The natural PB on the space $\Theta_0$ is given by the appropriate
term in (2.24),
$$
\{  \bar f, \bar h \}(\theta_0)=
\int_{S^1}\,
 \tr \left(\theta_0 \left[{\delta \bar f \over \delta \theta_0},
{\delta  \bar h \over \delta \theta_0}\right]
-{\delta \bar f \over \delta \theta_0}
 \left({\delta \bar h \over \delta \theta_0}\right)'\right)
\eqno(4.26)$$
for $\bar f$, $\bar h$ smooth functions on $\Theta_0$.
We have the following result.

\smallskip
\noindent
{\bf Proposition 4.1.} {\it The set $M_K\subset {\cal A}$
of operators $K$ of the form (4.22) is
a Poisson submanifold of ${\cal A}$ with respect to the
quadratic Gelfand-Dickey PB (4.5).
The mapping $\eta$ (4.25) is a Poisson mapping with respect to the free
current algebra PB (4.26) on $\Theta_0$ and the quadratic Gelfand-Dickey
PB on $M_K$.}
\smallskip

\noindent
{\it Proof.}
Let $f$ and $h$ be arbitrary (smooth) functions on ${\cal A}$.
The statement of the proposition is equivalent to the equality
$$
\{ f, h\}^{(2)}\circ \eta =\{ f \circ \eta, h\circ \eta \},
\eqno(4.27)$$
where the l.h.s.\ is determined by the quadratic Gelfand-Dickey
PB (4.5) and the r.h.s.\  is determined  by  the free current algebra
 PB (4.26).
This equality can be verified by a straightforward computation.
Since the computation is rather long, we relegated it to
Appendix A.
{\it Q.E.D.}
 \smallskip

Thanks to Proposition 4.1, we are now ready to describe the relationship
between the PBs on ${\cal M}_c/{\cal N}$ induced from
the PBs (1.2-3) on ${\cal M}$ by the Drinfeld-Sokolov  reduction
and the Gelfand-Dickey PBs  on $M\subset {\cal A}$.

\smallskip
\noindent
{\bf Theorem 4.2.}~{\it The set $M\subset {\cal A}$ of operators
$L$ (4.13)
is a Poisson submanifold with respect to the quadratic Gelfand-Dickey
PB (4.5).
The mapping $\bar\pi: {\cal M}_c/{\cal N}\rightarrow M$ (4.15b)
defined by the elimination procedure  is a Poisson mapping,
where $M$ is endowed with the quadratic Gelfand-Dickey PB and
${\cal M}_c/{\cal N}$ is endowed with the reduced second PB
resulting from the current algebra
PB (1.2) on ${\cal M}$ by means of the Drinfeld-Sokolov reduction.}
\smallskip

\noindent
{\it Proof.}
We have seen that $M_K\subset {\cal A}$ is a Poisson submanifold
with respect to the PB (4.5).
It is well-known (see \q{\FHM} Section 2.2.)  that the other factors
$\{ \Delta(\pa + \theta_i) \,\vert \,
\theta_i\in\widetilde{gl}_p\,\}\subset {\cal A}$
appearing in $L_{\Theta}$ (4.21) are also Poisson submanifolds
with respect to the PB (4.5), which coincides with
the free current algebra  on these submanifolds:
$$
\{  \bar f, \bar h \}^{(2)}(\theta_i)=
\int_{S^1}\,
 \tr \left(\theta_i \left[{\delta  \bar f \over \delta \theta_i},
{\delta \bar h \over \delta \theta_i}\right]
-{\delta \bar f \over \delta \theta_i}
 \left({\delta \bar h \over \delta \theta_i}\right)'\right) ,
\eqno(4.28)$$
for $\bar f$, $\bar h$ smooth functions of $\theta_i$.
Recall the ``product property'' of the quadratic bracket
according to which the product of
Poisson submanifolds is also a Poisson submanifold.
The statement of the theorem follows from this  on account of (4.23)
and the fact that in the
$\Theta$-gauge (2.21)
the reduced second PB  is given by the current
algebra (2.24).
{\it Q.E.D.}
  \smallskip

So far we have dealt with the reduced second PB on the
bihamiltonian manifold   ${\cal M}_c/{\cal N}$.
Remember that the reduced first PB on ${\cal M}_c/{\cal N}$,
which results from the PB (1.3) on ${\cal M}$, is the Lie derivative
of the reduced  second PB  with respect to the infinitesimal
generator of the one parameter
group action on ${\cal M}_c/{\cal N}$ induced by the
one parameter group action (2.18) on ${\cal M}_c$.
The manifold $M$ of Lax operators (4.13)
is also a bihamiltonian manifold since the linear Gelfand-Dickey
PB $\{\cdot , \cdot \}^{(1)}$ (4.8) on ${\cal A}$ can be restricted
to $M\subset {\cal A}$.
To see this recall that
 the linear PB
 $\{\cdot ,\cdot \}^{(1)}$
(4.8) on ${\cal A}$ is the Lie derivative of the
quadratic PB $\{\cdot ,\cdot \}^{(2)}$ (4.5) on ${\cal A}$
with respect to the infinitesimal generator of the one parameter
group action (4.7) on ${\cal A}$  and notice that this group maps
$M\subset {\cal A}$ to itself.
This together  with the first statement of Theorem 4.2 implies that
$M\subset {\cal A}$  is in fact a Poisson submanifold also with
respect to the linear PB (4.8).
Theorem 4.2 and Theorem 4.3 below state  that
the mapping $\bar \pi$ (4.15b) is a Poisson mapping of
bihamiltonian manifolds.

\smallskip
\noindent
{\bf Theorem 4.3.}~{\it The manifold $M\subset {\cal A}$ is a
Poisson submanifold with respect to the linear Gelfand-Dickey
PB (4.8) on ${\cal A}$  and the mapping
$\bar \pi: {\cal M}_c/{\cal N}\rightarrow M$ (4.15b) is a Poisson
 mapping with respect to the reduced first PB (1.3) on ${\cal M}_c/{\cal N}$
and the linear Gelfand-Dickey PB (4.8) on $M$.}

\smallskip
\noindent
{\it Proof.}
Using the
identification ${\cal M}_c/{\cal N}\simeq {\cal M}_{\rm DS}$,
it is enough to show
that the mapping $\bar \pi: {\cal M}_{\rm DS} \rightarrow M$ (4.15)
intertwines the one parameter group action (2.18) on ${\cal M}_{\rm DS}$
and the one parameter group action (4.7) on $M$.
By (4.14),
this  follows from the elimination procedure that converts
${\cal L}\in {\cal M}_{\rm DS}$ appearing in (4.9) into $L\in M$
appearing in (4.12).
{\it Q.E.D.}
\smallskip

The above results may be used to determine the reduced first and second
PBs between such functions on
${\cal M}_{\rm red}={\cal M}_c/{\cal N}$
which are of the form ${\cal F} \circ \bar \pi$, ${\cal H}\circ \bar \pi$
with functions ${\cal F}$, ${\cal H}$ on $M$.
We now  wish  to present the explicit formulae for the PBs
of arbitrary functions on ${\cal M}_{\rm red}$.
We can parametrize ${\cal M}_{\rm red}\simeq {\cal M}_{\rm DS}$
by the variables $u_i, z_\pm, w$ or equivalently by
the variables $\ell, z_\pm, w$ where $\ell$ is the positive
part of $L$,
$$
\ell=\Delta^r \pa^r +\sum_{i=1}^r u_i \pa^{r-i},
\qquad
L=\ell + z_+ ({\bf 1}_s\pa +w)^{-1} z_-.
\eqno(4.29)$$
With the aid of the usual functional derivatives,
the arbitrary variation $\delta H$ of a function $H$ on ${\cal M}_{\rm red}$
may be written as
$$
\delta H = \int_{S^1} {\rm tr}\left( \sum_{i=1}^r
{\delta H\over \delta u_i}\delta u_i +
{\delta H\over \delta z_+}\delta z_+
+{\delta H\over \delta z_-}\delta z_-
+{\delta H\over \delta w}\delta w\right),
\eqno(4.30{\rm a})$$
or equivalently  as
$$
\delta H={\rm Tr}\left({\delta H\over \delta \ell}\delta \ell\right)
+
\int_{S^1} {\rm tr}\left(
{\delta H\over \delta z_+}\delta z_+
+{\delta H\over \delta z_-}\delta z_-
+{\delta H\over \delta w}\delta w\right)
\eqno(4.30{\rm b})$$
using the Adler trace (4.3) and the definition
$$
{\delta H\over \delta \ell}=\sum_{i=1}^r \pa^{i-r-1}
{\delta H\over \delta u_i}.
\eqno(4.30{\rm c})$$
We can write  the reduced first and second PBs,
denoted by $\{ F,H \}^*_i$ $(i=1,2)$,
of  the arbitrary functions $F, H$ as follows:
$$
\{ F, H\}^*_i= {\bf X}_H^i(F) =
{\rm Tr}\left({\delta F\over \delta \ell}{\bf X}_H^i( \ell)\right)
+
\int_{S^1} {\rm tr}\left(
{\delta F\over \delta z_+}{\bf X}_H^i( z_+)
+{\delta F\over \delta z_-}{\bf X}_H^i(z_-)
+{\delta F\over \delta w}{\bf X}_H^i(w)\right),
\eqno(4.31)$$
where ${\bf X}_H^i$ is the corresponding Hamiltonian vector
field associated to the function $H$ and
${\bf X}_H^i(G)= \langle \delta G, {\bf X}_H^i\rangle$ is the
derivative of the function $G$ with respect to ${\bf X}_H^i$.
An arbitrary PDO $A$ can be expanded either in the right form
$A=\sum_k A_k \pa^k$ or in the left  form $A=\sum_k \pa^k \tilde A_k$.
Using these expansions we define
$$
P_0(A):=A_0,
\qquad
P_0^\dagger(A):=\tilde A_0.
\eqno(4.32)$$
It will be also convenient to rewrite the PDO $(\pa +w)$ as
$$
(\pa +w) = W^{-1} \pa\, W \qquad\hbox{with}\qquad  w=W^{-1}W',
\eqno(4.33)$$
where the $GL_s$ valued function $W$ on ${\Real}$ is uniquely associated
to $w$ by (4.33) and the condition $W(0)={\bf 1}_s$.
In terms of these notations
we can now write down ${\bf X}_H^i$.

\smallskip
\noindent
{\bf Theorem 4.4.}~{\it The Hamiltonian vector field
${\bf X}_H^2$  associated to a function $H$ on
${\cal M}_c/{\cal N}$
by means of the reduced second  PB is given by
$$\eqalign{
&{\bf X}_H^2(\ell)=\left(\ell {\delta H\over \delta \ell}\right)_+ \ell
-\ell \left({\delta H\over \delta \ell} \ell\right)_+
+\left( \ell {\delta H\over \delta z_-} (\pa +w)^{-1} z_-\right)_+
-\left( z_+(\pa +w)^{-1}{\delta H\over \delta z_+}\ell\right)_+ ,\cr
&{\bf X}_H^2(z_+)=P_0\left(\ell\left({\delta H\over \delta\ell}z_+
+{\delta H\over \delta z_-}\right)W^{-1}\right)W-z_+{\delta H\over \delta w},
\cr
&{\bf X}_H^2(z_-)=-W^{-1}P_0^\dagger\left(W\left(z_-{\delta H\over \delta\ell}
+{\delta H\over \delta z_+}\right)\ell\right)+{\delta H\over \delta w}z_-,\cr
&{\bf X}_H^2(w)={\delta H\over \delta z_+}z_+ - z_-{\delta H\over \delta z_-}
+\left[ {\delta H\over \delta w}, w\right] -
\left({\delta H\over \delta w}\right)'.\cr}
\eqno(4.34)$$
The Hamiltonian vector field ${\bf X}_H^1$ corresponding to the
reduced first PB reads
$$
{\bf X}^1_H(\ell)=\left[\ell, {\delta H\over \delta \ell}\right]_+,
\quad
{\bf X}^1_H(z_\pm )=\pm {\delta H\over \delta z_\mp},\quad
{\bf X}^1_H(w)=0.
\eqno(4.35)$$}

\noindent
{\it Proof.} On account of Theorem 4.2,
in order to verify (4.34) it is enough to compute
the Hamiltonian vector fields separately for functions on
${\cal M}_c/{\cal N}$ that have the
special form $\int_{S^1}{\rm tr}\left( f_{\pm} z_\pm\right)$ or
$\int_{S^1} {\rm tr} \left(\alpha w\right)$
with some  matrix valued test functions $f_\pm$, $\alpha$.
This computation is presented in Appendix B.
After writing down the formula of the reduced second PB from
(4.34), it is easy to compute its Lie derivative with respect to the
vector field $V$ on ${\cal M}_{\rm red}$ given by
$$
V(\ell)={\bf 1}_p, \quad V(z_\pm)=V(w)=0.
\eqno(4.36)$$
We know from (2.18) that this gives the formula of the reduced first PB
and we find
$$
\{ F, H\}^*_1=-{\rm Tr}
\left( \ell \left[{\delta F\over \delta \ell},
{\delta H\over \delta \ell}\right]\right)
+\int {\rm tr}\left(
{\delta F\over \delta z_+}{\delta H\over \delta z_-}
-{\delta F\over \delta z_-}{\delta H\over \delta z_+}\right),
\eqno(4.37)$$
which is equivalent to (4.35). For computational details,
see  Appendix B. {\it Q.E.D}
\smallskip

Note that the introduction of the ``integrating factor''
$W$ in the above is only a notational trick which we used to get compact
formulae. For instance,
$$\eqalign{
&P_0\left(\ell {\delta H\over \delta z_-} W^{-1}\right)W=
(-X)^{-r}\tilde {\cal D}^{r}\left({\delta H\over \delta z_-} \right)+
\sum_{k=1}^r u_k \tilde {\cal D}^{r-k}\left({\delta H\over \delta z_-} \right),
\cr
&W^{-1}P_0^\dagger\left(W
{\delta H\over \delta z_+}\ell\right)=
{\cal D}^{r}\left({\delta H\over \delta z_+} X^{-r}\right)+
\sum_{k=1}^r (-1)^{r-k} {\cal D}^{r-k}
\left({\delta H\over \delta z_+} u_k\right),\cr}
\eqno(4.38{\rm a})$$
where for arbitrary $s\times p$ and $p\times s$ matrix
valued functions $\beta$ and $\tilde \beta$ on $S^1$  we define
their covariant
derivatives
$$
{\cal D}(\beta):=(\beta'+w\beta ),
\qquad
{\tilde {\cal D}}(\tilde \beta):=({\tilde \beta}' -\tilde \beta w ).
\eqno(4.38{\rm b})$$
All other terms containing $W$ can be rewritten in terms of $w$
in an analogous fashion.

Let us now consider a function $H$ that depends on $\ell, z_\pm, w$
only through the Lax operator $L$ in (4.29),
$$
H(\ell, z_+, z_-, w)={\cal H}(L),
\eqno(4.39)$$
i.e., $H={\cal H}\circ \bar \pi$ for some function $\cal H$ on $M$.
Naturally,  in this case we  have the equality
$$
\delta H ={\rm Tr}\left( {\delta {\cal H}\over \delta L} \delta L\right).
\eqno(4.40)$$
Comparing  (4.40)
and (4.30b) using  (4.29) leads to
$$
\left(\delta {\cal H}\over \delta L\right)_-={\delta H\over \delta \ell},
\eqno(4.41{\rm a})$$
up to terms that do not contribute in (4.40), and to
$$
{\delta H\over \delta z_+}=
W^{-1}P_0^\dagger\left(Wz_-\left(\delta {\cal H}\over \delta L\right)_+\right),
\qquad
{\delta H\over \delta z_-}=
P_0\left(\left({\delta {\cal H}\over \delta L}\right)_+z_+W^{-1}\right)W,
\eqno(4.41{\rm b})$$
$$
{\delta H\over \delta w}=-{\rm res}
\left(W^{-1}\pa^{-1}Wz_-\left({\delta {\cal H}\over
\delta L}\right)_+ z_+ W^{-1}
\pa^{-1} W\right).
\eqno(4.41{\rm c})$$
The PDO   ${\delta {\cal H}\over \delta L}$
may be regarded as an arbitrary  solution of these requirements.
Using these requirements and certain identities, e.g.~the identities
$$\eqalign{
&\left(L\left({\delta {\cal H}\over \delta L}\right)_+\right)_-=
z_+ (\pa +w)^{-1} W^{-1} P_0^\dagger\left(Wz_-
\left({\delta {\cal H}\over \delta L}\right)_+\right),\cr
&\left(\left({\delta {\cal H}\over \delta L}\right)_+L\right)_-=
 P_0\left(
\left({\delta {\cal H}\over \delta L}\right)_+ z_+W^{-1}\right)
 W(\pa +w)^{-1}z_-,\cr}
\eqno(4.42)$$
for $H$ in (4.39) we  can rewrite the
Hamiltonian vectors fields ${\bf X}_H^i$ of Theorem 4.4 as
$$\eqalign{
&{\bf X}_H^2(L)=\left( L{\delta {\cal H}\over\delta L} \right)_+L
-L\left( {\delta {\cal H}\over\delta L}L \right)_+,\cr
&{\bf X}_H^2(z_+) =
P_0\left(
L{\delta {\cal H}\over\delta L}z_+W^{-1}\right)W,\cr
&{\bf X}_H^2( z_-)= -W^{-1}
P_0^\dagger\left( Wz_-{\delta {\cal H}\over\delta L}L \right),\cr
&{\bf X}_H^2(w)=0,\cr}
\eqno(4.43)$$
and respectively as
$$\eqalign{
&{\bf X}^1_H(L)=
\left[L, \left({\delta {\cal H}\over \delta L}\right)_-\right]_+
-\left[L, \left({\delta {\cal H}\over \delta L}\right)_+\right]_-,\cr
&{\bf X}_H^1(z_+)=
P_0\left( {\delta {\cal H}\over \delta L} z_+W^{-1}\right)W,\cr
&{\bf X}_H^1(z_-)=-
W^{-1}P_0^\dagger\left(Wz_- {\delta {\cal H}\over \delta L} \right),\cr
&{\bf X}_H^1(w)=0.\cr}
\eqno(4.44)$$
These formulae are  consistent with the
claims of Theorems 4.2 and 4.3.
They will be used at the  end of Section 5 to  determine
the  evolution equations of the KdV type hierarchy
that results from the  Drinfeld-Sokolov  reduction.

\smallskip
\noindent{\it Remark 4.2.}
The compact presentation of the formulae for the PBs in (4.43), (4.44),
(4.34) and (4.35) was suggested by the formulae in \q{\OS}. It may
be verified that our PBs reduce to those in \q{\OS}
 in the scalar case $p=s=1$ upon constraining to $w=0$.
\smallskip

\noindent{\it Remark 4.3.}
Notice that for $r=p=s=1$  the AKS hierarchy on $\M$
is the  $gl_2$ version of the well-known AKNS hierarchy.
For $r=1$ and  arbitrary $p$, $s$ (see Remark 2.2.),
it is reasonable to call
the system on $\M$ a generalized AKNS hierarchy.
For the generalized AKNS hierarchy
the pseudo-differential Lax operator associated to
$\L\in \M$  becomes just the operator $\Delta K$ (4.22).
For this reason we can  call
$\Delta K$ in the factorization (4.21)
the ``AKNS factor''.
(It is an easy exercise to directly verify the equivalence
between the respective  formulae (1.2-3) and (4.34-35) for $r=1$.)
In the simplest case $r=p=s=1$ the connection between the AKNS
hierarchy on $\M$ and the constrained KP hierarchy on $M$ was
observed in \q{\Cheng,\OS,\FW} too.
The nonlinear  Schr\"odinger
(NLS) hierarchy results from constraining the AKNS hierarchy,
and it has many generalizations \q{\FK}.
The connection between generalized AKNS
($N$-wave) and NLS systems and constrained (matrix) KP systems
given by the above result and the results  in Section 5
can be extended to more general cases than those treated in
this paper.
\smallskip

To summarize,
in this section we  investigated the properties of the
mapping $\pi$ (4.15)
and found that  the push forward
of the bihamiltonian structure
on ${\cal M}_c/{\cal N}$ induced by the
Drinfeld-Sokolov  reduction is the Gelfand-Dickey bihamiltonian
structure on $M={\bar \pi}\left({\cal M}_c/{\cal N}\right)$.
We  presented the explicit formulae of the PBs on ${\cal M}_c/{\cal N}$.
We have also established the factorization (4.21) of the Lax
operator $L\in M$ using the Miura map
$\mu: \Theta\rightarrow {\cal M}_{\rm DS}$
mentioned in the paragraph following (2.24).
This Miura map together with (2.24),
and formula (4.34) of Theorem 4.4 provide
alternative means for computing the reduced second PB.
The commuting Hamiltonians
furnished by the local monodromy invariants
of ${\cal L}\in {\cal M}_c$ define
generalized KdV and   modified KdV
systems in terms of the  gauge slices
${\cal M}_{\rm DS}$ and $\Theta$, respectively.

\bigskip
\centerline{\bf 5.~Local monodromy invariants and residues of
fractional powers}
\medskip

In Section 4 we established a relationship between the
Poisson brackets on the reduced phase space
${\cal M}_{\rm red}\simeq {\cal M}_c/{\cal N}$ and
the Gelfand-Dickey Poisson brackets on $M\subset {\cal A}$.
Our next task is to characterize the Hamiltonians
generated by the  {\it local} monodromy invariants of ${\cal L}\in {\cal M}_c$.
These Hamiltonians, which define the commuting hierarchy of evolution equations
on ${\cal M}_{\rm red}$, turn out to admit a description purely in terms
of the Lax operator $L\in M$ attached to ${\cal L}$ by
the elimination procedure, $L=\pi({\cal L})$.
Namely, the Hamiltonians defined by the
local monodromy invariants of ${\cal L}$ will be identified
in terms of  integrals of componentwise residues of fractional powers
of the diagonal PDO $\hat L$  obtained
by diagonalizing $L$ in the PDO algebra ${\cal A}$.
This identification will result from computing the local monodromy invariants
of ${\cal L}\in {\cal M}_{\rm DS}$
in two alternative ways: first using the procedure of (1.10-11) outlined
in Section 1 and then using the diagonalization of $L$ combined with a
reverse of the elimination procedure.
The same method was used in \q{\FHM},
but the presence of the singlets in the partition (2.1)
gives rise to  complications requiring
a non-trivial refinement of the argument.

\bigskip
\centerline{\bf 5.1.~Local monodromy invariants and solutions of exponential
 type}
\medskip

We wish to compute the local invariants of the monodromy matrix
$T$ associated to the linear problem
$$
{\cal L}\Psi=0 \quad \Longleftrightarrow
\quad (\pa_x + j(x) + \Lambda) \Psi(x) =0,
\eqno(5.1)$$
where $j(x+2\pi )=j(x)$ since ${\cal L}=(\pa + j +\Lambda)\in {\cal M}_c$.
If $\Psi: {\Real} \rightarrow GL_n$ is a solution of (5.1),
which means that the columns of the matrix $\Psi$ are a complete set
of solutions, then the monodromy matrix is given by
$$
T=\Psi(2\pi) \Psi^{-1}(0).
\eqno(5.2)$$
Following the procedure outlined in Section 1,
perform the transformation
$$
{\cal L}\mapsto  \tilde {\cal L}=
e^\Xi {\cal L} e^{-\Xi}=(\pa + h +\Lambda),
\qquad
\tilde \Psi = e^\Xi \Psi,
\eqno(5.3)$$
where
$$
\Xi\in \left({\rm Im}({\rm ad\,} \Lambda)\right)_{<0},\qquad
h \in \left({\rm Ker}({\rm ad\,} \Lambda)\right)_{<1},
\eqno(5.4)$$
and the subscripts refer to the grading $d$ in (2.7).
The fact that $\Xi$ and $h$ are uniquely determined differential polynomial
expressions in the components of $j$ implies that $\Xi(j(x))$
and $h(j(x))$ are periodic functions of $x\in {\Real}$.
It follows that the invariants of the monodromy matrix $T$
that we are interested in are the same as the invariants of the
transformed monodromy matrix
$$
\tilde T :=\tilde\Psi (2\pi) \tilde\Psi^{-1}(0)=G^{-1}(0) T G(0),
\eqno(5.5)$$
with the definition
$$
G(x):=\exp\left(-\Xi(j(x))\right).
\eqno(5.6)$$
Using the notation (2.2-3),  $h$ in (5.3) may be written as
$$
h(j) =\left(\matrix{A(j)&0\cr 0&D(j)\cr}\right),
\eqno(5.7{\rm a})$$
where $A(j)$ and $D(j)$ are uniquely determined series of the form
$$
A(j)=\sum_{k=0}^\infty \sum_{i=1}^p h_{k,i}(j)\, \Lambda_r^{-k}\otimes e_{i,i},
\qquad
D(j)=\sum_{k=0}^\infty \lambda^{-k}\, e_{r+1,r+1}\otimes D_k(j),
\eqno(5.7{\rm b})$$
with  $h_{k,i}(j(x))\in {\Complex}$ and $D_k(j(x))\in gl_s$.
A basis of the centre of ${\rm Ker}({\rm ad\,} \Lambda)\subset \ell(gl_n)$
is given in (2.5) and (2.11).
The  Hamiltonians defined by the procedure in (1.10-11) are  then
$$
H_{k,i}(j):=\int_0^{2\pi} dx\, h_{k,i}(j(x)),
\qquad  i=1,\ldots, p,\,\,\, k=0,1,2,\ldots\,
\eqno(5.8)$$
and
$$
E_k(j):=\int_{0}^{2\pi} dx\, \tr D_k(j(x)).
\eqno(5.9)$$
Taking the trace of eq.~(5.3) using the identity
$\Lambda_r^{-kr}=\lambda^{-k} {\bf 1}_r$,
we  obtain the equality
$$
E_{k}(j)=\delta_{k,0} \int_{0}^{2\pi}dx\, \tr j(x)
-r\sum_{i=1}^p H_{kr,i}(j).
\eqno(5.10)$$
Since  $j\mapsto \int_{0}^{2\pi}dx\, \tr j(x)$ defines  a Casimir function
with respect to both Poisson brackets on ${\cal M}_{\rm red}$,
this equality means that the complete set of independent
Hamiltonians associated to the centre of ${\rm Ker}({\rm ad\,} \Lambda)$
is given by the $H_{k,i}(j)$ above.
General arguments that go back to the
r-matrix (AKS) construction
(see e.g.~\q{\RSTS}) guarantee
that the Hamiltonians  $H_{k,i}$ are in involution
(commute among themselves)  since
they can be interpreted as particular monodromy invariants.
To explain this interpretation,   notice
that the transformed linear problem $\tilde {\cal L}\tilde\Psi =0$
has the solution
$$
\tilde \Psi(x)=
\left(\matrix{\tilde \Psi_{11}(x)&0\cr 0&\tilde\Psi_{22}(x)\cr}\right),
\eqno(5.11)$$
where
$$
\tilde \Psi_{11}(x)=\exp\left(- x \Lambda_r\otimes \Gamma
 -\sum_{k=0}^\infty \sum_{i=1}^p\int_0^{x}d\xi\, h_{k,i}(j(\xi))\,
\Lambda_r^{-k}\otimes e_{i,i}\right),
\eqno(5.12)$$
and
$$
\tilde \Psi_{22}(x)={\cal P}\exp\left(-\int_0^x d\xi\,
 \sum_{k=0}^\infty \lambda^{-k} D_k(j(\xi)) \right).
\eqno(5.13)$$
The corresponding monodromy matrix is
$$
\tilde T = \left(\matrix{\tilde \Psi_{11}(2\pi)&0\cr 0&\tilde
\Psi_{22}(2\pi)\cr}\right).
\eqno(5.14)$$
To diagonalize $\tilde T$,  consider
an $r$th root $\zeta$ of $\lambda$, $\zeta^r=\lambda$, and define
$$
\zeta_a:=\zeta \omega^a
\quad\hbox{with}\quad
\omega:=\exp\left(2i\pi/ r\right).
\eqno(5.15)$$
The matrix $\Lambda_r$ is conjugate to
$$
\tilde \Lambda_r: ={\rm diag\,}\left(\zeta_1, \zeta_2 ,\ldots,
\zeta_r\right).
\eqno(5.16)$$
In fact, we  have
$$
\Lambda_r = S \tilde\Lambda_r S^{-1},
\eqno(5.17{\rm a})$$
with
$$
S_{ab}={1\over\sqrt{r}}\left(\zeta_b\right)^{a-1},
\qquad
\left(S^{-1}\right)_{ab}={1\over \sqrt{r}}\left(\zeta_a\right)^{1-b}.
\eqno(5.17{\rm b})$$
Using this conjugation the  upper block
$\tilde \Psi_{11}(2\pi)$ of $\tilde T$ becomes  diagonal,
$$
\tilde \Psi_{11}(2\pi)=
\left(S\otimes {\bf 1}_p\right)
\exp\left(-2\pi \tilde\Lambda_r\otimes \Gamma
 -\sum_{k=0}^\infty \sum_{i=1}^p H_{k,i}(j)\, \tilde \Lambda_r^{-k}
\otimes e_{i,i}\right) \left(S^{-1} \otimes {\bf 1}_p\right).
\eqno(5.18)$$
Hence, up to the constant $-2\pi \tilde\Lambda_r\otimes \Gamma$,
 the Hamiltonians $H_{k,i}(j)$
can be identified as expansion coefficients defining the expansions
of logarithms of certain eigenvalues of the monodromy matrix
around $\zeta \approx \infty$.
As usual, this expansion has to be interpreted as an asymptotic
--- or formal --- series.
It is clear from (5.13) that the spectral invariants
determined by the ``lower block''
$\tilde \Psi_{22}(2\pi)$ of the monodromy matrix are in general,
except the functionals $E_{k}(j)$ given above, {\it non-local}
functionals of $j$.

We have seen that  the local monodromy invariants
$H_{k,i}(j)$ associated to the centre of
${\rm Ker}({\rm ad\,} \Lambda)$
are determined purely in terms of the upper block $\tilde\Psi_{11}(2\pi )$
of the transformed monodromy matrix $\tilde T$.
To see what this means in terms of the original linear problem (5.1)
consider the solution $\Psi$ given by
$$
\Psi(x):=G(x)\tilde \Psi(x)\pmatrix{S\otimes {\bf 1}_p&0\cr 0&{\bf 1}_s\cr}.
\eqno(5.19)$$
Using a block notation similar to (2.2-3),
$$
\Psi(x)=\pmatrix{\Psi_{11}(x)&\Psi_{12}(x)\cr \Psi_{21}(x)&\Psi_{22}(x)\cr},
\qquad
G(x)=e^{-\Xi(j(x))}=\pmatrix{ G_{11}(x)& G_{12}(x)\cr
G_{21}(x)& G_{22}(x)\cr},
\eqno(5.20)$$
we have
$$
\pmatrix{\Psi_{11}\cr\Psi_{21}\cr}=
\pmatrix{ G_{11}\tilde\Psi_{11} S \otimes {\bf 1}_p\cr
G_{21} \tilde\Psi_{11} S\otimes {\bf 1}_p\cr}
\eqno(5.21{\rm a})$$
and
$$
\pmatrix{\Psi_{12}\cr\Psi_{22}\cr}=
\pmatrix{ G_{12}\tilde\Psi_{22}\cr
G_{22} \tilde\Psi_{22}\cr}.
\eqno(5.21{\rm b})$$
It follows from (5.4) that the infinite series $\Xi(j(x))$
contains only {\it non-positive} powers of $\lambda$
and therefore the entries of the matrix $G$ are given by similar
series.
This together with  (5.12-13), (5.17) implies that
 the  column vector solutions of the linear problem
${\cal L}\psi=0$ comprising the matrices in (5.21a) and
in (5.21b) are qualitatively different.
They are different in the sense that
--- apart from the constant  $S(\zeta)\otimes {\bf 1}_p$ that we
included in the definition of $\Psi$  for later convenience ---
the columns of the matrix in (5.21a)  have the form of descending
series in non-positive powers
of $\zeta$ multiplied by a leading term of the type
$e^{-x\zeta_b \Gamma}$ while
the columns of the matrix in (5.21b) do not contain such a
leading term only a descending
series in non-positive powers of $\zeta$.
We refer to the  series solutions
containing a leading term $e^{-x\zeta_b \Gamma}$ as
``solutions of exponential type''.
The solutions
of exponential type
{\it contain all information about the
local monodromy invariants} since
the matrix $\tilde \Psi_{11}(2\pi)$ in (5.18), whose eigenvalues
 generate  the Hamiltonians $H_{k,i}(j)$, is conjugate to the matrix
$$
\Psi_{11}(2\pi ) \left(\Psi_{11}(0)\right)^{-1}=
G_{11}(0) \tilde\Psi_{11}(2\pi)\left(G_{11}(0)\right)^{-1}.
\eqno(5.22)$$
It is convenient to write the
$(rp+s)\times (rp)$ matrix in (5.21a)
in the following detailed form:
$$
\pmatrix{\Psi_{11}\cr\Psi_{21}\cr}
=\pmatrix{
\psi_{1}^1&\cdots&\psi_{1}^r\cr
\psi_{2}^1&\cdots&\psi_{2}^r\cr
\vdots&\phantom{\cdots}&\vdots\cr
\psi_{r}^1&\cdots&\psi_{r}^r\cr
\phi^1&\cdots&\phi^r\cr}, \qquad
( \psi_a^b \in gl_p, \,\, \phi^b\in {\rm mat}(s\times p)).
\eqno(5.23)$$
Putting the above formulae together we have
$$\eqalign{
&\psi_a^b(x,\zeta)=
{1\over\sqrt{r}}\sum_{c=1}^r G_{11}^{ac}(x,\lambda)
\left(\zeta_b\right)^{c-1}\exp\left(-{\cal D}(x,\zeta_b)-x\zeta_b
\Gamma\right),\cr
 &\phi^b(x,\zeta)=
{1\over\sqrt{r}}\sum_{c=1}^r G_{21}^{c}(x,\lambda)
\left(\zeta_b\right)^{c-1}\exp\left(-{\cal D}(x,\zeta_b)-x\zeta_b
\Gamma\right),
\cr}
\eqno(5.24{\rm a})$$
where $G_{11}^{ac}$ and $G_{21}^c$ are $p\times p$ and $s\times p$
matrices, respectively, and ${\cal D}(x,\zeta)$ is the $p\times p$ diagonal
matrix series
$$
{\cal D}(x, \zeta)=\sum_{k=0}^\infty \zeta^{-k}
\int_0^x d\xi\, {\rm diag}\bigl( h_{k,1}\left(j(\xi)\bigr), \ldots,
h_{k,p}\left(j(\xi)\right)\right).
\eqno(5.24{\rm b})$$
Observe that all the gauge invariant components $\psi_1^b(x,\zeta)$
can be  obtained from $\psi_1^r(x,\zeta)$ simply  replacing
the argument $\zeta=\zeta_r$ by $\zeta_b$.
By the reverse of the elimination procedure (see (4.11)),
this means that the complete set of solutions of exponential type can
be recovered from $\psi_1^r(x,\zeta)$.
In particular, the local monodromy invariants $H_{k,i}$ in (5.8) can be read
 off from the relation
$$
\psi_1^r(x+2\pi,\zeta)=\psi_1^r(x,\zeta)\exp\bigl(-2\pi \zeta \Gamma
-\sum_{k=0}^\infty \zeta^{-k}{\rm diag}\left(H_{k,1},\ldots, H_{k,p}\right)
\bigr),
\eqno(5.25)$$
which is a consequence of (5.24).
This relation will play a crucial r\^ole in finding
the link between  the local monodromy invariants and the residues
of fractional powers, which is the ultimate aim of the present   section.

In this subsection we  obtained the matrix solution (5.23)
of exponential type  to the linear problem (5.1) by transforming
${\cal L}$ to $\tilde{\cal L}$ and imposing on the solution
$\tilde\Psi$ (5.11) of $\tilde{\cal L}\tilde\Psi=0$ the condition
 $\tilde\Psi={\bf 1}_n$ at $x=0$, see (5.12-13).
We noticed  that the matrix solution in (5.23)
is determined completely by the block $\psi_1^r$.
We then observed that the block $\psi_1^r(x,\zeta)$
directly encodes   the Hamiltonians of our interest,
the Hamiltonians $H_{k,i}$ given in (5.8),
through the relation (5.25).
In the next subsection we will
consider the consequence of looking at $\psi_1^r$ from a different point of
view, namely as a solution of  equation (4.12), which was obtained via the
elimination procedure. The result will be an explicit connection between
the family $\{H_{k,i}\}$ of local monodromy invariants and the components of
the residues of the fractional powers of the diagonalized form of $L$.

\vfill\eject

\bigskip
\centerline{\bf 5.2.~Solutions of exponential type and residues of
fractional powers}
\medskip

Consider the  $p\times p$ matrix PDO
$$
L=\Delta^r \pa^r + u_1 \pa^{r-1} +u_2 \pa^{r-2}
+ \cdots
+u_{r-1} \pa +  u_r + z_+ ({\bf 1}_s \pa + w)^{-1} z_-,
\quad
\Delta=-\Gamma^{-1},
\eqno(5.26)$$
attached to ${\cal L}\in {\cal M}_c$ by the elimination procedure.
The strategy of this subsection will be to determine
$\psi_1^r$ in (5.24) as a solution of the linear problem
$$
L \psi_1=\lambda \psi_1
\eqno(5.27{\rm a})$$
given by a (asymptotic  or formal) series of the form
$$
\psi_1(x,\zeta)=\left(\sum_{k=0}^\infty \chi_k(x)\zeta^{-k}\right)
e^{-x\zeta \Gamma}
\quad\hbox{with}\quad\chi_k\in\cinf({\Real},gl_p),\quad
 \det\left(\chi_0(x)\right)\neq 0.
\eqno(5.27{\rm b})$$
The elimination procedure implies that $\psi^r_1(x,\zeta)$
in (5.24) is a solution of (5.27a).
We shall see below that $\psi_1^r(x,\zeta)$ can be expanded in
the form given in (5.27b) and that  the solution of (5.27a), (5.27b)
is essentially unique.

That $\psi_1^r$ in (5.24) can be expanded  in
the form given in (5.27b) can be seen by
inspection.
The key step is to  check that the series $\Xi$ defining
$G=e^{-\Xi}={\bf 1}_n-\Xi +{1\over 2} \Xi^2\cdots$ in (5.20)
contains only negative powers of
$\lambda$ in its first row due to the grading condition (5.4),
which implies that
the leading term of $\psi^r_1$ comes from the unit matrix
contained in the $c=1$ contribution in the sum in (5.24a).
Computing the first term of the ``abelianised current'' $h(j)$ in (5.7),
one obtains
$$
h_{0,i}(j)= {1\over r} \left( (-\Gamma)^r u_1\right)_{ii},
\eqno(5.28{\rm a})$$
where $u_1$ is the gauge invariant component  of $j$ entering
the Lax operator $L$ (5.26) attached to ${\cal L}=(\pa + j+\Lambda)$.
It follows that when rewritten as a series  of the form (5.27b),
the leading term $\chi_0$ of $\psi_1^r$ in (5.24) is given by
$$\chi_0(x)
=\exp\left( -{1\over r}(-\Gamma)^{r} \int_0^x d\xi\,
\left(u_1(\xi)\right)_{\rm diag}\right),
\eqno(5.28{\rm b})$$
and indeed has non-zero determinant.
Incidentally, the constant factor $S(\zeta)$ was inserted in the definition
(5.19) to set
the leading power of $\zeta$ in (5.27b), which  multiplies the product of
$\chi_0$ (5.28b) and $e^{-x\zeta \Gamma }$, to be $\zeta^0$.

Below our aim is  to determine $\psi_1(x,\zeta)$ from equations
(5.27a) and (5.27b).
To make precise the meaning of equation (5.27a), which has been derived
by formally applying the elimination procedure,
we note that an arbitrary $p\times p$ matrix PDO
$\alpha =\sum_i \alpha_i(x) \pa^i$ acts on a series
of the form $\psi_1(x,\zeta)$ in (5.27b) as follows.
Defining the action of $\pa^i$ for {\it any} integer $i$ on
$e^{-x\zeta \Gamma}$ by
$\left(\pa^i e^{-x\zeta \Gamma}\right):=(-\zeta \Gamma)^i e^{-x\zeta \Gamma}$
one first associates the PDO $\chi$ to $\psi_1$ (5.27b) by writing
$$
\psi_1(x,\zeta) =\left( \chi e^{-x\zeta \Gamma}\right),
\quad\hbox{i.e.}\quad
\chi(x,\pa)=\sum_{k=0}^\infty \chi_k(x)\Delta^{-k}\pa^{-k}.
\eqno(5.29{\rm a})$$
Then $\left(\alpha \psi_1\right)(x,\zeta):=
\left(\beta e^{-x\zeta \Gamma}\right)$
with $\beta=\alpha \chi$ being defined by the
composition rule of PDOs (as in (4.2)).
To avoid confusion,  we stress that here the coefficienst of the PDOs
$\chi, \alpha, \beta$ are not required to be periodic functions on ${\Real}$.
When equation (5.27a) is understood in this sense it is easily seen
to be equivalent to the ``dressing equation''
$$
\chi^{-1} L \chi = \Delta^r \pa^r.
\eqno(5.29{\rm b})$$
This reformulation  of (5.27) is well-known, see \q{\Wi}
and references therein.
Using the  reformulation (5.29)  and the fact that $\Delta^r$ has
 distinct, non-zero
eigenvalues, it is not hard to
verify\footnote{${}^{2}$}{\ninerm A  proof
can be found in \q{\Wils} for the case when $(u_1)_{\rm diag}=0$ and
$\chi_0={\bf 1}_p$, but these assumptions can be dropped.}
that (5.27a) uniquely determines
the series solution $\psi_1(x,\zeta)$ of the form (5.27b)
up to multiplication
on the right
by a  constant ($x$-independent) series $c(\zeta)$ of the form
$$
c(\zeta)=\sum_{k=0}^\infty c_k \zeta^{-k}
\quad\hbox{with}\quad \det\left(c_0\right)\neq 0,
\eqno(5.30)$$
where all the $c_k$ are  $p\times p$ {\it diagonal} matrices.

Thanks to the above uniqueness property, the following procedure
may be used to determine the series $\psi_1(x,\zeta)$.
First we determine a $p\times p$ matrix PDO $g$ of the
form
$$
g = {\bf 1}_p + \sum_{i=1}^\infty g_i\pa^{-i},
\eqno(5.31)$$
with periodic coefficients, $g_i(x+2\pi)=g_i(x)$, such that
$$
L = g\hat Lg^{-1}
\eqno(5.32)
$$
where $\hat L$ is {\it diagonal}, i.e.,
$$
\hat L = \Delta^r\pa^r + \sum_{i=1}^\infty a_i\pa^{r-i},
\qquad\hbox{$a_i:$  all diagonal.}
\eqno(5.33)$$
Since $\Delta^r$ is a diagonal matrix with distinct, non-zero entries,
then if we require the $g_i$'s to be off-diagonal matrices
we can recursively determine both the $g_i$'s and the $a_i$'s by
comparing the two sides of $Lg=g\hat L$ term-by-term, according to
powers of $\pa$.
The solution is given by unique differential polynomial expressions
in the coefficients defining the expansion of $L$  in $\pa$.
For instance, we have
$$
a_1=\left( u_1\right)_{\rm diag}.
\eqno(5.34)$$
Then we
consider a $p\times p$  matrix  series
$$
\hat\psi_1(x,\zeta) =
\left(\sum_{k=0}^\infty \zeta^{-k} \hat\chi_k(x) \right)e^{-x \zeta \Gamma}
\quad\hbox{ with }\quad \det\left(\hat \chi_0(x)\right)\neq 0,
\eqno(5.35)$$
which satisfies the equation
$$
(\hat L\hat\psi_1)(x,\zeta) = \lambda \hat\psi_1(x,\zeta),
\qquad ( \lambda=\zeta^r ).
\eqno(5.36)$$
At the end of the procedure, we find the desired  solution of (5.27) from
$$
\psi_1(x,\zeta):=\left( g \hat \psi_1\right)(x,\zeta).
\eqno(5.37)$$
We know that up to a  diagonal constant matrix
$c(\zeta)$ of the form  (5.30)
the series solution of (5.27)
determined by this procedure must coincide with $\psi_1^r(x,\zeta)$  given
by equation (5.24).
To proceed further we need the following identity.

\smallskip
\noindent
{\bf  Proposition 5.1.}
{\it The operator ${\bf 1}_p \pa$ can be expressed as
$$
{\bf 1}_p\pa =
-\Gamma\hat L^{1/r} + \sum_{k=0}^\infty
{\cal F}_k\Delta^k \left(\hat L^{1/r}\right)^{-k},
\eqno(5.38)
$$
with $\hat L^{1/r}$ being  defined by
$$
(\hat L^{1/r})^r = \hat L,
\qquad
\hat L^{1/r} = \Delta\pa + \sum_{i=0}^\infty b_i\pa^{-i}
\quad\hbox{for some}\quad b_i,
\eqno(5.39)$$
and  uniquely determined $p\times p$ diagonal matrix valued
functions ${\cal F}_k$,  which satisfy
$$
{\cal F}_0 = -{1\over r}(-\Gamma)^ra_1\quad\hbox{ and }\quad
\int_0^{2\pi}dx\,\Bigl(k{\cal F}_k +
(-\Gamma)^k\, {\rm res}\bigl(\hat L^{k/r}\bigr)\Bigr)(x)  = 0
\quad\hbox{for }k>0.
\eqno(5.40)
$$}
The above proposition, which was crucial in \q{\FHM}
for obtaining results analogous to
those under consideration here, is taken from \q{\DS}
and is originally due to Cherednik \q{\Che}
(see also \q{\Wi,\Fla}).
Since  $\hat \psi_1$ is uniquely determined by (5.35-36) up
to multiplication by a diagonal constant matrix $c(\zeta)$
of the form given in (5.30),
equation (5.36) implies
$$
(\hat L^{1/r}\hat\psi_1)(x,\zeta)=
\zeta\hat\psi_1(x,\zeta).
\eqno(5.41)$$
This together with (5.38) leads to
$$\eqalignno{
 \hat\psi_1'(x,\zeta)
&=
\bigl(-\zeta \Gamma + \sum_{k=0}^\infty {\cal F}_k(x)\Delta^k
\zeta ^{-k}\bigr)\hat\psi_1(x,\zeta)\cr
\Rightarrow\hat\psi_1(x,\zeta)
&=\exp
\Bigl(-x\zeta \Gamma + \sum_{k=0}^\infty \Delta^k \zeta^{-k}
\int_0^x d\xi\, {\cal F}_k(\xi)\Bigr)
\hat\psi_1(0,\zeta),
&(5.42)\cr}$$
where $\hat\psi_1(0,\zeta)$ is an arbitrary diagonal matrix series
of the form given in (5.30).
Combining this with (5.34) and (5.40) results in the crucial relation
$$
\hat\psi_1(x+2\pi,\zeta) = \hat\psi_1(x,\zeta)\tau(\zeta)
\eqno(5.43{\rm a})$$
with
$$
\tau(\zeta)=
\exp\left(-2\pi\zeta \Gamma -
{(-\Gamma)^r\over r}\int_{0}^{2\pi}dx\,
\left(u_1(x)\right)_{\rm diag}\,
-\sum_{k=1}^\infty {\zeta^{-k}\over k}
 \int_{0}^{2\pi}dx\, {\rm res}
\left(\hat L^{k/r}\right)(x) \right).
\eqno(5.43{\rm b})$$
Observe now that $\psi_1(x,\zeta)$ defined in (5.37) satisfies
$$
\psi_1(x+2\pi,\zeta )=\psi_1(x,\zeta) \tau(\zeta)
\eqno(5.44)$$
with the same $\tau(\zeta)$, because the coefficients $g_i(x)$
defining the PDO $g$ are periodic functions of $x$.
This immediately leads to the following result.

\smallskip
\noindent
{\bf  Theorem  5.2.}
{\it The set of commuting Hamiltonians provided  by the local
monodromy invariants $H_{k,i}(j)$
given in (5.8)
is exhausted by the Hamiltonians $H_{0,i}(j)$ together
with the Hamiltonians defined
by the residues of componentwise fractional powers of the PDO
$\hat L$ (5.33)  obtained by diagonalizing the Lax operator $L$ (5.26)
attached to ${\cal L}=(\pa +j + \Lambda)$  by the elimination procedure.
More precisely,
$$\eqalign{
&{\rm diag}\left(H_{0,1},\ldots, H_{0,p}\right)=
{(-\Gamma)^r\over r} \int_{0}^{2\pi}d x\,
\left(u_1(x)\right)_{\rm diag} \,, \cr
&{\rm diag}\left(H_{k,1},\ldots, H_{k,p}\right)=
 {1\over k} \int_{0}^{2\pi}dx\,{\rm res}
\Bigl(\hat L^{k/r}\Bigr)(x)\, ,
\quad\hbox{for}\quad k>0.\cr}
\eqno(5.45)$$}
\smallskip
\noindent
{\it Proof.}
The statement follows by comparing (5.25) with (5.44) taking into account
that $\psi_1(x,\zeta)=\psi_1^r(x,\zeta) c(\zeta)$ where $c(\zeta)$
is a diagonal constant matrix of the form (5.30).
{\it Q.E.D.}
\smallskip

Finally, we shall write down the evolution equations generated
by the above Hamiltonians on ${\cal M}_{\rm red}={\cal M}_c/{\cal N}$.
For this it is convenient to consider an arbitrary
 $p\times p$ diagonal constant matrix $Q$ and  associate to it
the Hamiltonian
$$
H^Q_k :=r\sum_{i=1}^p Q_{ii} H_{k,i}.
\eqno(5.46)$$
Observe that $H^Q_k$ has the form (4.39) since
$$
H_k^Q(\ell, z_+, z_-, w)={\cal H}_k^Q(L)
\eqno(5.47{\rm a})$$
with
$$\eqalign{
&{\cal H}^Q_0(L)=\int_0^{2\pi} dx\,{\rm tr}\left(
(-\Gamma)^r Q u_1(x)\right)\,,\cr
&{\cal H}_k^Q(L)={r\over k}{\rm Tr}\left(Q {\hat L}^{k/r}\right),
\qquad k\geq 1.\cr}
\eqno(5.47{\rm b})$$
Using (4.40) the  gradients of these functions  are found to be
$$
{\delta {\cal H}_0^Q \over \delta L}=(-\Gamma)^r Q \pa^{-r},
\qquad
{\delta {\cal H}_k^Q\over \delta L}=g Q g^{-1} L^{(k-r)/r},
\quad k\geq 1,
\eqno(5.48)$$
where $g$ is the PDO  (5.31) which  diagonalizes $L$ and
$$
L^{l/ r}:=g \hat L^{l/ r} g^{-1}, \qquad\forall\, l.
\eqno(5.49)$$
Note that $L^{l/ r}$ commutes with $gQg^{-1}$.
We let ${\bf X}_{k, Q}^i$ denote the
 Hamiltonian vector field associated to $H_k^Q$
by means of the first and second PBs on ${\cal M}_{\rm red}$
given in Theorem 4.4.
Inserting the gradients (5.48) into formulae
(4.43) and (4.44) we obtain the following result.

\smallskip
\noindent
{\bf Corollary 5.3.}
{\it The Hamiltonian vector fields ${\bf X}^i_{k, Q}$
defining the  local hierarchy of compatible evolution equations on
${\cal M}_c/{\cal N}$ take the form
$$\eqalign{
& {\bf X}^2_{k,Q}(L)= {\bf X}^1_{k+r, Q}(L)=
\left[ \left( g Q g^{-1} L^{k/ r}\right)_+ , L\right],\cr
&{\bf X}^2_{k,Q}(z_+)= {\bf X}^1_{k+r, Q}(z_+)=
P_0\left( g Q g^{-1} L^{k/ r} z_+ W^{-1}\right) W,\cr
&{\bf X}^2_{k,Q}(z_-)= {\bf X}^1_{k+r, Q}(z_-)=
-W^{-1}P_0^\dagger \left( W z_-g Q g^{-1} L^{k/r} \right),\cr
&
{\bf X}^2_{k,Q}(w)= {\bf X}^1_{k+r, Q}(w)= 0,
\qquad \forall k=0,1,\ldots\,.\cr}
\eqno(5.50)$$
In particular, the flows are bihamiltonian.
In terms of the notation
$$
\left( g Q g^{-1} L^{k/r}\right)_+ := \sum_{i=0}^k A_{k,i} \pa^i
\eqno(5.51)$$
and the covariant derivatives defined in (4.38b)
the second and third equations in (5.50) can be rewritten as
$$\eqalign{
& {\bf X}^2_{k,Q}(z_+)={\bf X}^1_{k+r,Q}(z_+)=
 \sum_{i=0}^k A_{k,i} {\tilde {\cal D}}^i\left(z_+\right)\cr
& {\bf X}^2_{k,Q}(z_-)={\bf X}^1_{k+r,Q}(z_-)=
-\sum_{i=0}^k (-1)^i {\cal D}^i\left( z_- A_{k,i}\right).\cr}
\eqno(5.52)$$}
\smallskip

This completes our general analysis of the KdV type hierarchy
resulting from the generalized Drinfeld-Sokolov
 reduction defined in Section 2.
The evolution equations associated to  the vector fields in (5.50)
define  natural ``covariantized matrix generalizations''  of the
constrained  KP
hierarchy  considered previously in the literature (see \q{\Cheng,\OS,\Di} and
references therein)
in the scalar case $p=s=1$ with the constraint $w=0$.
Notice that  $w$ and the diagonal components of  $u_1$
(the subleading term of $L$ in (5.26))
do not evolve with respect to the flows determined by the vector
fields in (5.50).
The reason for this is the fact
that  $(u_1)_{ii}$ for $i=1,\ldots, p$ and the components of $w$
are Casimir functions with respect
to the reduced first PB, and generators of residual symmetries with
respect to the reduced second PB.
Indeed, evaluating the gauge invariant current component
$\Phi_i$ given by (3.3a) in the DS gauge (2.19),
 we see that $\Phi_i$ is  proportional with $(u_1)_{ii}$.

\smallskip
\noindent
{\it Remark 5.1.}
In the above we focused on the {\it local} monodromy invariants,
but it is clear that there exist also {\it non-local} monodromy
invariants,  generated  by the eigenvalues of the lower
block $\tilde \Psi_{22}(2\pi)$
of $\tilde T$ in (5.13-14).
The possible role of these non-local monodromy invariants
concerning  the integrability
of the system should be further studied. In particular, it is not
obvious to us  whether or not these are to be regarded as independent from
the commuting local Hamiltonians that we  described.
(An argument valid in finite dimensions would imply
that the gradients of all invariant functions
on the dual of a Lie algebra belong to  the centre of the centralizer.)
It  would be also interesting to know whether or not,
apart from  functions of the current components $\Phi_i$ and $w$ in
(3.3a) and (3.3b),
there  exist further (local)
commuting Hamiltonians that are not generated by monodromy
invariants.

\bigskip
\centerline{\bf 6.~Examples for  $r=2$
and  factorizations of the AKNS factor}
\medskip

\def \pa {\partial}
\def \paw {(\partial+w)^{-1}}
\def \M {{\cal M}}
\def \D {{\cal D}}
\def \tD {\tilde{\cal D}}
\def \bxxh {{\bf X}^2_H}
\def \half {{1/2}}
\def \L {{\cal L}}
\def \tr {{\rm tr}\,}
\def \Tr {{\rm Tr}\,}

The purpose of this section is  to  elaborate the simplest examples
for $r=2$ in order  to illustrate the preceding results  and to
make the abstract developments  more concrete.
A particularly interesting result that will be obtained from
considering the $r=2$ case is  a further factorization of the
AKNS factor $\Delta K$ (4.22) appearing in
the modified KdV Lax operator (4.21).
In fact, we shall derive the  factorization
$$
 K= \left(  {\bf 1}_p\pa +a-b ({\bf 1}_s\pa +d)^{-1} c \right) =
({\bf 1}_p\pa +\vt_{-1}) ( {\bf 1}_p -
\Delta \beta ({\bf 1}_s\pa +\vt_0 +\gamma \Delta \beta )^{-1}\gamma),
\eqno(6.1)$$
and a similar one containing a factor $({\bf 1}_p \pa +\vt_1)$ on the right,
in terms of the new variables
$$
\beta \in \widetilde{{\rm mat}}(p\times s),
\quad
\gamma\in\widetilde{{\rm mat}}(s\times p),
\quad
\vartheta_i\in\widetilde{ gl}_p,
\,\,\,
i=\pm 1, \quad
\vartheta_0\in\widetilde{gl}_s.
\eqno(6.2)$$
Substituting this factorization of $K$ in (4.21) leads to a
second modification of the generalized KdV  hierarchy for any $r$.
The variables in (6.2) will be seen to parametrize
a new gauge,   the so called $\Theta'$-gauge, which will arise
as a gauge section in an
alternative version of the Drinfeld-Sokolov  reduction
leading to the same reduced phase space  ${\cal M}_c/{\cal N}$ (2.16-17).
In the first part of this section,
we shall obtain the explicit form of the reduced second PB in terms of
the $\Theta'$-gauge as well as the $\Theta$-gauge,
the $\bar \Theta$-gauge and the DS gauge
defined in Section 2, and shall also present the local Poisson maps
given by  gauge transformations between these gauges.
The simplest evolution equations of
the generalized (modified) KdV hierarchies for $r=2$ will be worked out
in the second half of the section.


\bigskip
\centerline{\bf 6.1.~The reduced second PB, Miura maps and factorizations}
\medskip

Below we shall use 3 different  gradings of $\G:=gl_n$, $n=2p+s$, defined by
the adjoint actions of the following 3 matrices:
$$
I_0={1\over 2}{\rm diag}\left({\bf 1}_p, -{\bf 1}_p, {\bf 0}_s\right),
\quad
H={\rm diag}\left({\bf 1}_p, {\bf 0}_p, {\bf 0}_s\right),
\quad
\bar H= {\rm diag}\left({\bf 0}_p, -{\bf 1}_p, {\bf 0}_s\right).
\eqno(6.3)$$
For example, we have
$$
\G=\G^{I_0}_{-1} + \G^{I_0}_{-{1/2}} + \G^{I_0}_0 + \G^{I_0}_{1/2} +
\G^{I_0}_{ 1},
\eqno(6.4)$$
where $\G^{I_0}_k$ is the eigensubspace of ${\rm ad} I_0$
with eigenvalue $k$.
Using a $3\times 3$ block-matrix notation, the grades assigned to the
entries of $m\in \G$ are given by
$$
m^{I_0}=\pmatrix{0&1&{1/2}\cr -1&0&-1/2\cr -1/2&1/2&0\cr},
\quad
m^{H}=\pmatrix{0&1&1\cr -1&0& 0\cr -1&0&0\cr},
\quad
m^{\bar H}=\pmatrix{0&1&0\cr -1&0&-1\cr 0&1 &0\cr},
\eqno(6.5)$$
in correspondence with $I_0$, $H$, $\bar H$ above.
Previously we made use of the $H$-grading to define the
generalized Drinfeld-Sokolov  reduction.
Now we define an alternative version of it using the $I_0$-grading.
In this version of the reduction procedure we first introduce
the new constrained manifold $\mc'\subset{\cal M}$ given by
$$
\mc':=\{ \,
{\cal L}= \pa + \hatj + \Lambda \,\vert\,
\hatj:
S^1 \rightarrow \G^{I_0}_{\leq {1/2}}\,\},
\eqno(6.6{\rm a})$$
where the explicit form of $\Lambda$ and $\hatj$ is
$$
\Lambda=\pmatrix{0&\Gamma &0\cr \lambda \Gamma &0&0\cr 0&0&{\bf 0}_s\cr},
\qquad
\hatj=\pmatrix{*&0&*\cr *&*&*\cr *&*&*\cr}.
\eqno(6.6{\rm b})$$
Then we factorize $\mc'$
by the new ``gauge group''  ${\cal N}'$  generated by $\G^{I_0}_{<0}$.
An element  $g\in {\cal N}'$ acts on $\M_c'$ via the
gauge transformation
$$
g: {\cal L}\mapsto g {\cal L} g^{-1},
\qquad
g=\pmatrix{{\bf 1}_p&0&0\cr *&{\bf 1}_p & *\cr *&0& {\bf 1}_s\cr}:
S^1\rightarrow \exp\left( \G^{I_0}_{<0}\right).
\eqno(6.7)$$
Comparing the $H$-grading with the $I_0$-grading  one sees that
$\mc\subset \mc'$
and $\N\subset\N'$,
where $\mc$ and $\N$ are given in (2.16-17).
This enlargement of the constrained manifold
and the gauge group has been defined  in such a way that the factor space
is unchanged,
$$
\mr:=\mc/\N \simeq \mc'/\N'.
\eqno(6.8)$$
This follows since
there exist unique gauge transformations
that bring an arbitrary element of the  constrained manifold
to the DS gauge ${\cal M}_{\rm DS}$,
$$
{\cal L}_{\rm DS}=\pa +
\pmatrix{0&0&0\cr v_2&v_1&\zeta_+\cr \zeta_- &0& w\cr}+\Lambda
\qquad\hbox{for}\qquad {\cal L}_{\rm DS}\in \mds,
\eqno(6.9)$$
in either of the two reduction procedures, as one may verify
by computing the  (differential polynomial)
formulae of these gauge transformations.

The interest  of the new reduction procedure is that it admits
the  gauge section $\Theta'\subset \mc'$ whose general element
${\cal L}_{\Theta'}\in \Theta'$ takes the form
$$
{\cal L}_{\Theta'}=\pa  +
\pmatrix{\vt_1&0&\beta\cr 0&\vt_{-1}&0\cr 0 &\gamma& \vt_0\cr} +\Lambda,
\eqno(6.10)$$
that is, $\Theta'$ is defined by restricting $\hatj$
to take its value in  $(\G^{I_0}_0 + \G^{I_0}_{1/2})$.
We also have the $\Theta$-gauge and the $\bar \Theta$-gauge
defined  by the submanifolds $\Theta\subset {\cal M}_c'$,
$\bar \Theta\subset {\cal M}_c'$ where the constrained current
$\hatj$ is restricted to $\G^H_0$ and $\G^{\bar H}_0$, respectively.
We may parametrize the general elements ${\cal L}_\Theta\in \Theta$ and
${\cal L}_{\bar \Theta}\in \bar \Theta$ as follows:
$$
{\cal L}_{\Theta}=\pa  +
\pmatrix{\theta_1&0&0\cr 0&a&b\cr 0 &c& d\cr} +\Lambda
\qquad\hbox{and}\qquad
{\cal L}_{\bar \Theta}=\pa  +
\pmatrix{\bar a&0&\bar b\cr 0&\bar \theta_{-1}&0\cr \bar c &0& \bar d\cr}
+\Lambda.
\eqno(6.11)$$

Let now $F$, $H$ be  gauge invariant functions
on $\mc'$ and consider the function
$$
P:=\{ F,H\}_2,
\eqno(6.12)$$
determined with the aid of the usual extension-computation-restriction
algorithm, which is also a gauge invariant function on $\mc'$.
Of course, the functions $F$, $H$, $P$ can be recovered from their
restrictions to any of the above mentioned   gauge slices,
denoted respectively as $F_{\Theta'}$, $F_\Theta$, $F_{\bar \Theta}$,
$F_{\rm DS}$ and similarly for $H$ and $P$.
The function $P_{{\Theta}'}$ may be determined in terms of
$F_{\Theta'}$, $H_{\Theta'}$ using the following simple formula:
$$\eqalign{
&P_{\Theta'}(\vt_{-1},\vt_0, \vt_1, \beta,\gamma)=
\sum_{i=-1,0,1} \int_{S^1}\tr\left(
\vt_i \left[{\delta F_{\Theta'} \over \delta \vt_i},
{\delta  H_{\Theta'} \over \delta \vt_i}\right]
-{\delta  F_{\Theta'} \over \delta \vt_i}
\left({\delta H_{\Theta'} \over \delta\vt_i}\right)'\right)   \cr
&\qquad\qquad\qquad\qquad\qquad\qquad
-\int_{S^1}\, {\rm tr}\left( \Gamma\left(
{\delta  F_{\Theta'} \over \delta \gamma} {\delta H_{\Theta'} \over
\delta\beta}
-{\delta  H_{\Theta'} \over \delta \gamma} {\delta
F_{\Theta'} \over \delta\beta}
\right)\right).\cr}
\eqno(6.13)$$
The formula  on the right  hand side
describes the reduced second PB in the variables parametrizing $\Theta'$.
Here the functional derivatives are defined in the standard way using
the scalar product provided by the ordinary matrix trace.
The derivation of this formula is rather straightforward and  we omit it.
The formula of the reduced second
PB in terms of the variables parametrizing the $\Theta$-gauge
can be read off from (2.24),
$$
P_\Theta( \theta_0,\theta_1)=
\sum_{i=0,1}\int_{S^1}
 \tr\left(\theta_i \left[{\delta {F_\Theta} \over \delta \theta_i},
{\delta {H_\Theta} \over \delta \theta_i}\right]
-{\delta {F_\Theta} \over \delta \theta_i}
\left({\delta {H_\Theta} \over \delta \theta_i}\right)' \right),
\quad\,
\theta_0:=\pmatrix{a&b\cr c&d\cr},
\eqno(6.14)$$
and in terms of the $\bar \Theta$-gauge the formula is similar,
$$
P_{\bar \Theta}( \bar\theta_0,\bar\theta_{-1})=
\sum_{i=0,-1}\int_{S^1}
\tr\left(\bar\theta_i \left[{\delta {F_{\bar\Theta}} \over \delta
\bar \theta_i},
{\delta {H_{\bar \Theta}} \over \delta \bar \theta_i}\right]
-{\delta {F_{\bar \Theta}} \over \delta \bar\theta_i}
\left({\delta {H_{\bar \Theta}} \over \delta \bar \theta_i}\right)' \right),
\quad
\bar\theta_0:=\pmatrix{\bar a&\bar b\cr \bar c&\bar d\cr}.
\eqno(6.15)$$
Now we wish to present the PB in the DS gauge.
It is convenient to do so in terms of the variables
$u, v, y_\pm, w$  that appear in the Lax operator
$$
L=\Delta^2\left(\pa^2 +u\pa +v +y_+(\pa+w)^{-1}y_-\right),
\eqno(6.16{\rm a})$$
attached to ${\cal L}_{\rm DS}$ (6.9)
by the elimination procedure, i.e.,
$$
u=-\Gamma v_1\Delta,\quad
v=-\Gamma v_2,\quad
y_+=\Gamma \zeta_+,
\quad
y_-=\zeta_-,
\eqno(6.16{\rm b})$$
since then the entries of the constant matrix $\Delta=-\Gamma^{-1}$
will not appear explicitly in the PB.
To simplify the formula,
which can be obtained  by a direct computation
or by specifying (4.34), we introduce the notation
$$
\xi_u = {\delta H_{\rm DS}\over\delta u},\quad
\xi_v = {\delta H_{\rm DS}\over\delta v},\quad
\xi_+ = {\delta H_{\rm DS}\over\delta y_+},\quad
\xi_- = {\delta H_{\rm DS}\over\delta y_-},\quad
\xi_w = {\delta H_{\rm DS}\over\delta w}.
\eqno(6.17{\rm a})$$
The function $P_{\rm DS}(u,v,y_\pm, w)$
defined by restricting  $P$ in (6.12) to the DS gauge
is given by
$$
P_{\rm DS}=
\int_{S^1} \tr\left(  {\delta F_{\rm DS}\over \delta u}{\bf X}_H^2(u)
+{\delta F_{\rm DS}\over \delta v}{\bf X}_H^2(v)
+\sum_{\pm}{\delta F_{\rm DS}\over \delta y_\pm}{\bf X}_H^2(y_\pm)
+{\delta F_{\rm DS}\over \delta w}{\bf X}_H^2(w) \right),
\eqno(6.17{\rm b})$$
where the formula of the
Hamiltonian vector field ${\bf X}_H^2$ is found to be
$$
\eqalignno{
\bxxh(u) &= [\xi_u,u] - 2\xi_u' + [\xi_v,v] + \xi_v'' - (\xi_v u)'
+\xi_- y_- - y_+\xi_+,\cr
\bxxh(v) &= [\xi_u,v] - \xi_u'' - u\xi_u' + u\xi_v v - v\xi_v u + u\xi_v''
- (\xi_v u)'' - u(\xi_v u)' + v\xi_v' + (\xi_v v)' \cr
&\qquad  + \xi_v''' + y_+\D(\xi_+) - y_+\xi_+ u + (\xi_- y_-)' +
\tD(\xi_-)y_- + u\xi_- y_- +[\xi_v, y_+y_-],\cr
\bxxh(y_+) &= \xi_u y_+ + u\xi_v y_+ + \tD(\xi_v y_+) + \tD^2(\xi_-) +
u\tD(\xi_-) + v\xi_- -
y_+\xi_w  ,\cr
\bxxh(y_-) &= -y_-\xi_u + \D(y_-\xi_v) + y_-\xi_v' - y_-\xi_v u -
\D^2(\xi_+)
+ \D(\xi_+ u) - \xi_+ v + \xi_w y_-,\cr
\bxxh(w) &= \xi_+ y_+ - y_-\xi_- + [\xi_w, w] - \xi_w',
&(6.17{\rm c})\cr}
$$
with the covariant derivatives defined in (4.38b).

Using (6.7) and (6.9-11),
a  straightforward inspection shows that there exist local gauge
transformations
(for which the entries of $g\in {\cal N}'$ acting on
${\cal L}=\pa +\hatj +\Lambda$ are differential polynomials in the
components of $\hatj$) implementing the following changes of gauge:

\noindent
i) The mapping $\nu:\Theta'\rightarrow \Theta$, where
$\L_\Theta=\nu(\L_{\Theta'})=g_\nu \L_{\Theta'}g_\nu^{-1}$ is given
explicitly  as
$$
\theta_1=\vartheta_1,\quad
a=\vartheta_{-1}-\Delta\beta\gamma,\quad
d=\vartheta_0+\gamma \Delta\beta,\quad
b=\vartheta_{-1}\Delta\beta -\Delta\beta\vartheta_0 -
\Delta\beta\gamma \Delta\beta+\Delta\beta', \quad
c=\gamma.
\eqno(6.18)$$
\noindent
ii) Analogously, the mapping $\bar\nu:\Theta'\rightarrow \bar\Theta$, where
$\L_{\bar \Theta}=\bar \nu(\L_{\Theta'})=
g_{\bar \nu} \L_{\Theta'}g_{\bar \nu}^{-1}$ is given by
$$
\bar\theta_{-1}=\vt_{-1},\quad
\bar a=\vt_1 -\beta\gamma\Delta,\quad
\bar d=\vt_0 +\gamma \Delta\beta,\quad
\bar b=\beta,\quad
\bar c=\gamma\Delta \vt_1 -\vt_0 \gamma\Delta -
\gamma\Delta \beta \gamma\Delta -\gamma'\Delta.
\eqno(6.19)$$
iii) The mapping $\mu:\Theta\rightarrow \mds$, where
$\L_{\rm DS}=\mu(\L_{\Theta})=g_\mu \L_{\Theta}g_\mu^{-1}$ is defined by
$$
v_1=a-\Delta \theta_1 \Gamma,\quad
v_2=a\Delta\theta_1-bc\Delta+\Delta\theta_1',\quad
\zeta_+=b,\quad
\zeta_-=c\Delta\theta_1-dc\Delta-c'\Delta,\quad
w=d.
\eqno(6.20)$$
iv) Similarly, the mapping $\bar\mu:\bar\Theta\rightarrow \mds$, where
$\L_{\rm DS}=\bar\mu(\L_{\bar\Theta})=
g_{\bar \mu} \L_{\bar\Theta}g_{\bar \mu}^{-1}$ reads
$$
v_1=\bar \theta_{-1} -\Delta \bar a \Gamma,\quad
v_2=\bar \theta_{-1}\Delta \bar a-\Delta \bar b\bar c +\Delta{\bar a}',\quad
\zeta_+=\bar \theta_{-1}\Delta \bar b -\Delta \bar b \bar d +\Delta {\bar b}',
\quad
\zeta_-=\bar c,\quad
w=\bar d.
\eqno(6.21)$$
The construction ensures
that these ``generalized Miura maps'' are in fact Poisson maps
with respect to the PBs given above that realize the reduced second PB
in terms of the respective gauges.
In accordance with what is expected from Miura maps,
the inverses of these maps are not single valued and are non-local.
Incidentally, the map  $\nu: \Theta' \rightarrow \Theta$
(similarly the map $\bar \nu$)
provides a realization
of the $gl_{p+s}$ current  algebra appearing as a term
in the PB (6.14) on $\Theta$ in terms of
the $gl_p \oplus gl_s$ current algebra
and the ``symplectic bosons'' $\beta$, $\gamma$ appearing
in the PB  (6.13) on $\Theta'$.
This realization of the $gl_{p+s}$ current algebra is
analogous to the well-known Wakimoto realization of
the $gl_2$ current algebra (see \q{\BF} for an
elaboration of this observation).

Now we come to the relationship between the above Miura maps
and the factorizations of the Lax operator.
Observe first that the linear problem ${\cal L}\psi=0$
 has similar  properties
for $\L\in\mc'$ as for ${\cal L}\in {\cal M}_c$.
It is covariant under
$$
g: \L\mapsto e^f \L e^{-f},
\quad
\psi \mapsto g \psi
\quad\hbox{for any}\quad
 g\in {\cal N}',
\eqno(6.22)$$
and the  component $\psi_1$ of $\psi=(\psi_1^t,\psi_2^t,\phi^t)^t$
is invariant.
Thus we may apply the elimination procedure and
on account of (6.8) the
resulting image of $\mc'/\N'$ will turn out to be
the same manifold of PDOs which was obtained
in Sect.~4.
At the end of the chain of Miura maps
$\Theta' {\buildrel \nu \over \rightarrow} \Theta
{\buildrel \mu \over \rightarrow} \mds$
is the DS gauge to which the elimination procedure associates
the PDO $L$ (6.16).
We have also seen in Sect.~4 that applying the elimination
in the $\Theta$-gauge leads to a factorization of this Lax operator,
$$
L=L_{\Theta}=\Delta K \Delta (\pa +\theta_1)
\quad\hbox{with}\quad
K=\left(  {\bf 1}_p\pa +a-b ({\bf 1}_s\pa +d)^{-1} c \right).
\eqno(6.23)$$
We now wish to show that performing the elimination procedure
in the $\Theta'$-gauge yields a further factorization according to
$$
L=L_{\Theta}=L_{\Theta'}=\Delta
({\bf 1}_p\pa +\vt_{-1})
( {\bf 1}_p - \Delta \beta ({\bf 1}_s\pa +\vt_0 +\gamma \Delta \beta )^{-1}
\gamma)
\Delta ({\bf 1}_p\pa +\vt_1),
\eqno(6.24)$$
which amounts to the factorization (6.1) of the AKNS factor $\Delta K$
since $\theta_1=\vt_1$ by eq.~(6.18).
For  $\L_{\Theta'}\in \Theta'$ (6.10)
the linear problem takes the following form:
$$
\left(\matrix{
({\bf 1}_p\pa +\vt_1)&\Gamma &\beta\cr
\lambda \Gamma &({\bf 1}_p\pa + \vt_{-1})&0\cr
0&\gamma&({\bf 1}_s\pa + \vt_0)\cr}\right)
\left(\matrix{ \psi_1\cr \psi_{2} \cr\phi\cr}\right)=0.
\eqno(6.25)$$
More explicitly,
$$\eqalign{
({\bf 1}_p\pa + \vt_1) \psi_1 + \beta \phi + \Gamma \psi_{2}&=0,\cr
({\bf 1}_p\pa +\vt_{-1})\psi_{2} +\lambda \Gamma \psi_1 &=0,\cr
({\bf 1}_s\pa +\vt_0) \phi +\gamma \psi_{2} &=0.\cr}
\eqno(6.26)$$
In order to obtain the desired equation,
$$
L_{\Theta'} \psi_1 = \lambda \psi_1,
\eqno(6.27)$$
{}from the middle equation in (6.26) we  must express
$\psi_{2}$ in terms of $\psi_1$ using the other two equations.
To do this, we first ``formally integrate'' the last equation
in (6.26) to yield
$$
\phi = -({\bf 1}_s\pa +\vt_0)^{-1} \gamma \psi_{2}.
\eqno(6.28)$$
Substituting this back, the first equation in (6.26) becomes
$$
({\bf 1}_p\pa + \vt_1)\psi_1 + (\Gamma -\beta ({\bf 1}_s
\pa + \vt_0)^{-1}\gamma ) \psi_{2}=0.
\eqno(6.29)$$
Finally, formally solving (6.29) for $\psi_{2}$, and then using the
middle relation in (6.26), we obtain (6.27) with
$$
L_{\Theta'}=\left(\Delta ({\bf 1}_p\pa + \vt_{-1})\right)
\,\left({\bf 1}_p+\Delta\beta ({\bf 1}_s\pa +\vt_0)^{-1} \gamma\right)^{-1}\,
\left(\Delta ({\bf 1}_p\pa +\vt_1)\right).
\eqno(6.30)$$
This is the same as $L_{\Theta'}$ in (6.24) on account of the PDO identity
$$
\left({\bf 1}_p+\Delta\beta (\pa +\vt_0)^{-1}\gamma\right)^{-1}=
\left({\bf 1}_p -
\Delta \beta ({\bf 1}_s\pa +\vt_0 +\gamma \Delta \beta )^{-1}\gamma\right).
\eqno(6.31)$$
The factorization (6.24) now follows
since  the covariance properties of the linear problem imply
 $L=L_\Theta=L_{\Theta'}$ for the Lax operators associated
to gauge equivalent points of $\M_c'$.

Considering the alternative chain of Miura maps
$\Theta' {\buildrel \bar\nu \over \rightarrow} \bar\Theta
{\buildrel \bar\mu \over \rightarrow} \mds$ in a similar manner,
we obtain
$$
L=L_{\bar \Theta}=L_{\Theta'}
\quad\hbox{where}\quad
L_{\bar \Theta}=
\Delta ({\bf 1}_p\pa +\bar\theta_{-1})
\Delta \left({\bf 1}_p\pa +\bar a-\bar b ({\bf 1}_s\pa +\bar d)^{-1} \bar c
\right).
\eqno(6.32)$$
This is equivalent to the alternative  factorization of the AKNS factor
$$
\Delta \left({\bf 1}_p\pa +\bar a-\bar b ({\bf 1}_s\pa +\bar d)^{-1} \bar c
\right)
=
\left({\bf 1}_p -
\Delta \beta ({\bf 1}_s\pa +\vt_0 +\gamma \Delta \beta )^{-1}\gamma\right)
\Delta \left({\bf 1}_p\pa +\vt_1\right).
\eqno(6.33)$$
It is worth noting that all the factors in equations (6.1) and (6.33)
correspond to Poisson submanifolds in the PDO space ${\cal A}$
endowed with the quadratic Gelfand-Dickey PB (4.5).


\bigskip
\centerline{\bf 6.2.~Some explicit examples}
\medskip

\def \pa {\partial}
\def \paw {(\partial+w)^{-1}}
\def \M {{\cal M}}
\def \D {{\cal D}}
\def \tD {\tilde{\cal D}}
\def \bxxh {{\bf X}^2_H}
\def \half {{1\over 2}}
\def \L {{\cal L}}
\def \tr {{\rm tr}\,}
\def \Tr {{\rm Tr}\,}

We here work out the first few  Hamiltonians and
evolution equations for the generalized (modified) KdV hierarchies with $r=2$.
For simplicity, we now restrict ourselves to the scalar
case $p=1$, where $\Delta$ can be set equal to $1$.
To compute the commuting Hamiltonians
we need the expansion of $L$ in powers of $\pa$,
and from (6.16) using (4.33) we have
$$
L=\pa^2+u\pa+v+\sum_{k=1}^\infty L_k\pa^{-k}
\quad\hbox{with}\quad
L_k=(-1)^{k-1} y_+ {\cal D}^{k-1}(y_-).
\eqno(6.34)$$
The commuting Hamiltonians may be obtained from the residues of
powers of $L^{1/2}$ in accordance with (5.47), where
in the scalar case $\hat L=L$ and  $Q=1$.
If we  write
$$
L^{1/2} = \pa + {1\over 2}u + \sum_{k=1}^\infty \ell_k\pa^{-k} ,
\eqno(6.35)
$$
then for the first three terms we  find
$$
\ell_1 = {1\over 2}v - {1\over 8}u^2 - {1\over 4}u',\quad
\ell_2 = \half L_1 - \half u\ell_1 - \half\ell_1',\quad
\ell_3 = {1\over 2}L_2 - {1\over 2}\ell_2 u + {1\over 4}u'\ell_1
- {1\over 2}\ell_1^2 - {1\over 2}\ell_2'.
\eqno(6.36)$$
Substituting  the above formulae we get
$$\eqalignno{
{\cal H}_0(L)&=\int u ,&(6.37{\rm a})\cr
{\cal H}_1(L) &= 2\int {\rm res}\left(L^{1/2}\right) = 2\int \ell_1
=\int \left(v-{1\over 4}u^2\right), &(6.37{\rm b})\cr
{\cal H}_2(L)&= \int {\rm res}\left(L\right) = \int L_1= \int y_+y_-,
&(6.37{\rm c}) \cr
 {\cal H}_3(L)&={2\over 3}
 \int {\rm res}\left(L^{3/2}\right)={2\over 3}\int
\left({1\over 2}uL_1 + L_2 + u\ell_1' + v\ell_1 + u\ell_2 + \ell_3\right)
&(6.37{\rm d})\cr
&= \int\left( {1\over 2}u y_+ y_- -y_+{\cal D}(y_-) +
{1\over 4}uv'
+ {1\over {16}}(u')^2 + {1\over 4}v^2 - {1\over 8}u^2v
+ {1\over {64}}u^4\right), \cr
{\cal H}_4(L)&={1\over 2}\int {\rm res}\left(L^2\right)
= \int \left(L_3 +uL_2 + uL_1'  + vL_1\right)&(6.37{\rm e})\cr
&=\int\left(y_+{\cal D}^2(y_-)-uy_+{\cal D}(y_-)+u(y_+y_-)'+vy_+y_-\right).
\cr}$$
These formulae describe the first few Hamiltonians in terms of the
generalized KdV fields   $u, v, y_\pm, w$
associated to the DS gauge.
The expressions for the Hamiltonians of the  modified
KdV systems  corresponding to the gauges $\Theta$  and  $\Theta'$
can be found as the pull backs of the above Hamiltonians by means
of the Miura maps $\mu$ in (6.20) and $\nu$ in (6.18).
For example, in terms of the variables $\theta_1, a, b, c, d$,
the first three Hamiltonians of the
modified KdV hierarchy associated to the $\Theta$-gauge read
$$
\mu^* {\cal H}_0 = \int (a+\theta_1),\quad
\mu^*{\cal H}_1 =-\int\left( bc +{1\over 4}(a-\theta_1)^2\right),\quad
\mu^*{\cal H}_2=\int b\left(c'+cd-c\theta_1\right).
\eqno(6.38)$$
In terms of the variables $\beta, \gamma, \vartheta_i$ ($i=0, \pm 1$),
the corresponding Hamiltonians of the second modification
associated to the $\Theta'$-gauge are given by
$$\eqalignno{
\nu^*\mu^* {\cal H}_0&=\int \left(\vt_{1}+\vt_{-1}-\beta\gamma\right),
\phantom{XXXXXXXXXXXXXXXXXXXXXXXXX}
&(6.39{\rm a})\cr
\nu^*\mu^*{\cal H}_1&=-{1\over 4}\int\left( \left(\vt_1-\vt_{-1}\right)^2
+2 \left(\vt_1  +\vt_{-1} - 2\vt_0\right)\beta\gamma
 +4 \beta'\gamma -3\beta^2\gamma^2  \right),
&(6.39{\rm b})\cr
\nu^*\mu^*{\cal H}_2&=
\int\left(\vt_{-1}\beta -\vt_0\beta -\gamma\beta^2 +\beta'\right)
\left( \gamma' +\gamma\left(\vt_0 -\vt_1 +\beta\gamma\right)\right).
&(6.39{\rm c})\cr}$$
The expressions of the higher Hamiltonians soon become very long.

As an illustration, let us give the evolution equations
in the different gauges determined by the Hamiltonian functions
${\cal H}_1$, $\mu^*{\cal H}_1$ and  $\nu^*\mu^*{\cal H}_1$
with respect to the appropriate PBs given in Sect.~6.1.
In the DS gauge we find the flow
$$
\dot w =\dot u=0,\quad
\dot v = v' - \half(u''+uu'),\quad
\dot y_+ = \tD(y_+)+\half uy_+,\quad
\dot y_- = \D(y_-)-\half uy_-.
\eqno(6.40)$$
In the $\Theta$-gauge and in the $\Theta'$-gauge we have
$$
\dot d=0,\quad
\dot a =-\dot\theta_1= {1\over 2}(a-\theta_1)',\quad
\dot b ={ 1\over 2}ab + {1\over 2}b\theta_1 - db + b',\quad
\dot c = -{1\over 2}ac - {1\over 2}\theta_1 c + dc + c' ,
\eqno(6.41)$$
and respectively
$$\eqalignno{
&\dot\vartheta_0=-(\beta\gamma)',\qquad
\dot\vartheta_{\pm 1}=\pm \half (\vartheta_{1}-\vartheta_{-1})
+{1\over 2}(\beta\gamma)',&(6.42)\cr
&\dot\beta
=\half\beta(\vartheta_1+\vartheta_{-1}-2\vartheta_0-3\beta\gamma)+\beta',
\quad
\dot\gamma
=-\half\gamma(\vartheta_1+\vartheta_{-1}-2\vartheta_0-3\beta\gamma)+\gamma'.
\cr}$$
It can be checked that the Miura  maps
$\Theta' {\buildrel \nu \over \rightarrow} \Theta
{\buildrel \mu \over \rightarrow} \mds$
map these flows one to another accordingly.
In the same way one can in principle compute the evolution
equations for the other Hamiltonians.
Since the formulae are rather complicated we only give
the next two evolution  equations in the DS gauge.
The simplest way to obtain these  is to combine  the formulae in (5.50)
with the identities $(L)_+=\pa^2+u\pa+v$ and
$$
(L^{3/2})_+=\pa^3+{3\over 2}u\pa^2 +
{3\over 2}(v+{1\over 4}u^2+{1\over 2}u')\pa
+{3\over 4}(v'+2y_+y_-+uv)-{1\over 16}u^3+{1\over 8}u''.
\eqno(6.43)$$
The flow determined by the Hamiltonian
${\cal H}_2$ (6.37c) through  the PB (6.17c) satisfies
$$
\dot w=\dot u=0,\,\,\,
\dot v=2(y_+y_-)',\,\,\,
\dot y_+=\tD^2(y_+)+u\tD(y_+)+vy_+,\,\,\,
\dot y_-=-\D^2(y_-)+\D(uy_-)-vy_-.
\eqno(6.44)$$
For the flow determined by ${\cal H}_3$ (6.37d)
by means of the PB (6.17c) we get
$\dot w=\dot u=0$ together with the equation
$$
\eqalign{
\dot v&= {1\over4}v'''
+{3\over2}vv'+{3\over2}(y_+''y_--y_+y_-'')
-{3\over4}(u'v'+u''v +uu'v +{1\over 2}u^2v')\cr
&\qquad  + {1\over 16}((u^3)''-2u''''+3u^3u'-2uu''')
-3(wy_+y_-)'+{3\over2}(u y_+y_-)' ,\cr
\dot y_+&=\tD^3(y_+) +{3\over2}u\tD^2(y_+)
 +{3\over2}v\tD(y_+)
+{3\over4}u'\tD(y_+)
+{3\over 8}u^2\tD(y_+)
\cr
&\qquad
+{3\over2}(y_+y_-)y_+ +{3\over4}v'y_+
+{3\over4}uvy_+ -{1\over 16}u^3y_++{1\over8}u''y_+,\cr
\dot y_-&=\D^3(y_-)
-{3\over2}\D^2(uy_-)
+{3\over2}\D(vy_-)
+{3\over 4}\D(u'y_-)
+{3\over 8}\D(u^2y_-)
\cr
&\qquad
-{3\over2}(y_+y_-)y_- -{3\over4}v'y_-
-{{3\over4}}uvy_- +{1\over16}u^3y_- - {1\over8}u''y_-.}
\eqno(6.45)$$
This generalization of the KdV
equation reduces to one of the equations in \q{\OS} when
$w=0$ and $u=0$.

In the matrix case $p>1$ the Lax operator $L$ has
to be diagonalized according to (5.32) to compute the Hamiltonians from
the residues of fractional powers of $\hat L$.
The formulae that can be obtained straightforwardly
do not appear particularly enlightening
to us and they are too long to justify presenting them.


\bigskip
\centerline{\bf 7.~Discussion: discrete reductions and
 generalized KdV hierarchies}
\medskip

In this paper we have derived a Gelfand-Dickey type PDO model
of the hierarchy resulting from  generalized Drinfeld-Sokolov
reduction in the case of $\ell(gl_n)$ with the partition
$$
(\overbrace{ r, \ldots, r}^{p\rm\; times},
\overbrace{ 1, \ldots, 1}^{s\rm\; times}),
\qquad n=pr +s,
\eqno(7.1)$$
using the grade $1$ semisimple element $\Lambda$ in (0.6).
The reduced phase space turned out to be the space of quadruples
$(\ell, z_+, z_-, w)$ where $\ell$ is a $p\times p$ matrix $r$-KdV
type operator which is coupled to the fields $z_\pm, w$.
The compatible PBs and the commuting Hamiltonians  obtained from the
 Hamiltonian symmetry reduction are given by Theorem 4.4  and
Theorem 5.2., respectively,  with the
corresponding  evolution equations being described in
Corollary 5.3.
These results extend the  results of \q{\FHM} on the
matrix KdV system for which  $s=0$.

Of course the Drinfeld-Sokolov method works in a more general
Lie algebraic context than the case of $\ell(gl_n)$ and a semisimple
$\Lambda$ of minimal positive grade that we have considered.
However, it is worth noting that  some of the systems that would result from
the Drinfeld-Sokolov construction applied to
$\ell(\G)$ for $\G$ a classical Lie algebra
can be also obtained from ``discrete reduction'' applied to a
generalized KdV system associated to $gl_n$.
In the rest of this section we wish to briefly explain how this comes
about.

Let us consider  the discrete symmetries of the system described by
Theorems 4.4 and 5.2.
More precisely, let us look for  symmetries given
by some  involutive map $\sigma$ on the  reduced phase space
${\cal M}_{\rm red}={\cal M}_c/{\cal N}=\{(\ell, z_+, z_-, w)\}$,
$$
\sigma : {\cal M}_{\rm red}\rightarrow {\cal M}_{\rm red},
\qquad
\sigma^2 = {\rm id},
\eqno(7.2)$$
which  leaves the PBs invariant,
$$
\{ f\circ \sigma , h\circ \sigma\}_i^*=\{ f, h\}_i^*\circ \sigma,
\qquad i=1,2,
\eqno(7.3)$$
for arbitrary functions $f,h$ on ${\cal M}_{\rm red}$.
We  will take the following  ansatz for  $\sigma$.
Let $m\in GL_p$ and let $q\in GL_s$, i.e.,
$m$ and $q$ are constant, invertible, respectively
$p\times p$ and  $s\times s$  matrices.
Define  the map
$\sigma_{m,q}: {\cal M}_{\rm red}\rightarrow {\cal M}_{\rm red}$ by
$$
\sigma_{m,q}: \pmatrix{\ell\cr z_+\cr z_-\cr w\cr}\mapsto
\pmatrix{ m\ell^\dagger m^{-1} \cr -m z_-^t q^{-1} \cr q z_+^t m^{-1}\cr
-q w^t q^{-1}\cr},
\eqno(7.4)$$
where $\ell^\dagger$ is given  by the standard
adjoint operation on the PDO space ${\cal A}$,
$$
\ell^\dagger = (-1)^r\Delta^r \pa^r +\sum_{i=1}^r (-1)^{r-i} \pa^{r-i} u_i^t
\qquad\hbox{for}\qquad
\ell=\Delta^r \pa^r +\sum_{i=1}^r u_i \pa^{r-i}.
\eqno(7.5)$$
It is not hard to verify that  $\sigma_{m,q}$ is a Poisson map with
respect to the PBs given by Theorem 4.4 whenever it maps the phase space
${\cal M}_{\rm red}$ to itself, which is ensured by the condition
$$
m \Delta^r m^{-1} = (-1)^r \Delta^r.
\eqno(7.6{\rm a})$$
The involutivity  of $\sigma_{m,q}$ leads to the conditions
$$
m^t = \epsilon_m m,\quad \epsilon_m=\pm 1,
\qquad
q^t = \epsilon_q q,\quad \epsilon_q=\pm 1,
\quad\hbox{with}\quad \epsilon_m \epsilon_q=-1.
\eqno(7.6{\rm b})$$
Notice that if $\epsilon_m=-1$ then $p$ must be ${\it even}$ and
when $\epsilon_q=-1$ then $s$ must be even.
For any natural numbers  $a$ and $b$
define  the $a\times a$ and $2b\times 2b$ matrices $\eta_a$ and
$\Omega_{2b}$ by
$$
\eta_a=\sum_{i=1}^a e_{i, a+1-i},
\qquad
\Omega_{2b}
=\sum_{i=1}^b e_{i, 2b +1-i} - \sum_{i=b+1}^{2b} e_{i, 2b+1-i},
\eqno(7.7{\rm a})$$
and let $\xi_{a}$ denote an  arbitrary $a\times a$ diagonal,
invertible matrix  subject to
$$
\eta_a \xi_{a} \eta_a = -\xi_{a},
\eqno(7.7{\rm b})$$
which means that $\xi_a$ is anti-symmetric under  transpose with
respect to the anti-diagonal.
We have the following solutions for $\sigma_{m,q}$.

\noindent
Type i):   $r=2\rho$  even, $\forall\, p$,
$m$  is  diagonal and  $q$ is arbitrary with $\epsilon_q =-1$ ($s=2l$ even).

\noindent
Type ii):   $r=(2\rho +1)$  odd, $p=2k$  even,
$\Delta$ is such that
$\eta_p \Delta\eta_p=-\Delta$, $m=\xi_{p} \Omega_p$ and
$q$ is arbitrary with $\epsilon_q=-1$ ($s=2l$ even).

\noindent
Type iii): $r=(2\rho +1)$ odd, $p=2k$ even,
$\Delta$ is such that
$\eta_p \Delta\eta_p=-\Delta$,  $m=\xi_p \eta_p$ and
 $q$ is arbitrary with $\epsilon_q=+1$ ($\forall\,s$).
\smallskip

\noindent
Note that $\eta_p \Delta\eta_p =-\Delta$
 requires $\Delta$ to have the form
$\Delta={\rm diag}
\left(\Delta_1,\ldots,\Delta_k, -\Delta_k,\ldots, -\Delta_1\right)$,
where $p=2k$ and  $\Delta_i^r\neq \pm \Delta_j^r\neq 0$ for $i\neq j$ since
$\Delta^r$ must have distinct, non-zero eigenvalues
(cf.~eq.~(2.9) for $\Gamma=-\Delta^{-1}$).
The symmetries of type i), ii), iii) are
of course also available in the case of  the corresponding $r$-KdV systems,
where $s=0$ and the phase space is simply the space of operators
$\ell$ on which $\sigma_{m,q}$ acts by $\ell \mapsto  m\ell^\dagger m^{-1}$.

Given an involutive symmetry $\sigma=\sigma_{m,q}$, one finds
that
$$
\sigma :L\mapsto m L^\dagger m^{-1}
\quad\hbox{for}\quad
L=\ell + z_+ ({\bf 1}_s\pa +w)^{-1} z_-,
\eqno(7.8)$$
which implies that the commuting Hamiltonians given by Theorem 5.2
admit a basis consisting of invariant and anti-invariant
(that change sign) linear combinations with respect to $\sigma$.
On account of (7.3), if $H\circ \sigma =H$
then the
Hamiltonian vector fields ${\bf X}_H^i$ are tangent to the
fixed point set ${\cal M}_{\rm red}^\sigma \subset {\cal M}_{\rm red}$
of $\sigma$.

The flows of a  ``discrete reduced hierarchy'' may therefore
be defined by restricting the flows generated on ${\cal M}_{\rm red}$
by the $\sigma$-invariant linear combinations of the Hamiltonians
in Theorem 5.2 to the fixed point set ${\cal M}_{\rm red}^\sigma$.
These flows are bihamiltonian with respect to the restricted
Hamiltonians and a naturally induced bihamiltonian structure on
${\cal M}_{\rm red}^\sigma$.
The induced PBs on ${\cal M}_{\rm red}^\sigma$
are defined by restricting the original PBs
of functions of $\sigma$-invariant linear combinations of the components
of $\ell, z_+, z_-, w$ --- which may be regarded as coordinates
on ${\cal M}_{\rm red}^\sigma$ ---
to ${\cal M}^\sigma_{\rm red}$.
For fixed $p, r, s$ and a given symmetry
type i), ii) or  iii) the various possible choices of $m$ and $q$
defining $\sigma_{m,q}$  are equivalent from the point of view of
the discrete reduction.  In fact,  the fixed
point sets corresponding to two different  choices are always related
by a Poisson map of ${\cal M}_{\rm red}$ given by
$$
\pmatrix{\ell\cr z_+\cr z_-\cr w\cr}\mapsto
\pmatrix{ \bar m\ell {\bar  m}^{-1} \cr\bar m z_+  {\bar q}^{-1} \cr
{\bar q} z_-{\bar m}^{-1}\cr
\bar q w {\bar q}^{-1}\cr},
\eqno(7.9)$$
with some constant matrices $\bar m\in GL_p$ and $\bar q\in GL_s$.

As explained in particular cases in \q{\DF},
the  above discrete reductions are
actually induced by the reductions of $gl_n$ to a simple complex
Lie algebra of $B$, $C$ or $D$ type.
Correspondingly,
many of the generalized  KdV hierarchies that may be associated to
certain conjugacy classes in the Weyl group ${\bf W}(\G)$ for $\G$
a classical  simple Lie algebra by generalized Drinfeld-Sokolov reduction
 (see \q{\McI,\GHM,\BGHM,\FGMG})  can be  also obtained
by discrete reducing either the matrix $r$-KdV hierarchy   or its extended
version associated to $gl_n$.
These  KdV type hierarchies  are associated
to a regular or a non-degenerate,
in the sense that it admits DS gauge fixing,
graded semisimple element $\Lambda$ of minimal positive grade
from  a graded Heisenberg subalgebra of $\ell(\G)$ by means of
the construction described in Section 1.
(The construction described in
Section 1 is the special case  of the construction described in
\q{\McI,\GHM} for which the  ``coarser grading'' of $\ell(\G)$
is chosen to be the homogeneous grading.)
Since the graded Heisenberg
subalgebras are classified \q{\KP} by the conjugacy
classes \q{\Car} in ${\bf W}(\G)$,
we can label these generalized KdV hierarchies by the
respective conjugacy classes  in ${\bf W}(\G)$.
The conjugacy classes  that occur here
can be parametrized (as in \q{\Car,\DF})  by certain ``signed partitions''.
Using this notation, we find that  the
discrete reduction  operates
on the generalized  KdV systems associated to $gl_n$ according to the
following reduction rules:
$$\eqalign{
&\sigma_{\Delta, \Omega_{2l}}:
(\overbrace{ 2\rho, \ldots, 2\rho}^{p\rm\; times},
\overbrace{1,\ldots,1}^{2l\;\rm times})
\in {\bf W}(gl_{2(p\rho +l)})
\quad \Longrightarrow \quad
(\overbrace{ \bar \rho,\ldots, \bar \rho}^{p\;\rm times},
\overbrace{1,\ldots,1}^{l\;\rm times})\in
{\bf W}(C_{p\rho+l}),\cr
&\sigma_{\Delta \Omega_{2k}, \Omega_{2l}}:
(\overbrace{ r, \ldots, r}^{2k\rm\; times},
\overbrace{1,\ldots,1}^{2l\;\rm times})
\in {\bf W}(gl_{2(kr+l)})
\quad \Longrightarrow \quad
(\overbrace{  r,\ldots,  r}^{k\;\rm times},
\overbrace{1,\ldots,1}^{l\;\rm times}
)\in {\bf W}(C_{kr+l}), \cr
&\sigma_{\Delta \eta_{2k}, \eta_{2l+1}}:
(\overbrace{ r, \ldots, r}^{2k\rm\; times},
\overbrace{1,\ldots,1}^{2l+1\;\rm times} )
\in {\bf W}(gl_{2(kr+l)+1})
\  \Longrightarrow \
(\overbrace{  r,\ldots,  r}^{k\;\rm times},
\overbrace{1,\ldots,1}^{l\;\rm times}
)\in
{\bf W}(B_{kr+l}),\cr
&\sigma_{\Delta \eta_{2k}, \eta_{2l}}:
(\overbrace{ r, \ldots, r}^{2k\rm\; times},
\overbrace{1,\ldots,1}^{2l\;\rm times})
\in {\bf W}(gl_{2(kr+l)})
\quad \Longrightarrow \quad
(\overbrace{  r,\ldots,  r}^{k\;\rm times},
\overbrace{1,\ldots,1}^{l\;\rm times}
)\in {\bf W}(D_{kr+l}),\cr}
\eqno(7.10)$$
where $l\geq 0$ is arbitrary,  $r=2\rho+1$ is odd.
By detailed inspection,
this result has been established in \q{\DF} for $l=0$
in all cases.
Since the case of arbitrary $l$ can be treated in a similar way,
we  omit the proof\footnote{${}^{3}$}{\ninerm
The $l=1$ case of the $D_{kr+l}$ series is exceptional since
in this case the Drinfeld-Sokolov reduction applied to $D_{kr+1}$
gives more commuting
Hamiltonians than the  discrete reduction of
the corresponding  hierarchy based on $gl_{2kr+2}$.
The relevant grade one  semisimple element
$\Lambda \in \ell(gl_{2kr+2})\cap \ell(D_{kr+1})$ is a
regular element of $\ell(D_{kr+1})$ but is not a regular
element of $\ell(gl_{2kr+2})$.
The abelian subalgebra
${\rm Ker}({\rm ad\,} \Lambda )\subset \ell(D_{kr+1})$ contains
extra generators in addition to those generators of the
centre of the centralizer of $\Lambda$ in $\ell(gl_{2kr+2})$ that
belong to $\ell(D_{kr+1})\subset \ell(gl_{2kr+2})$.
These yield  the extra Hamiltonians
by means of the Drinfeld-Sokolov construction.
The form of these Hamiltonians in terms of the reduced phase space
 variables is not known to us. }.
It is worth noting that
the above reduction rules are also valid in the generalized  AKNS case
(see Remarks 2.2.~and 4.3.) as well as for
the non-abelian Toda systems corresponding to the generalized KdV systems.

Finally, we wish to remark  that {\it not} all generalized
(non-principal)  KdV  hierarchies
based on a classical Lie algebra are discrete reductions
of hierarchies  associated to $gl_n$.
This is apparent from  the classification of graded regular
elements \q{\Sp,\DF}, since
certain  ``primitive'' cases exist for $\G=D_{2p}$ in
correspondence with the conjugacy class
$(\bar p,\bar p)$ in  ${\bf W}(D_{2p})$.
The two negative cycles $(\bar p,\bar p)$
 also appear as building blocks of more complicated
regular conjugacy classes in  ${\bf W}(D_n)$ and in ${\bf W}(B_n)$.
A PDO description of the  KdV systems associated to these
conjugacy classes in the Weyl group by generalized Drinfeld-Sokolov
reduction is not known.

\bigskip
\bigskip
\noindent
{\bf Acknowledgements.}
Reference \q{\Sp} was pointed out by an  anonymous referee of \q{\DF}.
We wish  to thank  I.R.~McIntosh for making available ref.~\q{\McI}
and we are grateful to F.~Delduc, L.A.~Dickey,
 L.~Gallot and F.~Toppan for discussions, encouragement and
comments on the manuscript. Part of the work for this paper was carried
out whilst LF was a visitor at ENSLAPP in Lyon. We wish to thank the CNRS
for supporting the visit of IM to ENSLAPP and especially to thank
P.~Sorba and the members of ENSLAPP for their kind hospitality.

\vfill\eject

\centerline{\bf Appendix A: The Poisson submanifold $M_K\subset {\cal A}$}
\medskip

\def\H{{\cal H}}

In this appendix we consider the mapping
$\eta: \Theta_0 \rightarrow M_K\subset {\cal A}$ given in (4.25)
and show that $M_K\subset {\cal A}$ is a Poisson submanifold
with respect to the quadratic Gelfand-Dickey PB and $\eta$ is a
Poisson mapping as stated by Proposition 4.1 in Sect.~4.

It will be convenient to reformulate the statement of Proposition
4.1 in terms of Hamiltonian vector fields.
Let ${\cal H}$ be an arbitrary function on ${\cal A}$,
 ${\bf X}_\H$ the corresponding  Hamiltonian vector field on
${\cal A}$ defined by means of the quadratic Gelfand-Dickey PB,
$$
{\bf X}_{\cal H}(L)=\left( L {\delta \H\over \delta L}\right)_+ L
-L\left( {\delta \H\over \delta L} L\right)_+,
\qquad \forall\, L\in {\cal A},
\eqno(A.1)$$
and $\tilde {\bf X}_{\cal H}$ the restriction of
${\bf X}_\H$ to $M_K\subset {\cal A}$.
Then consider the function $H$ on $\Theta_0$ given by
$$
H=\H\circ \eta,
\eqno(A.2)$$
and denote by ${\bf Y}_H$ the corresponding Hamiltonian vector field
on $\Theta_0$ defined by means of the current algebra PB (4.26).
If
$$
\theta_0=\pmatrix{a&b\cr c&d\cr}
\eqno(A.3)$$
denotes the arbitrary element of the space
$\Theta_0=\widetilde{gl}_{p+s}$ as in (2.22b), (2.23)
 and
the gradient
$$
{\delta H\over \delta \theta_0}=\pmatrix{
{\delta H\over \delta a}& {\delta H\over \delta c}\cr
{\delta H\over \delta b} & {\delta H\over \delta d}\cr}
\eqno(A.4)$$
is defined in the usual way, one has
$$
{\bf Y}_H(\theta_0)=\left[ {\delta H\over \delta \theta_0} , \theta_0\right]
-\left({\delta H\over \delta \theta_0}\right)'\,.
\eqno(A.5)$$
Clearly, the claim of Proposition 4.1 requires
the vector field $\tilde {\bf X}_\H$
to be equal to the push forward of the vector field ${\bf Y}_H$ by $\eta$,
 i.e., we must
verify the equality
$$
\tilde {\bf X}_\H(\eta(\theta_0))=\eta_*\left( {\bf Y}_H(\theta_0)\right)
\qquad \forall\, \theta_0\in \Theta_0.
\eqno(A.6)$$
In particular, the required equality implies that $\tilde {\bf X}_\H$
is tangent to $M_K$ (which is just the statement that
 $M_K\subset {\cal A}$ is a Poisson submanifold) and that
$\eta_*({\bf Y}_H)$ gives a well-defined vector field on $M_K$.

The formula for  $\eta$,
$$
K=\eta(\theta_0)=(\pa +a) - b(\pa +d)^{-1} c,
\eqno(A.7)$$
and (A.5) permits us to write down the explicit form of the r.h.s.\ of
(A.6) as follows:
$$
\eta_*\left( {\bf Y}_H(\theta_0)\right)={\bf Y}_H(a)
-{\bf Y}_H(b)(\pa +d)^{-1}c - b (\pa +d)^{-1}  {\bf Y}_H(c),
\eqno(A.8)$$
where we have
$$\eqalign{
&{\bf Y}_H(a)=\left[{\delta H\over \delta a}, a\right]
-\left({\delta H\over \delta a}\right)' +{\delta H\over \delta c}c
-b{\delta H\over \delta b},\cr
&{\bf Y}_H(b)={\delta H\over \delta a} b
-a{\delta H\over \delta c} - b {\delta H\over \delta d}
+ {\delta H\over \delta c} d -\left({\delta H\over \delta c}\right)',\cr
&{\bf Y}_H(c)={\delta H \over \delta b}a +{\delta H\over \delta d}c
-c {\delta H\over \delta a}
- d {\delta H\over \delta b} -\left({\delta H\over \delta b}\right)'.\cr}
\eqno(A.9)$$
To derive (A.8) we have used that
for a function $H$ of the form (A.2)
$$
{\bf Y}_H(d)=\left[ {\delta H\over \delta d}, d\right]
-\left({\delta H\over \delta d}\right)' +{\delta H\over \delta b} b
-c {\delta H\over \delta c}=0.
\eqno(A.10)$$
This follows from the fact that
$$
\eta (\theta_0)= \eta( \hat \theta_0)
\eqno(A.11{\rm a})$$
for any
$$
\hat \theta_0 = \pmatrix{{\bf 1}_p &0\cr 0 & e^D\cr}
\theta_0
\pmatrix{{\bf 1}_p &0\cr 0 & e^{-D}\cr} +
\pmatrix{{\bf 1}_p &0\cr 0 & e^D\cr} \pmatrix{{\bf 1}_p &0\cr 0 & e^{-D}\cr}',
\qquad  D\in\widetilde{gl}_s.
\eqno(A.11{\rm b})$$
Relation (A.11) is easily verified directly,  or it
may be traced back to the symmetry  in (3.1b).
Writing  the arbitrary variation $\delta H$ of $H$ in (A.2)
in  the alternative ways,
$$\eqalign{
&\delta H= \int_{S^1}
{\rm tr}\left(\delta \theta_0 { \delta H\over \delta \theta_0} \right)
={\rm Tr}\left( \delta K {\delta \H\over \delta L}(K)\right)=\cr
&\ \int_{S^1} {\rm tr\,}{\rm res} \left(
\left( \delta a - \delta b (\pa+d)^{-1} c -b(\pa+d)^{-1} \delta c
+b (\pa +d)^{-1} \delta d (\pa +d)^{-1} c\right)
{\delta \H\over \delta L}(K)
\right) ,\cr}
\eqno(A.12)$$
leads to the relations
$$\eqalign{
&{\delta H\over \delta a}={\rm res}
\left( {\delta \H\over \delta L}(K)\right),
\qquad
{\delta H\over \delta d}=
{\rm res}\left( (\pa+d)^{-1} c {\delta \H\over \delta L}(K)
 b (\pa+d)^{-1}\right),\cr
&{\delta H\over \delta b}=-{\rm res}\left((\pa +d)^{-1}c
{\delta \H\over \delta L}(K)\right),
\qquad
{\delta H\over \delta c}=
-{\rm res}\left( {\delta \H\over \delta L}(K) b(\pa +d)^{-1}\right).\cr}
\eqno(A.13)$$
Using (A.1), (A.8), (A.9), (A.10) and (A.13) it is in principle
possible to verify (A.6) directly for an arbitrary function $\H$,
but we find it easier to
do so for a conveniently chosen complete set of functions on
${\cal A}$, which is enough by the general properties of the PB.

The complete set of functions for which we are now going to verify
(A.6)  consists of the functions $\H_k$ of the form
$$
\H_k (L)={\rm Tr}\left(L \xi \pa^k \right),
\qquad \forall k\in {\Zed},\quad
\forall \xi\in \widetilde{gl}_p,
\eqno(A.14)$$
for which ${\delta \H_k\over \delta L}=\xi \pa^k$.
In order to simplify the computation (throughout which
we shall arbitrarily fix $\xi$), we make use of the
 ``integrating factor'' trick similarly to that in
 Sect.~4, that is we write
$$
(\pa +d)=W^{-1} \pa W, \qquad d=W^{-1} W',
\eqno(A.15)$$
in analogy to (4.33).  We then  readily find,
$$
(\xi \pa^k K)_+ =0=(K\xi \pa^k)_+
\qquad \hbox{for}\quad k<-1,
\eqno(A.16)
$$
$$
(\xi\pa^{-1}K)_+ =\xi=(K\xi\pa^{-1})_+
\qquad \hbox{for}\quad  k=-1,
\eqno(A.17)$$
and for $k\geq 0$,
$$\eqalign{
&(\xi\pa^k K)_+ =\xi\pa^k K+\xi (bW^{-1})^{(k)}\pa^{-1}Wc
=\xi\pa^k K+\xi \tilde {\cal D}^k(b)(\pa +d)^{-1}c ,\cr
&(K\xi \pa^k)_+=K\xi \pa^k+(-1)^k b W^{-1}\pa^{-1}(Wc\xi)^{(k)}
=K\xi \pa^k+(-1)^k b (\pa+d)^{-1}{\cal D}^k(c\xi),\cr}
\eqno(A.18)$$
where we  used the
covariant derivatives defined similarly to (4.38b),
using $d$ in place of $w$. (Incidentally, $w=d$ in the $\Theta$-gauge.)
These relations and (A.1) give the l.h.s.~of (A.6) as follows:
$$
\tilde {\bf X}_{\H_k}(K)=0 \qquad\hbox{for}\quad k<-1,
\eqno(A.19)$$
$$
\tilde {\bf X}_{\H_{-1}}(K)=[\xi , K]
=[\xi, a] -\xi' -\xi b(\pa+d)^{-1}c+b(\pa+d)^{-1}b\xi,
\eqno(A.20)$$
and for $k\geq 0$,
$$\eqalign{
&\tilde {\bf  X}_{\H_k}(K) = (K\xi\pa^k)_+K-K(\xi\pa^kK)_+
=\left( (-1)^k b {\cal D}^k\left(c\xi\right)-
\xi \tilde {\cal D}^k\left(b\right)c\right) - \cr
&\,\, \left( a\xi \tilde {\cal D}^k\left(b\right)+
\tilde {\cal D}\left( \xi \tilde {\cal D}^k\left(b\right)\right)
\right)(\pa +d)^{-1} c
- b(\pa+d)^{-1} \Bigl( (-1)^k {\cal D}^{k+1}\left(c\xi\right)
-(-1)^k{\cal D}^k\left(c\xi\right) a \Bigr)\cr
&\,\, +b(\pa+d)^{-1}\left(
c \xi \tilde{\cal D}^k(b)-
(-1)^k{\cal D}^k(c\xi) b\right)(\pa +d)^{-1}c. \cr}
\eqno(A.21)$$

We now set $H_k:=\H_k\circ \eta$ and verify
the equality (A.6).
This equality is trivial for $k<-1$ and it is also
clear for $k=-1$ by comparing (A.20) and (A.8-9) for $H_{-1}$ using that
$$
{\delta H_{-1}\over \delta a}=\xi,
\quad
{\delta H_{-1}\over \delta b}=
{\delta H_{-1}\over \delta c}={\delta H_{-1}\over \delta d}=0,
\eqno(A.22)$$
as follows from (A.13).
To deal with the case $k\geq 0$, we note from (A.13) that
$$
{\delta H_k\over \delta a}=0,\quad
{\delta H_k\over \delta b}=-(-1)^k {\cal D}^k(c\xi),\quad
{\delta H_k\over \delta c}=-\xi \tilde {\cal D}^k(b),
\quad\hbox{for $k\geq 0$.}
\eqno(A.23)$$
The explicit form of ${\delta H_k\over \delta d}$ will not be
needed, only its property  (A.10).
Thanks to this and (A.23) we can rewrite the last term of (A.21),
$$\eqalign{
&b(\pa+d)^{-1}\left(
c \xi \tilde{\cal D}^k(b)-
(-1)^k{\cal D}^k(c\xi ) b\right)(\pa +d)^{-1}c=
b(\pa +d)^{-1}
\left[ \pa +d, {\delta H_{k}\over  \delta d}\right](\pa+d)^{-1}c\cr
&\qquad\qquad  = b {\delta H_k\over \delta d} (\pa +d)^{-1} c -
b(\pa +d)^{-1} {\delta H_k\over \delta d} c.\cr}
\eqno(A.24)$$
Plugging this back into (A.21) and using (A.23) again,
we finally obtain
$$\eqalign{
&\tilde {\bf X}_{\H_k}(K)
= \left({\delta H_k\over \delta c}c -b{\delta H_k\over \delta b}\right)
+\left( a{\delta H_k\over \delta c} + b {\delta H_k\over \delta d}
- {\delta H_k\over \delta c}d +
\left({\delta H_k\over \delta c}\right)'\right)(\pa +d)^{-1}c\cr
&\qquad\qquad
-b(\pa +d)^{-1}\left( {\delta H_k \over \delta b}a +{\delta H_k
\over \delta d}c
 - d {\delta H_k\over \delta b}
-\left({\delta H_k\over \delta b}\right)'\right)
\qquad\hbox{for $k\geq 0$}.\cr}
\eqno(A.25)$$
This immediately yields (A.6) by taking into account
(A.8-9) and that ${\delta H_k \over \delta a}=0$ for $k\geq 0$.
This then completes the proof of  Proposition 4.1, whose claim
 is  equivalent to relation (A.6).


\bigskip
\centerline{\bf Appendix B: The formula of the
 reduced PB in the DS gauge}
\medskip

\def \tr {{\rm{tr}}}  \def \integral {\int_0^{2\pi}}   \def \D {{\cal D}}
\def \cinf {C^\infty}  \def \pa {\partial}  \def \paw  {(\pa + w)^{-1}}


The purpose of this appendix is to present the computation
leading to  formula (4.34) of Sect.~4 that
describes  the reduced current algebra PB on
${\cal M}_{\rm red}\simeq {\cal M}_{\rm DS}$.
We recall that the constrained current
$J\in\widetilde{gl}_{pr+s}$ defining  a generic point of
${\cal M}_{\rm DS}$ has the form
$$
J= j_{p,r,s} + C_+=
\sum_{i=1}^r e_{r,i}\otimes v_{r-i+1}
 + e_{r,r+1} \otimes \zeta_+ + e_{r+1,1} \otimes \zeta_- +
e_{r+1,r+1}\otimes w + C_+,
\eqno(B.1{\rm a})$$
where the explicit matrix form of $j_{p,r,s}$ is given
by  equation (0.8) with  the variables
$$
v_i\in \widetilde{gl}_p,\quad
w\in \widetilde{gl}_s,\quad
\zeta_+\in\widetilde{ {\rm mat}}\left(p\times s\right),\quad
\zeta_-\in\widetilde{ {\rm mat}}\left(s\times p\right),
\eqno(B.1{\rm b})$$
and   $C_+=\sum_{k=1}^{r-1} e_{k,k+1}\otimes  \Gamma$ is the constant matrix
appearing in $\Lambda_{p,r,s}=C_++\lambda C_-$ in (0.6).
We know already by Theorem 4.2  that the PBs of functions of
the components $v_1,\dots,v_r$ are given by the standard quadratic
Gelfand-Dickey PB on the space of operators $\ell$,
$$
\ell=L_+=\Delta^r\pa^r +\sum_{i=1}^r u_i \pa^{r-i},
\qquad u_i=\Delta v_i \Delta^{r-i}, \quad \Delta=-\Gamma^{-1}.
\eqno(B.2)$$
We need  then only to compute the other PB relations.
We choose to do this by computing the Hamiltonian
vector field ${\bf{X}}_H:={\bf X}_H^2$ for
$H=Q$, $P$, $R$ respectively where
$$
Q(J) = \integral\tr \left(f\zeta_+\right),\quad
P(J)=\integral\tr \left(\varphi \zeta_-\right),\quad
R(J)=\integral\tr \left(\alpha w\right)
\eqno(B.3)$$
with $f, \varphi, \alpha$ being  matrix valued test functions.
It is not hard to see,  for instance from the theory of
reduction by constraints,
that ${\bf X}_H$ takes the form
$$
{\bf X}_H(J) = [K_H,J] - K_H',
\qquad
K_H={\delta H\over \delta J} + B_H,
\eqno(B.4)
$$
where $[B_H, J] - B_H'$ is a linear combination
of the Hamiltonian vector fields associated to the (``second class'')
constraints that define the above special form of $J$.
The polynomial nature of the DS gauge ensures (see e.g.~\q{\FORTW}) that
once $J$ and ${\delta H \over \delta J}$ are given one can
 uniquely solve (B.4)
(where the form of ${\bf X}_H(J)$ must be consistent with that
of $J$ in (B.1))
for $B_H$, ${\bf X}_H$   in terms of differential
polynomial expressions in the components of $J$ and
${\delta H\over \delta J}$.

Let us determine in turn  ${\bf X}_Q$, ${\bf X}_P$, ${\bf X}_R$.
Inspecting the simplest examples leads us to search
for $K_Q$ in the form
$$
K_Q = \sum_{i=1}^r e_{r+1,r+1-i}\otimes f_i=\left(\matrix{
0&\cdots&0&0\cr
\vdots&\cdots&\vdots&\vdots\cr
0&\cdots&0&0\cr
f_r&\cdots&f_1&0\cr}\right),
\qquad {\hbox{with}}\quad f_1=f.
\eqno(B.5)
$$
Substituting from  $(B.5)$ in $(B.4)$ we get
$$
{\bf X}_Q(w)=f \zeta_+,\quad
{\bf X}_Q(\zeta_+)=0,\quad
{\bf X}_Q(\zeta_-)=fv_r-\D\left(f_r\right),\quad
{\bf X}_Q(v_i)=-\zeta_+f_i,
\eqno(B.6)
$$
and the recursion relation
$$
f_{i+1} =  fv_i\Delta - \D(f_i)\Delta ,
\eqno(B.7)
$$
where $\D(\beta)$ is given by $\D(\beta) = \beta' + w\beta$ for any
$\beta\in \widetilde{ {\rm mat}}(s\times p)$.
The solution of this recursion relation is found to be
$$
f_i = (-1)^{i-1}\D^{i-1}(f)\Delta ^{i-1}\, -
\, \sum_{k=1}^{i-1}(-1)^{i-k}\D^{i-1-k}(fv_k)\Delta^{i-k}.
\eqno(B.8)
$$
Substitution from $(B.8)$ in $(B.6)$  gives ${\bf X}_Q$
explicitly.

To do the analogous computation for $H=P$, it is advantageous to change
variables by transforming  to a different gauge section.
In fact, there exists a unique gauge transformation
$$
J\mapsto\tilde J = gJg^{-1} - g'g^{-1},\quad
g =
\left(\matrix{{\bf A}&0\cr 0&{\bf 1}_s}\right),
\eqno(B.9)
$$
where ${\bf A}$ is a block lower triangular matrix
with $p\times p$ unit matrices along the diagonal,
for which $\tilde J$ takes the form
$$\tilde J =\sum_{i=1}^re_{i,1}\otimes\tilde v_i + e_{r,r+1}\otimes \zeta_+
+ e_{r+1,1}\otimes \zeta_- + e_{r+1,r+1}\otimes w + C_+.
\eqno(B.10)$$
The $\tilde v_i$'s in (B.10)  are unique differential polynomials
in  the $v_i$'s in (B.1), which
may be determined from the  equality
$$
\Delta^r\pa^r + \sum_{i=1}^r u_i\pa^{r-i} + z_+\paw z_- = L =
\Delta^r\pa^r + \sum_{i=1}^r \pa^{r-i}\tilde u_i + z_+\paw z_-
\eqno(B.11)
$$
using the notations
$$
z_+=-\Delta \zeta_+,\quad z_-=\zeta_-,\quad
u_i=\Delta v_i\Delta^{r-i},
\quad
\tilde u_i
= \Delta^{r+1-i}\tilde v_i.
\eqno(B.12)$$
This equality  results from the elimination procedure
performed in the respective gauges (B.1) and (B.10).
In the latter gauge
we have a formula  for ${\bf X}_P$  precisely analogous to $(B.4)$,
$$
{\bf X}_P(\tilde J) = [\tilde K_P,\tilde J] - \tilde K_P',
\qquad
\tilde K_P={\delta P\over \delta \tilde J} + \tilde B_P,
\eqno(B.13)
$$
and $\tilde K_P$ turns out to have the form
$$
\tilde K_P =\sum_{i=1}^r e_{i,r+1}\otimes \varphi_i= \left(\matrix{
0&\cdots&0&\varphi_1\cr
\vdots&\vdots&\vdots&\vdots\cr
0&\cdots&0&\varphi_r\cr
0&\cdots&0&0\cr}\right),\qquad {\hbox{with}}\quad \varphi_1=\varphi .
\eqno(B.14)
$$
In fact,  substituting from $(B.14)$ in $(B.13)$ leads to
$$
{\bf X}_P(w)=-\zeta_-\varphi,\quad
{\bf X}_P(\zeta_+)=-\tilde\D(\varphi_r)-\tilde v_r\varphi,\quad
{\bf X}_P(\zeta_-)=0,\quad
{\bf X}_P(\tilde v_i)=\varphi_i \zeta_-
\eqno(B.15)
$$
with the recursion relation
$$
\varphi_{i+1} = \Delta\left(\tilde\D(\varphi_i)+\tilde v_i\varphi\right),
\eqno(B.16)
$$
where $\tilde\D(\tilde \beta )$ is given  by $\tilde\D(\tilde \beta ) =
\tilde \beta' - \tilde \beta w$
for any  $\tilde \beta \in\widetilde{{\rm mat}}(p\times s)$.
This yields
$$
\varphi_i = \Delta^{i-1}\tilde\D^{i-1}(\varphi)\, +\,
\sum_{k=1}^{i-1}\Delta^{i-k}\tilde\D^{i-k-1}(\tilde v_k\varphi).
\eqno(B.17)
$$
Plugging  this back into $(B.15)$ gives ${\bf X}_P$
explicitly.

In the case of $H=R$ we find that $K_R$
equals ${\delta R\over \delta J}$, i.e.,
$K_R=e_{r+1,r+1}\otimes\alpha$.
Therefore
$$
{\bf X}_R(w)=[\alpha,w]-\alpha',
\quad{\bf X}_R(\zeta_+)=-\zeta_+\alpha,
\quad{\bf X}_R(\zeta_-)=\alpha \zeta_-,
\quad{\bf X}_R(v_i)=0.
\eqno(B.18)$$

\smallskip

The remaining task is  to find a neater form of the above formulae.
First  we rewrite them in terms of  the operator $\ell$ in (B.2) as follows.

\noindent
{\bf Claim B1}:
Formula $(B.6)$ with $(B.8)$ is equivalent to
$$
{\bf X}_Q(\ell) = \left(\Delta\zeta_+\paw f\Gamma\ell\right)_+,\,\,\,
{\bf X}_Q(\zeta_+) = 0,\,\,
{\bf X}_Q(\zeta_-) = -\D (f_r) + fv_r,\,\,
{\bf X}_Q(w) = f\zeta_+.
\eqno(B.19)
$$
\noindent
{\bf Claim B2}:
Formula $(B.15)$ with $(B.17)$ is equivalent to
$$
{\bf X}_P(\ell)=\left(\ell\varphi\paw \zeta_-\right)_+,\quad
{\bf X}_P(\zeta_+)=-\tilde\D (\varphi_r)-\tilde v_r\varphi,\quad
{\bf X}_P(\zeta_-)=0,\,\,\,
{\bf X}_P(w)= -\zeta_-\varphi.
\eqno(B.20)
$$
\noindent
{\bf Claim B3}:
Formula $(B.18)$ is equivalent to
$$
{\bf X}_R(\ell)=0,
\quad{\bf X}_R(\zeta_+)=-\zeta_+\alpha,
\quad{\bf X}_R(\zeta_-)=\alpha \zeta_-,
\quad{\bf X}_R(w)=[\alpha,w]-\alpha'.
\eqno(B.21)$$

Claim  B3 is obvious from (B.18).
Claim B1  may be verified as follows.
We use our usual trick  to write
$$
\pa + w = W^{-1}\pa W\qquad{\hbox{with}}\qquad w=W^{-1}W'.
\eqno(B.22)
$$
Then we have
$$
-\Delta\zeta_+\paw f\Gamma\ell =
\Delta\zeta_+W^{-1}\pa^{-1}Wf\Delta^{r-1}\pa^r\,  -\,
\Delta\sum_{i=0}^{r-1}\zeta_+W^{-1}\pa^{-1}Wf\Gamma u_{r-i}\pa^i.
\eqno(B.23)$$
Using  the identity $\D(F)=W^{-1}(WF)'$ and the formula
$\pa^{-1}F = \sum_{i=0}^\infty (-1)^i F\pa^{-i-1}$
we can write
the  contribution to  non-negative powers of $\pa$ in
the right hand side of (B.23) in
the form
$$
\Delta\zeta_+\sum_{k=0}^{r-1}(-1)^k\D^k(f)\Delta^{r-1}\pa^{r-k-1}\, -\,
\Delta\zeta_+\sum_{i=1}^{r-1}\sum_{k=0}^{i-1}(-1)^k\D^k(f\Gamma
u_{r-i})\pa^{i-k-1}
\eqno(B.24)
$$
which simplifies to
$$
\sum_{k=1}^r \Delta \zeta_+f_k\Delta^{r-k}\pa^{r-k}=
-\sum_{k=1}^r {\bf X}_Q(u_k) \pa^{r-k},
\eqno(B.25)$$
as required by Claim B1.
Similarly, one may check Claim B2  using the
identity $\tilde\D\left(\tilde F\right) = (\tilde FW^{-1})'W$
and the expression of  $\ell=L_+$ in the variables
$\tilde u_i$ provided by (B.11) and (B.12).

Let us now consider the Hamiltonian $H=H_1+Q+P+R$ given by
$$
H(J) = \integral\tr\,{\rm res}\left(\ell\xi\right)\, +\,
\integral\tr\left( \zeta_+ f\right)\, +\,
\integral\tr\left( \zeta_-\varphi\right)\, +\,
\integral\tr \left(w\alpha\right),
\eqno(B.26)
$$
where $\xi$ is  a $p\times p$ matrix PDO of the form
$\xi=\sum_{i=1}^r \xi_i \pa^{i-r-1}$
with arbitrarily chosen  $\xi_i\in\widetilde{gl}_p$.
We wish to present the Hamiltonian vector field ${\bf X}_H$
associated to $H$ by means of the reduced current algebra PB
in terms of the variables $\ell$, $w$, and $z_\pm$ in (B.12).
To make contact with formula (4.34) of Sect.~4, we  now substitute
$$
f=-{\delta H\over \delta z_+} \Delta,
\quad
\varphi={\delta H\over \delta z_-},
\quad
\alpha={\delta H\over \delta w},
\quad
\xi={\delta H\over \delta \ell}.
\eqno(B.27)$$
Combining the above claims with Theorem 4.2 of Sect.~4 implies
the following formula:
$$\eqalign{
&{\bf X}_H (\ell ) = (\ell {\delta H\over \delta \ell})_+\ell
- \ell({\delta H\over \delta \ell} \ell)_+
+ \left(\ell{\delta H\over \delta z_-}\paw z_-\right)_+
- \left(z_+\paw {\delta H\over \delta z_+}\ell\right)_+,\cr
&{\bf X}_H(z_+)= P_0\left(\ell{\delta H\over \delta \ell} z_+W^{-1}\right)W
+\Delta^{r}\tilde\D^{r}\left({\delta H\over \delta z_-}\right)
+\sum_{k=1}^r \tilde\D^{r-k}\left(\tilde u_k {\delta H\over \delta z_-}\right)
- z_+ {\delta H\over \delta w},\cr
&{\bf X}_H(z_-)=
 -W^{-1}P_0^{\dagger}\left(Wz_-{\delta H\over \delta \ell} \ell\right)
-(-1)^r\D^r\left({\delta H\over \delta z_+}\right) \Delta^r \cr
&\qquad\qquad\qquad\qquad\qquad\qquad\qquad\quad
-\sum_{k=1}^r (-1)^{r-k}\D^{r-k}\left({\delta H\over \delta z_+}u_k\right)
 + {\delta H\over \delta w}  z_-,\cr
&{\bf X}_H(w) = {\delta H\over \delta z_+} z_+ -
 z_- {\delta H\over \delta z_-} + \left[{\delta H\over \delta w} , w\right]
 - \left({\delta H\over \delta w}\right)'.\cr}
\eqno(B.28)$$
Here the first term in ${\bf X}_H(z_\pm)$ has been found
using the skew-symmetry property of the PB.
Recall that the notations $P_0$ and $P_0^\dagger$ are defined
as follows.
For any PDO $\chi$,
expand $\chi$ so that all powers of $\pa$ are on the right, then
$P_0(\chi)$ is the coefficient of $\pa^0$; alternatively writing $\chi$
so that all powers of $\pa$ are on the left, $P_0^\dagger(\chi)$
is the coefficient of $\pa^0$.
This definition allows to convert (B.28)  into formula (4.34)
given in Theorem 4.4, which completes our derivation of this formula.
The reader may  derive (4.35) from  (4.34)
by computing the  Lie derivative mentioned  in the proof of
Theorem 4.4.
We only note that, like in the above,  it is advantageous to
make use of the linear functions of $J$  for this purpose.

\vfill\eject

\bigskip
\centerline{\bf References}
\medskip

\bibitem{\FHM}
L.\ Feh\'er, J.\ Harnad, I.\ Marshall,
{\it
Generalized Drinfeld-Sokolov reductions and KdV type hierarchies},
Commun.\ Math.\ Phys.\ {\bf 154} (1993) 181-214.

\bibitem{\GD}
I.M.\ Gelfand, L.A.\ Dickey,
{\it Fractional powers of operators and Hamiltonian systems},
Funct.\ Anal.\ Appl.\ {\bf 10:4} (1976) 13-29;
{\it The resolvent and Hamiltonian systems},
Funct.\ Anal.\ Appl.\ {\bf 11:2} (1977) 93-104.

\bibitem{\Man}
Yu.I.\ Manin,
{\it Algebraic aspects of nonlinear differential equations,}
Jour. Sov. Math. {\bf 11} (1979) 1-122.

\bibitem{\DS}
V.G.\ Drinfeld, V.V.\ Sokolov,
{\it Equations of Korteweg-de Vries type and simple Lie algebras},
Sov.\ Math.\ Dokl.\ {\bf 23} (1981) 457-462;
{\it Lie algebras and equations of KdV type},
J. Sov. Math. {\bf 30} (1985) 1975-2036.

\bibitem{\Wil}
G.\ Wilson,
{\it The modified Lax and two-dimensional Toda lattice equations
associated with simple Lie algebras},
Ergod.\ Th.\ and Dynam.\ Sys.\ {\bf 1} (1981) 361-380.

\bibitem{\McI}
I.R.\ McIntosh,
{\it An Algebraic Study of Zero Curvature Equations},
PhD Thesis, Dept.\ Math., Imperial College, London, 1988 (unpublished).

\bibitem{\GHM}
M.F.\ de Groot, T.J.\ Hollowood,  J.L.\ Miramontes,
{\it Generalized Drinfeld-Sokolov hierarchies},
Commun.\ Math.\ Phys.\ {\bf 145} (1992) 57-84.

\bibitem{\BGHM}
N.J.\ Burroughs, M.F.\ de Groot, T.J.\ Hollowood, J.L.\ Miramontes,
{\it Generalized Drinfeld-Sokolov hierarchies II:
 The Hamiltonian structures},
Commun.\ Math.\ Phys.\ {\bf 153} (1993) 187-215.

\bibitem{\FGMG}
C.R.\ Fern\'andez-Pousa, M.V.\ Gallas, J.L.\ Miramontes, J.S.\ Guill\'en,
{\it ${\cal W}$-algebras from soliton  equations
and Heisenberg subalgebras},
Santiago de Compostela preprint, US-FT/13-94, hep-th/9409016.

\bibitem{\RSTS}
A.G.\ Reyman, M.A.\ Semenov-Tian-Shansky,
{\it Compatible Poisson structures for Lax equations: An $r$-matrix
approach},
Phys.\ Lett.\ {\bf A130} (1988) 456-460.

\bibitem{\FORT} L.\ Feh\'er, L.\ O'Raifeartaigh, P.\ Ruelle, I.\ Tsutsui,
{\it On the completeness of the set of classical $\cal W$-algebras
obtained from DS reductions},
Commun.\ Math.\ Phys.\ {\bf 162} (1994) 399-431.

\bibitem{\Cheng}
Yi Cheng,
{\it Constraints of the Kadomtsev-Petviashvili hierarchy,}
J.\ Math.\ Phys.\ {\bf 33} 3774-3793.

\bibitem{\OS}
W.\ Oevel, W.\ Strampp,
{\it Constrained KP hierarchy and bi-Hamiltonian structures},
Commun.\ Math.\ Phys.\ {\bf 157} (1993) 51-81.

\bibitem{\Di}
L.A.~Dickey,
{\it On the constrained KP hierarchy}, Oklahoma preprint, hep-th/9407038.

\bibitem{\BEHHH}
R.\ Blumenhagen, W.\ Eholzer, A.\ Honecker, K.\ Hornfeck, R.\ H\" ubel,
{\it Coset realization of unifying ${\cal W}$-algebras,}
Bonn preprint, BONN-TH-94-11, hep-th/9406203.

\bibitem{\Kac}
V.G.\ Kac, {\sl Infinite Dimensional Lie Algebras},
Cambridge University Press, Cambridge, 1985.

\bibitem{\KP}
V.G.\ Kac, D.H.\ Peterson,
{\it 112 constructions of the basic representation of
the loop group of $E_8$},
pp.~276-298 in:~Proc.\ of Symposium on Anomalies, Geometry and Topology,
W.A.\ Bardeen, A.R.\ White (eds.),
World Scientific, Singapore, 1985.

\bibitem{\Sp}
T.A.\ Springer,
{\it Regular elements of finite reflection groups,}
Inventiones math.\ {\bf 25} (1974) 159-198.

\bibitem{\DF}
F.\ Delduc, L.\ Feh\' er,
{\it Conjugacy classes in the Weyl group admitting a regular eigenvector
and integrable hierarcies,}
Lyon preprint, ENSLAPP-L-493/94, hep-th/9410203;
revised version to appear.

\bibitem{\FORTW}
L.\ Feh\'er, L.\ O'Raifeartaigh, P.\ Ruelle, I.\ Tsutsui, A.\ Wipf,
{\it
On Hamiltonian reductions of the Wess-Zumino-Novikov-Witten theories},
Phys.\ Rep.\  {\bf 222(1)} (1992) 1-64.

\bibitem{\BTvD}
F.A.\ Bais,  T.\ Tjin, P.\ van Driel,
{\it Covariantly coupled chiral algebras},
Nucl.\ Phys.\ {\bf B357} (1991) 632-654.

\bibitem{\A}
M.\ Adler,
{\it On a trace functional for formal pseudo-differential operators
and the symplectic structure of the Korteweg-de Vries equations},
Invent.\ Math.\ {\bf 50:3} (1979) 219-248.

\bibitem{\Dic}
L.A.\ Dickey,
{\it Soliton Equations and Hamiltonian Systems},
Adv.\ Ser.\ Math.\ Phys., Vol.\ 12,
World Scientific, Singapore,  1991.

\bibitem{\De}
A.\ Deckmyn,
{\it On the generalized Miura transformation},
Phys.\ Lett.\ {\bf B298} (1993) 318-328.

\bibitem{\FW}
M.D.\ Freeman, P.\ West,
{\it On the quantum KP hierarchy and its relation to the
non-linear Schr\" odinger equation,}
Phys.\ Lett.\ {\bf B295} (1992) 59-66.

\bibitem{\FK}
A.P.\ Fordy, P.P.\ Kulish,
{\it Nonlinear Schr\"odinger equations and simple Lie algebras},
Commun.\ Math.\ Phys.\ {\bf 89} (1983) 427-443.

\bibitem{\Wi}
G.\ Wilson,
{\it On two constructions of conservation laws for Lax equations,}
Q.\ J.\ Math.\ Oxford (2) {\bf 32} (1981) 491-512.

\bibitem{\Wils}
G.\ Wilson,
{\it Commuting flows and conservation laws for Lax equations},
Math.\ Proc.\ Camb.\ Phil.\ Soc.\ {\bf 86} (1979) 131-143.

\bibitem{\Che}
I.V.\ Cherednik,
{\it Differential equations for Baker-Akhiezer functions of algebraic curves,}
Funct.\ Anal.\ Appl.\ {\bf 12:3} (1978) 45-54.

\bibitem{\Fla}
H.\ Flaschka,
{\it Construction of conservation laws for Lax equations: Comments on
a paper of G.\ Wilson,}
Quart.\ J.\ Math.\ Oxford (2) {\bf 34} (1983) 61-65.

\bibitem{\BF}
J.\ de Boer, L.\ Feh\' er, work in progress.

\bibitem{\Car}
R.\W.\ Carter,
{\it Conjugacy classes in the Weyl group},
Comp.\ Math.\ {\bf 25} (1972) 1-59.

\bye